\documentclass[]{svjour3}

\usepackage{graphicx,aas_macros,natbib}
\usepackage{hyperref}                 
\usepackage{amssymb,amsmath}
\begin{document}

\title{Old-Aged Stellar Population Distance Indicators}
\author{Rachael L.~Beaton, Giuseppe Bono, Vittorio Francesco Braga, Massimo Dall'Ora, Giuliana Fiorentino, In~Sung Jang, Clara E. Mart{\'i}nez-V{\'a}zquez, Noriyuki Matsunaga, Matteo Monelli, Jillian R.~Neeley, and Maurizio Salaris}
\authorrunning{Beaton et al.}
\institute{Rachael L. Beaton 
\at Hubble Fellow
\at Department of Astrophysical Sciences, Princeton University, 4 Ivy Lane, Princeton, NJ~08544,
\at The Observatories of the Carnegie Institution for Science, 813 Santa Barbara Street, Pasadena CA 91101, \email{rbeaton@princeton.edu}
\and 
Giuseppe Bono \at Department of Physics, University of Rome Tor Vergata 
\at INAF-Osservatorio Astronomico di Roma,
\and
Vittorio Francesco Braga \at Department of Physics, University of Rome Tor Vergata 
\at ASDC
\and 
Massimo Dall'Ora \at INAF-Osservatorio Astronomico di Capdoimonte,
\and 
Giuliana Fiorentino \at INAF—OAS  Osservatorio di Astrofisica \& Scienza dello Spazio di Bologna,
\and 
In Sung Jang \at Leibniz-Institut für Astrophysic Potsdam, D-14482 Potsdam, Germany,
\and 
Clara E. Mart{\'i}nez-V{\'a}zquez \at Cerro Tololo Inter-American Observatory, National Optical Astronomy Observatory, Casilla 603, La Serena, Chile,
\and 
Noriyuki Matsunaga \at Department of Astronomy, School of Science, The University of Tokyo, Japan,
\and 
Matteo Monelli 
\at IAC- Instituto de Astrof\'isica de Canarias, Calle V\'ia Lactea s/n, E-38205 La Laguna, Tenerife, Spain 
\at
Departmento de Astrof\'isica, Universidad de La Laguna, E-38206 La Laguna, Tenerife, Spain
\and 
Jillian R. Neeley \at Department of Physics, Florida Atlantic University, 777 Glades Rd, Boca Raton, FL 33431
\and 
Maurizio Salaris \at Astrophysics Research Institute, Liverpool John Moores University
146 Brownlow Hill, L3 5RF Liverpool, UK
}
\maketitle

\abstract{Old-aged stellar distance indicators are present in all Galactic structures (halo, bulge, disk) and in galaxies of all Hubble types and, thus, are immensely powerful tools for understanding our Universe. 
Here we present a comprehensive review for three primary standard candles from Population II:
(i)~RR Lyrae type variables (RRL), (ii)~type II Cepheid variables (T2C), and (iii)~the tip of the red giant branch (TRGB). 
 The discovery and use of these distance indicators is placed in historical context before describing their theoretical foundations and demonstrating their observational applications across multiple wavelengths.
The methods used to establish the absolute scale for each standard candle is described with a discussion of the observational systematics.
We conclude by looking forward to the suite of new observational facilities anticipated over the next decade; these  have both a broader wavelength coverage and larger apertures than current facilities. 
We anticipate future advancements in our theoretical understanding and observational application of these stellar populations as they apply to the Galactic and extragalactic distance scale.}

\clearpage
\tableofcontents

%
\clearpage
\section{Introduction}\label{sec:intro}

Standard candles drawn from old stellar populations have a significant advantage for distance scale as, with the exception of young Galactic star clusters, old stellar populations are found in every galactic structural component (disk, halo, bulge), galaxies of all Hubble types, and galaxies of all luminosities (from ultra-faint to ultra-luminous). 
Most importantly, from these distance indicators it is possible to map both Galactic and extra-galactic objects using tracers pulled from the same underlying stellar population, if not the same class of star. 
Moreover, due to the presence of old stars in most structural components of galaxies, it is possible to study nearby galaxies in three dimensions (e.g., measuring depths, orientations, etc.) and then to evaluate if there systematics in mean distances due to these complex structures \citep[see e.g.,][]{kunder_2018}.
In turn, this helps us to better understand how projection effects and line-of-sight depth could bias mean distances. 
Thus, by virtue of being ``old'' standard candles have immense potential. 

The goal of this chapter is to provide a comprehensive review for the primary standard candles drawn from Population II stars. 
The term Population II (Pop~II) is an old one that originated from \citet{baade_1944} in which the nebulous central regions of the Andromeda, M\,32, and NGC\,205 were first resolved into individual stars. 
\citet{baade_1944} realized that these stars more closely resembled those in Galactic globular clusters (GGCs) than the ``slow moving'' stars in the solar neighborhood (e.g., disk stars). 
Later work would frame the differences as a function of age and metallicity, as 
\citet{baade_review} concisely summarized at the Vatican conference. 
Interestingly, the nature of the variable stars and their proper classification into Pop~I or Pop~II was intimately entwined \citep{baade58b}.
We focus our attention in this chapter to relatively luminous tracers that can be used for a broad range of distances and can be considered ``primary'', in that there exist some absolute calibrations using parallaxes for these standard candles from before the onset of progressive \emph{Gaia} data releases (DR1 and DR2 at the time of writing).
These considerations result in three distance indicators: (i)~the RR Lyrae (RRL), (ii)~the Type II Cepheids (T2C), and (iii)~the tip of the red giant branch (TRGB) stars.
The basic properties of these standard candles are given in Table~\ref{tab:overview}. 

Both the RRL and the T2C are pulsational variables that occur when specific Pop~II sequences cross the classical instability strip and these stars adhere to specific period--luminosity (PL) relationships from which distances can be determined to individual stars. 
In contrast, the stars that comprise TRGB are non-variable in nature\footnote{These stars likely have some intrinsic variability -- as most stars do, but it is on a much smaller scale than the pulsational variables that have amplitudes $\sim$0.3 to \textgreater1~mag.}
and as a result, the distance measurement is performed using a population of stars, which makes it statistical in nature. 

\begin{figure} 
\centering
\includegraphics[width=0.48\columnwidth]{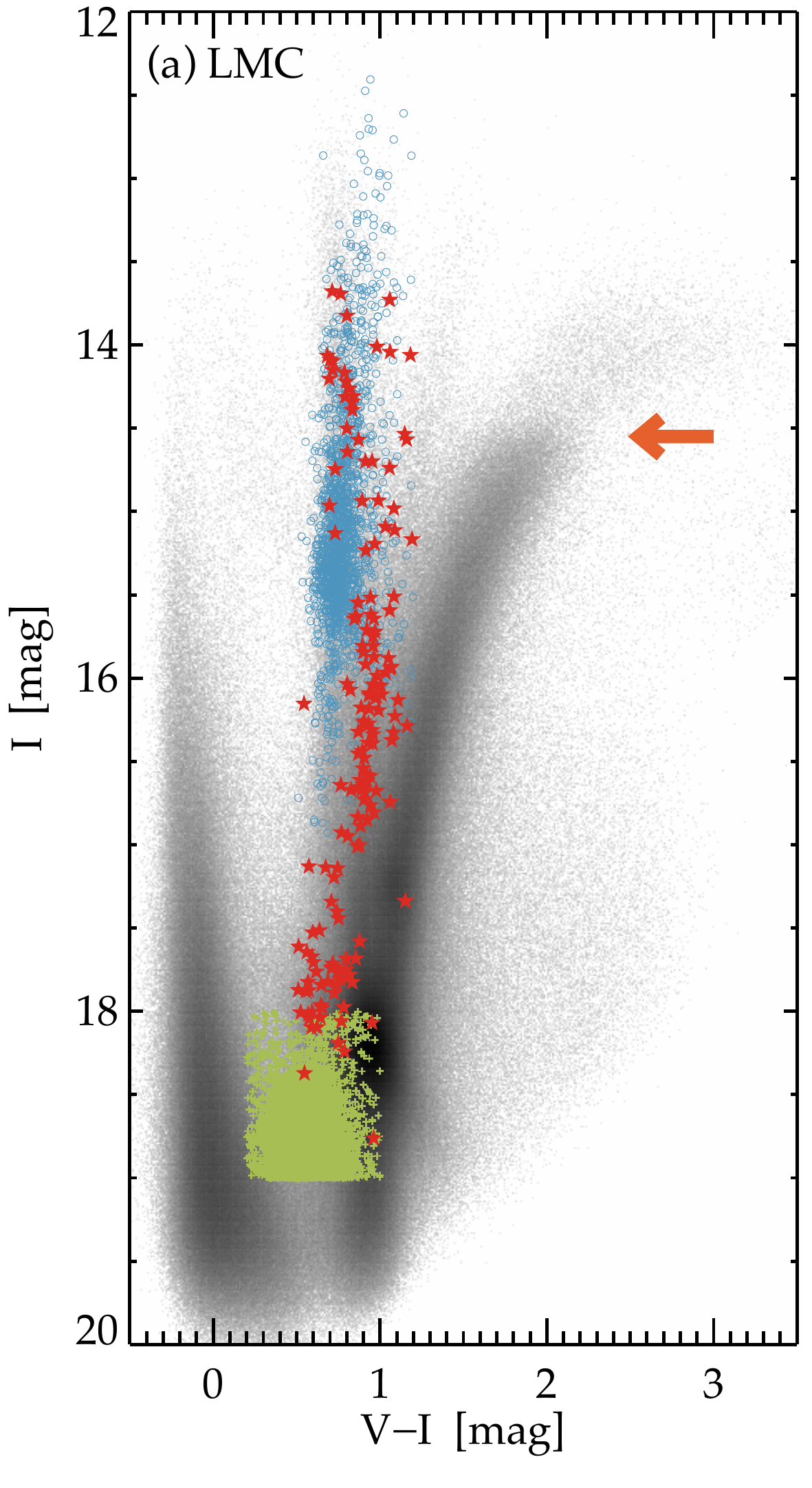}
\includegraphics[width=0.48\columnwidth]{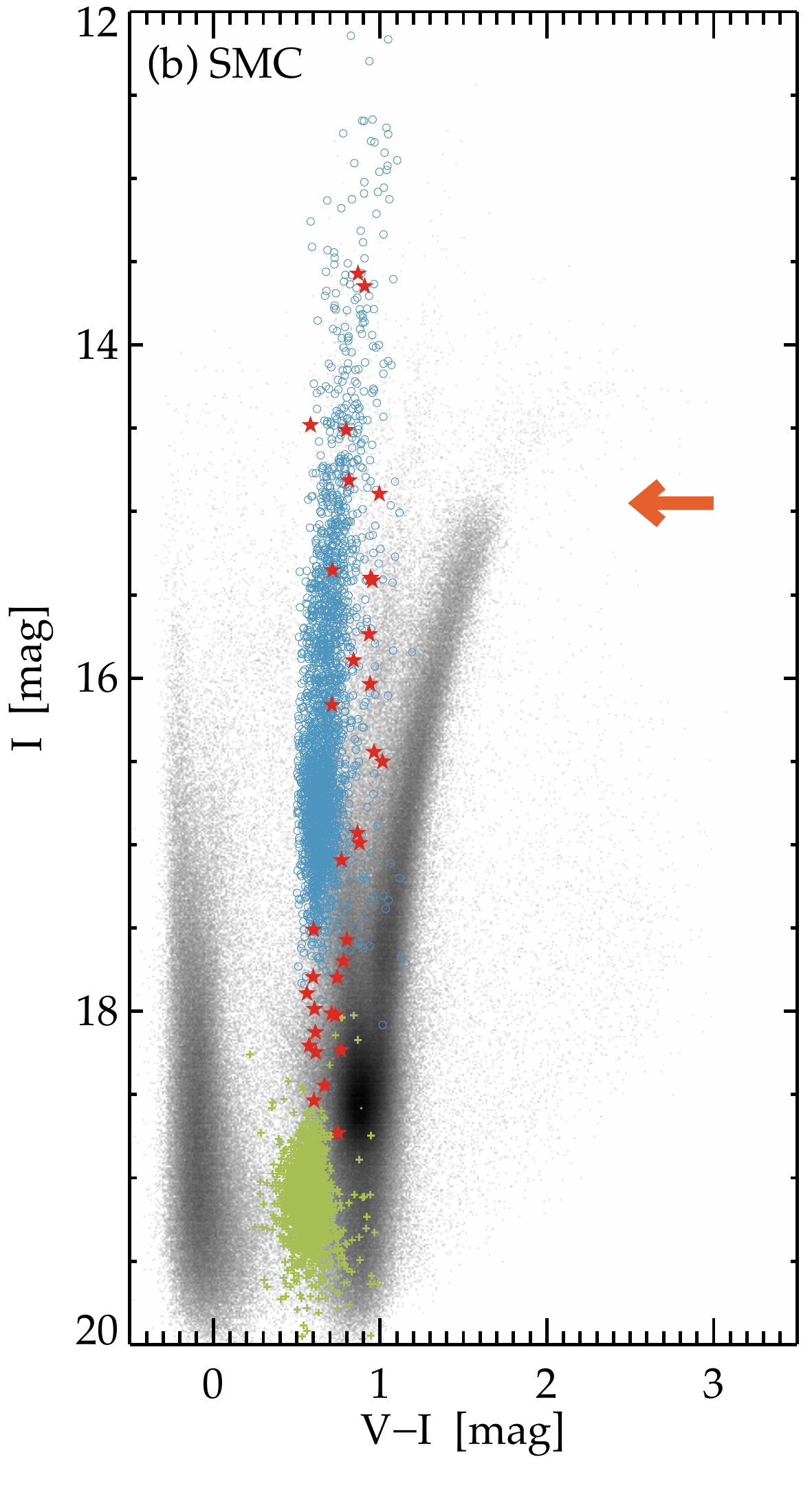}
\caption{ \label{fig:cmd} 
Optical color magnitude diagrams for the (a) LMC and (b) SMC from the Magellanic Clouds Photometric Survey \citep[black; MCPS][]{zaritsky_2002,zaritsky_2004} with variable star identifications from OGLE-III as follows RRLs (green pluses), T2Cs (red five-pointed stars), and the Classical Cepheids (blue open circles) over plotted \citep[][respectively]{soszynski_2016,Soszynski_2008,soszynski_2015}. 
The approximate location of the TRGB is indicated with a thick orange arrow. 
The axes for both panels are identical for ease of comparison. {\bf We note that no corrections for extinction (Galactic or internal to the LMC) have been applied and, thus, the color widths of the populations may be broader, to the red, than their intrinsic range.} }
\end{figure} 

Figures~\ref{fig:cmd}a and \ref{fig:cmd}b show the relative positions of the Classical Cepheids (blue), RRLs (green), T2Cs (red), and TRGB (orange arrow) for the Large and Small Magellanic Clouds (LMC, SMC) using variable star identifications from OGLE-III\footnote{The full variable star catalog can be queried here: \url{http://ogledb.astrouw.edu.pl/~ogle/OCVS/}} \citep[][]{Soszynski_2008,soszynski_2015,soszynski_2016} overplotted on a color magnitude density diagram from Magellanic Clouds Photometric Survey\footnote{Data is available here: \url{http://djuma.as.arizona.edu/~dennis/mcsurvey/}} \citep[greyscale][]{zaritsky_2002,zaritsky_2004}. 
Figures~\ref{fig:cmd}a and \ref{fig:cmd}b keenly demonstrate how the T2Cs could have been confused with the Classical Cepheids, as the two populations overlap in magnitude.
The three sub-classifications for the T2Cs are also visible by the distinct clumps in the LMC (Figure~\ref{fig:cmd}a) with these distinctions being less clear in the lower-metallicity SMC (Figure~\ref{fig:cmd}b). 
The difference in the population size for the three variable star classes is also apparent, with the T2Cs being less abundant than either the Classical Cepheids or the RRLs. 
Figures~\ref{fig:cmd}a and \ref{fig:cmd}b reinforce the mean magnitude differences given in Table \ref{tab:overview}, with the TRGB being more luminous than the bulk of the T2Cs and the T2Cs being brighter than the RRLs. 

As a class, the RRL have a long history in the optical, having been discovered in the 19th century in cluster diagrams, but recognition of their great potential in the infrared has come only recently \citep[most notably,][]{Longmore_1986}. 
The short period of RRLs make identifying them relatively simple with observations spanning only a few nights and with numerous Globular Clusters being sufficiently nearby for them to have been readily discovered by early photometric monitoring campaigns \citep[a detailed history is given in][]{smith_1995}.
In contrast, the T2Cs were only separated from the classical Cepheids in 1956 by \citeauthor{Baade_1956} and, despite being a solution to a difficulty in reconciling $H_0$ from the distance ladder and cosmological theory, have received little attention until the long-term monitoring from the OGLE project unveiled them {\it en masse} in the LMC \citep[Figure~\ref{fig:cmd}a,][]{Soszynski_2008}.
The recognition of the TRGB as a luminosity indicator came later still when the work of \citet{da90} provided high quality homogeneous photometry for a number of Galactic globular clusters (GGCs) transformed into absolute units by their RRL distances. 
The TRGB was first used to determine distances for galaxies by \citet{lee93}, who developed the analysis techniques necessary to detect the tip from color-magnitude diagrams.

Each of these distance indicators, thus, has a different volume of literature accompanying them and, as a result, have different depths of both theoretical understanding and observational applications. Often these vary not only by distance indicator, but also by wavelength regimes and objects (field stars, star clusters, galaxies) in which the techniques have been employed. Thus, the depth and breath of information provided in this review varies for each distance indicator.

\begin{table} 
\centering
\caption{Basic Properties of Population II Distance Indicators.\label{tab:overview}}
\begin{tabular}{cl cc c }
\hline \hline
Star     & Sub-Type & $M_V$ [mag] & $M_K$ [mag] & $P$ [days]  \\
\hline \hline
\multicolumn{5}{l}{{\sc RR Lyrae} (RRL)} \\ 
 & Fundamental Mode (RRab)    & $\sim$~+0.6 & $\sim$~-0.6 & 0.3~\textless~$P$ \textless~1.0  \\
 & First Overtone Mode (RRc)  & $\sim$~+0.6 & $\sim$~-0.4 & 0.2~\textless~$P$ \textless~0.5  \\
 \hline
\multicolumn{5}{l}{{\sc Type II Cepheids} (T2C)} \\ 
 & BL Herulis (BL~Her)  & $\sim$~-0.5 & $\sim$~-1.0 & 1~\textless $P$ \textless~4   \\
 & W Virginis (W~Vir)   & $\sim$~-1.0 & $\sim$~-4.0 & 4~\textless $P$ \textless~20 \\ 
 & RV Tauri (RV~Tau)    & $\sim$~-2.5 & $\sim$~-5.0 & 20~\textless $P$ \textless~80 \\
 \hline
\multicolumn{5}{l}{{\sc Tip of the Red Giant Branch Stars} (TRGB)} \\ 
 & Metal-Poor & $\sim$~-4.0 & $\sim$~-5.5 &    \\
 & Metal-Rich & $\sim$~-3.9 & $\sim$~-6.5 &   \\ 
\hline \hline
\end{tabular}
\end{table} 

Generally, a single book chapter cannot fully describe any one of these standard candles. 
Thus, we refer the reader to more detailed discussions that are in the literature.
Of particular note are the following books: \citet{smith_1995} on RRLs and \citet{catelan_2015} on pulsational variables of all kinds, including both RRLs and T2Cs. 
Additionally, \citet{prestonfest} is a set of online conference articles that present reviews of many aspects of RRL beyond those that will be discussed here, as well as other discussions relating to metal-poor, old stellar populations. 
\citet{cassisi_book} is an excellent resource on stellar evolution and stellar populations  that provides insight into all three of our distance indicators, but most especially the TRGB.
Lastly, \citet{beaton2016} provides comparison of RRL and TRGB methods with Cepheids in terms of the extragalactic distance scale that may help the reader understand the recent resurgence of interest Pop~II standard candles.

Our goal in this chapter is to place these Population II standard candles into the context of the distance scale by providing a sense of the current theoretical understanding and observational application of these tools. 
Where possible, we take a multi-wavelength approach discussing optical, near-infrared, and mid-infrared characteristics and applications.
In the sections that follow we discuss each of the distance indicators in turn,
 with parallel discussions of theory and practice for the RRL in Section~\ref{sec:rrl},
 a description and homogeneous PL relations for T2C in Section~\ref{sec:2cephs}, and both a physical description and application of the TRGB is given in Section~\ref{sec:trgb}.
The absolute scale for each distance indicator and inter-comparisons are described in Section \ref{sec:sys}.
We conclude in Section~\ref{sec:future} with an outlook for the future, in particular improvements to our physical understanding from \emph{Gaia} and the observational application with future large-aperture facilities.

\section{The RR Lyrae variables} \label{sec:rrl}

\begin{figure} 
\centering
\includegraphics[width=0.49\textwidth]{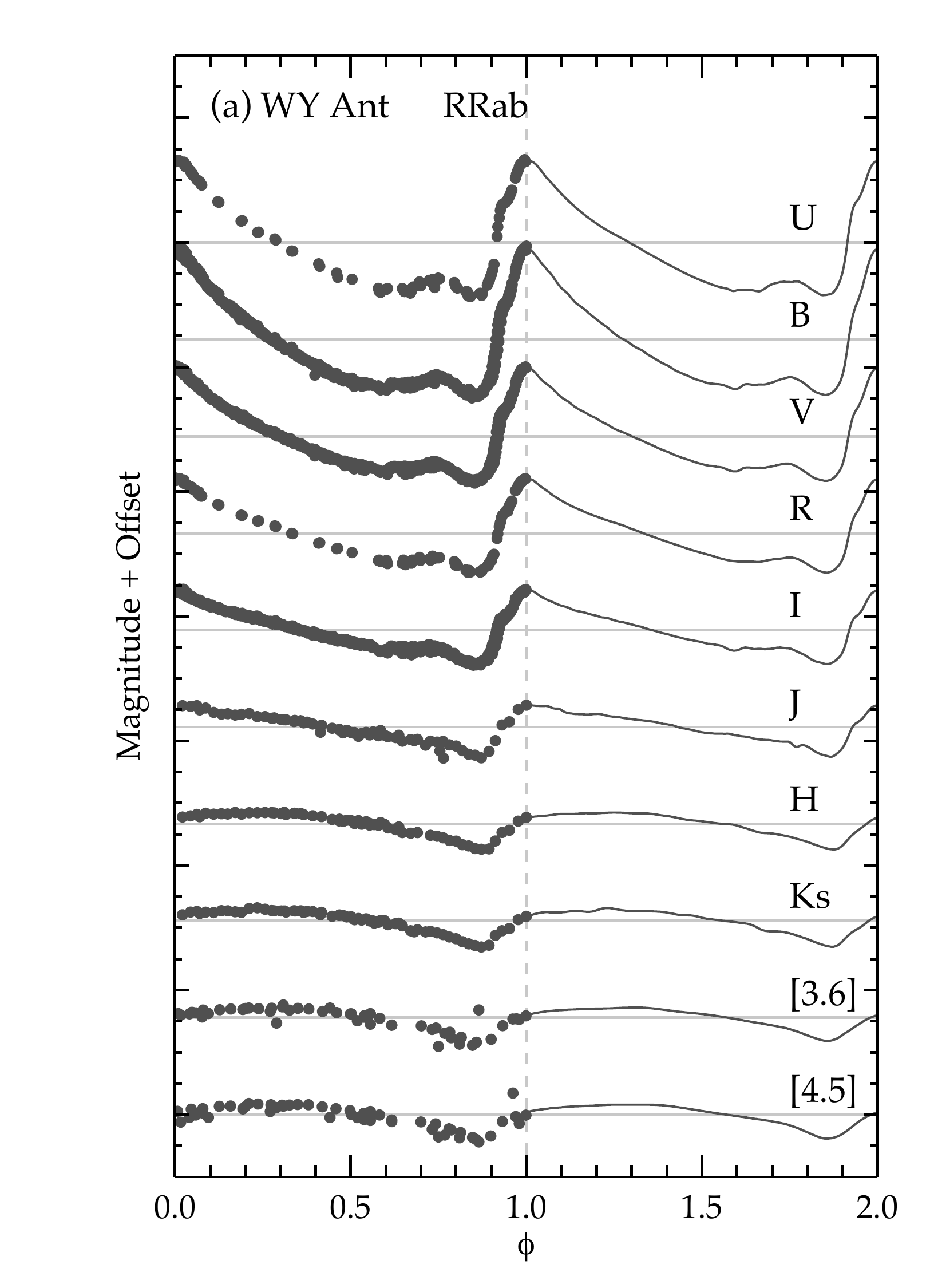}
\includegraphics[width=0.49\textwidth]{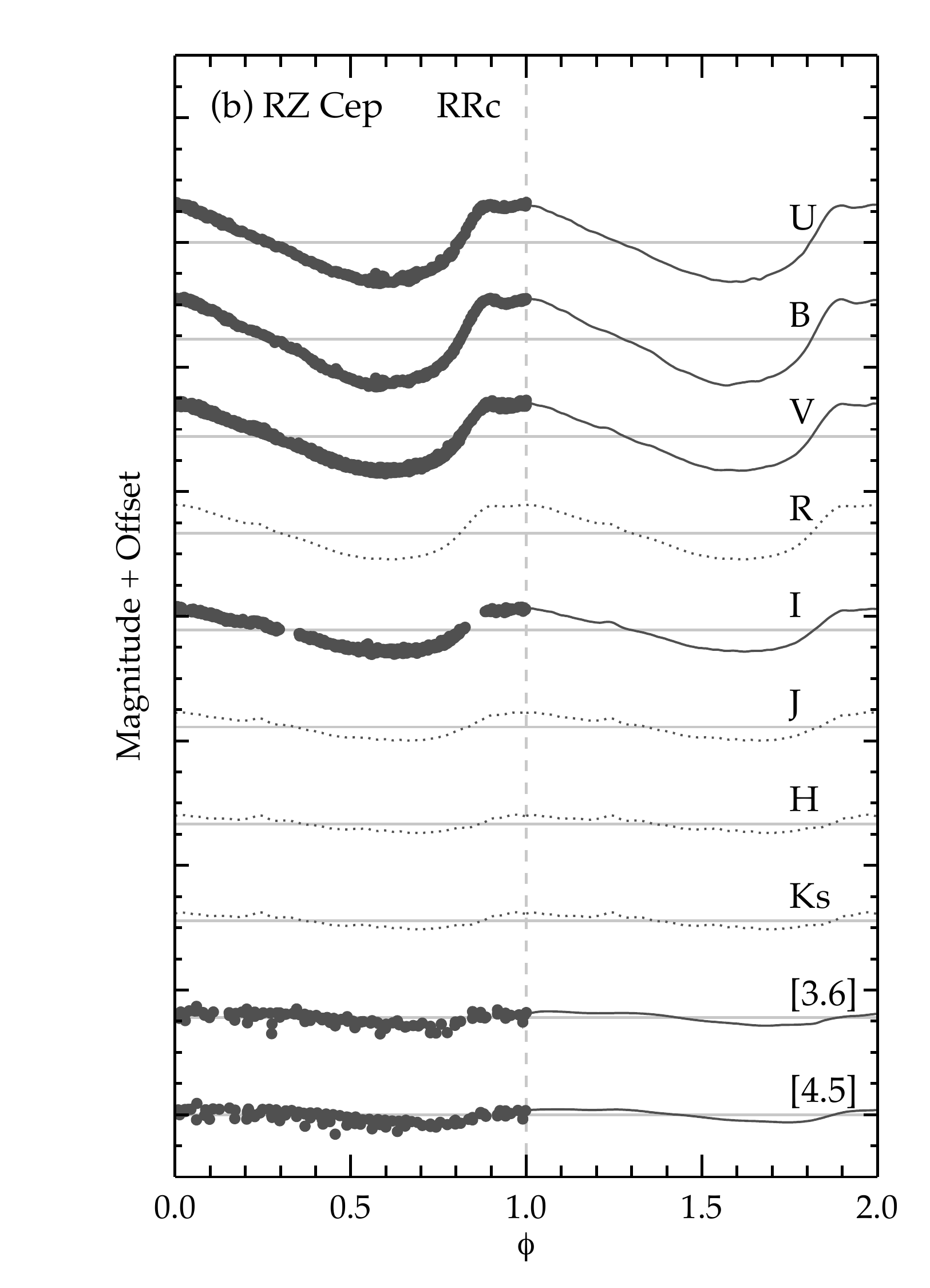}
\caption{\label{fig:lc}
 (a) Example multi-wavelength light curves for an RRab RRL, WY Ant and
 (b)Example multi-wavelength light curves for an RRc RRL, RZ Cep \citep[data from][]{monson2017}. 
 In each panel, the light curves are shown in ten photometric bands, which are from bottom to top $U$, $B$, $V$, $R$, $I_c$, $J$, $H$, $K$, $[3.6]$, and $[4.5]$. 
 The optical bands are the Johnson-Cousins system, the near-infrared bands are in the 2MASS system, and the mid-infrared bands are in the \emph{Spitzer}-IRAC system; a detailed discussion of the photometry homogenization is given in \citet{monson2017}. 
 In addition to the raw data points, a smoothed light curve is shown in black and a template light curve is shown in gray, if no data is present.
The horizontal line indicates the mean magnitude and the dashed vertical line indicates maximum light in the optical ($\phi=1$).
 }
\end{figure} 

The first variable in a GGC was discovered by Williamina Fleming and reported in \citet{pickering_1889}.
In 1893, Solon Bailey initiated a large scale program for imaging GGCs from the Harvard College Observatory in Arequipa, Peru with Williamina Fleming serving as his assistant. 
Very quickly several variables were discovered in the brightest GGCs in the Southern sky, in particular for $\omega$\,Centauri ($\omega$\,Cen).  
By the end of this project, Bailey had discovered over 500 variables in GGCs, which was equal in number to that found over the entire remainder of the sky!
A more detailed story of the discovery of RRLs is given in the introduction to \citet{smith_1995}.\footnote{We also refer the reader to \citet{sobel2016glass} for a description of the role of Williamina Fleming in these discoveries. More specifically, Williamina Fleming was first E.~C. Pickering's maid, became the first of the Harvard `computers,' and was involved in a large number of projects at the Harvard Observatory. Her accomplishments were honored when she became the first American woman elected to the Royal Astronomical Society.} 
RR Lyrae, itself, was discovered by W.~Flemming prior to 1899 and reported in \citet{pickering_1901}.

The observations reported in \citet{bailey_1902} define the nomenclature for RRLs that is used to this day.
RRLs come in two primary types; more specifically, those that pulsate in the fundamental mode (FU) known as RRab (a combination of the original Bailey types of a and b) and those that pulsate in the first overtone (FO) known as RRc (the Bailey type c stars).
The connection between the modes of pulsation and their Bailey types was first inferred by \citet{Schwarzschild_1940} using high quality photometric-plate light curves in the cluster M\,3; more specifically, \citet{Schwarzschild_1940} concluded that the the `c' stars needed to pulsate in a different mode to account for the strong deviation of this class from the period-density relation for the `a' and `b' type variables.
\citet{Schwarzschild_1940} further defined the color-edges of the instability strip concluding that no non-pulsating stars could exist in this temperature-luminosity range.
Additionally, there are the RRd type stars that show two pulsation modes and an ensemble of stars discovered from the OGLE program have more complex or atypical pulsation behavior \citep[e.g.,][and references therein]{soszynski_2016_mnras}.
The mean magnitudes of the RRL place them on the horizontal branch (HB) of GGCs, whereas at any given epoch, they will scatter above or below based on their amplitudes.

Some RRLs show an amplitude-modulation phenomenon known as the Blazhko effect \citep[first identified in][]{blazhko_1907} that currently has no consensus for its physical origin, though its impacts on the light curve are well documented.
A particularly interesting demonstration of the complexity of the Blazhko phenomenon can be found in \citet[][see also references therein]{chadid_2014} that presents data at very high cadence tracing a single star during Antarctic winter for 150 days.
The periods of the Blazhko effect can be from tens to hundreds of pulsation cycles for a given star \citep[see for instance, the study of][]{skarka_2014}.
The effect is well known and studied in the optical, but has recently been shown to persist into the $K$ band \citep{jurcsik_2018}.
The impact of the Blazhko effect on distances is that these stars have more uncertainty with respect to both their periods and mean magnitudes, with the impact being proportional to the level of amplitude-modulation. 
Thus, these stars, when identified, are not always used for PL fitting. 

\citet{bailey_1919} paved the road for use of RRLs as both tracers for old stellar population (Pop.~II) and as distance indicators. 
Subsequent investigations demonstrated that the RRL visual mean magnitude, within 
a given cluster, was nearly constant. 
The first strong evidence that RRL visual mean magnitude and the metallicity were correlated came in \citet{baade58c}, in which the populations of variables were compared between disk and halo star clusters and the Galactic Bulge.
Additional evidence accumulated over the next few years; in particular, (i) additional probes of the HB magnitude for GGCs with metallicity estimates by \citet{sandage_1960} and (ii) comparisons between variables found in more environments, in particular the Draco dwarf Spheroidal that showed both similarities and differences to stellar sequences in the GGCs \citep[][]{baade_swope_1961}.
Since those realizations, RRLs have been commonly used to estimate distances by means of a calibrated $V$-band magnitude versus metallicity relation \citep[e.g.,][and references therein]{sandage82}, subsequently rediscussed and calibrated several times, that is still widely used. 
The mean magnitude of RRLs in the $V$-band is nearly constant ($M_V \sim 0.6$ mag), with a dependence on their chemical composition, thus making RRLs solid standard candles.

Observations of RRLs in a number of GGCs supported the evidence that the topology of the Instability Strip (IS) changes with metallicity. 
In particular, the first overtone blue edge (FOBE) is virtually independent of the metallicity as originally suggested by linear radiative models \citep{baker65,cox63,iben71}. 
This finding and observations brought forward the opportunity to use the FOBE vs period relation as both a reddening and a distance indicator.
The theoretical scenario concerning the red edge of the IS was more complex, because it is caused by the increased efficiency of the convective transport when moving toward cooler effective temperatures. 
Pioneering nonlinear models that included a time dependent treatment of the convective transport \citep{deupree77b,stellingwerf82}, suggested that the red edge becomes cooler (redder) as the metallicity increases.  

The use of RRLs as distance indicators has had a quantum jump thanks to the empirical discovery by \cite{Longmore_1986} that RRLs obey a linear PL relation at near--infrared wavelengths. 
This discovery was later soundly supported by nonlinear convective models \cite{bono94a,bono01}. 
More recently, the advent of mid--infrared facilities on board of space telescopes, like \emph{WISE} and \emph{Spitzer}, led to the derivation of empirical PL relations at longer wavelengths (e.g., $\lambda \gtrsim$ than 3.6$\mu$m; Section~\ref{sec:mir}). 
Another popular diagnostic used to derive distances is the Wesenheit function \citep{vandenbergh75,madore_1982}, which is a reddening-free formulation of the less common PL--color relations (or PLC), to be discussed in Section~\ref{wes}. 

RRLs are readily recognizable from their light variation. 
Figure \ref{fig:lc} presents an example light curve for each of the two dominant sub-classes of RRL, an RRab (Figure \ref{fig:lc}a) and an RRc (Figure \ref{fig:lc}b) in ten photometric broadband filters from \citet{monson2017}.
Figure \ref{fig:lc} illustrates the large amplitudes and unique shapes in the optical for both types of stars, while also demonstrating how the amplitudes decrease strongly as a function of wavelength.
Indeed, the amplitude for RZ Cep is 0.1~mag in the IR compared to $\gtrsim$0.5~mag in the optical. 
The RRab stars, in particular, have a ``sawtooth'' shape in the optical, but become more sinusoidal at longer wavelengths as the impact of the temperature changes become less important. 
The RRc stars (right), in contrast, have a shape that changes comparatively little as a function of wavelength.

RRLs are nearly  ubiquitously present in Local Group galaxies.
Indeed, they have been identified in all the stellar systems that host an old ($\ge 10$ Gyr) stellar component. 
This evidence makes them excellent probes to investigate the structure and the old stellar populations at the early epochs of galaxy formation. 
They can be used to trace the components of our Galaxy (bulge, halo, thick disk) and to determine the distance and characterize the old population in Local Group (LG, distances within $\sim$1 Mpc) galaxies. 
Thus, RRLs provide a crucial first step to the extragalactic distance scale for Pop~II stellar systems, letting us control possible systematics affecting the commonly used distance scale based on classical Cepheids.

In the following sections, we focus our attention on the theoretical and semi--empirical background for the use of RRLs as distance indicators, with a special attention to some recent developments. 
Despite great progress in the determination of the RRL PL over the past decade, theoretical PL relations are often used for distance determination due to the lingering uncertainties associated with the absolute zero-point and, in particular, the effect of metallicity. Thus, the sections to follow describing theoretical efforts are quite detailed to motivate the strengths and weaknesses of the theoretical PLs.

\subsection{A physical description of RR Lyrae}\label{basic}

RRLs are radially pulsating low-mass ($0.6-0.8 M_\odot$) stars in their central helium-burning phase. 
The radial oscillations are an envelope phenomenon that takes place in a well--defined range of effective temperatures, therefore, they populate a relatively narrow region in the Hertzsprung-Russell diagram, which is the intersection between the so called {\it Cepheid} IS and the HB.
We will shortly describe the physical mechanisms driving the radial pulsation and we will analyze with some detail the different approaches to estimate distances using RRLs, highlighting their advantages and disadvantages.

The idea of a pulsating gaseous sphere was developed for the first time by \citet[][]{ritter_1879}\footnote{Several authors also cite Ritter's work from a series of papers published between 1878 and 1883 in Wiedemann's Annalen 5-20 that the authors of the present manuscript have been unable to find. Such references begin as early as \citet{shapley1914} and are cited as recently as \citet{smeyers_2010}.}, who found a simple relation between pulsation period and mean density, but it was only in \cite{shapley1914} that the radial pulsation hypothesis was advocated for Cepheid-like stars. 
The dispute between binarity and radial oscillations was settled in favor of the latter by the so-called Baade-Wesselink method.  
Radial pulsation is a phenomenon that involves the stellar envelope for certain values of the surface effective temperature (T$_{eff}$) and defines in the color magnitude diagram (CMD) a region in which the stars are unstable to pulsation, the IS. 
It is worth mentioning that a star radially pulsates only during its crossing(s) of the IS. 
As a result, the most numerous pulsators are those that have long evolutionary lifetimes within the IS, such as $\delta$ Scuti (central hydrogen burning) or RR Lyrae and Cepheids (central helium burning).

The physical mechanisms underlying this phenomenon are related to the cyclic variations of the opacity and of the equation of state for regions in which H and He are partially ionized, which are referred to as the $\kappa$ and $\gamma$ mechanisms.
The physics of radial oscillations was introduced by \cite{eddington26} and in the early models by \cite{cox1958} and \cite{zhevakin59}. 
Contrary to what happens in the rest of the envelope, these regions trap energy during the contraction and loose energy during the expansion, thus acting similar to a mechanical valve. 
The initial perturbation may originate from a stochastic fluctuation in the external layers of the envelope. 
The mechanical work can either be driven, if the envelope mass located on top of the ionization regions is large enough (below a critical effective temperature that defines the blue edge of the IS) and it can be quenched by the efficiency of convective transport that penetrates deep into the stellar envelope, toward effective temperatures lower than the red edge of the IS.

Simple linear adiabatic models, such as the one developed by \cite{eddington1918}, can predict the pulsation period and the pulsation mode, but they cannot predict the other observables such as the mean magnitudes, amplitudes, shapes of the light curve, and edges of the IS. 
This implies that non-adiabatic effects have to be considered to properly model the growth of the pulsational instability and the blue edge of the IS \citep{baker65,cox63,iben71}. 
The other fundamental pulsation observables, such as the pulsation amplitudes, the morphology of the light curves, and the topology of the instability strip (modal stability), can only be predicted with accuracy with the inclusion of non-linear terms in the hydrodynamical equations \citep{christy66,stellingwerf74} and by taking into account the coupling between pulsation and convection \citep{stellingwerf82,stellingwerf83,feuchtinger93}.

Current state-of-the-art models adopt a non-linear, time dependent formalism, which also takes into account the effects of convection and its non-linear coupling with the pulsation \citep[e.g.,][]{bono94b,bono97d,marconi03,dicriscienzo04,szabo04,Marconi15}. 
These models assume a spherical envelope (with no rotation or magnetic fields) and then solve the hydrodynamical equations; these are the conservation of mass and momentum plus a treatment for the radiative transfer that includes convection. 
These equations are solved as a function of time until they approach the limit cycle stability.
Although this is one of the most comprehensive approaches, we need to highlight that there is still room for improvement. In particular, the treatment of the convective transfer is non-local and time dependent. 
Despite this limitation, these models can account for all the observables, including the red boundary of the IS. Moreover, they appear capable of explaining more complicated observables including the double--mode pulsators and the Blazhko effect \citep{szabo14}.

For the sake of the subsequent discussion, we note that \citet{VanAlbada1971} formalized the relation between the pulsation periods and the structural parameters of mass ($M$), luminosity ($L$), effective temperature ($T_{eff}$), to which \citet{Marconi15} adds composition ($Z$). 
The relations take the following form:

  \begin{equation} \label{eq:pulsationeq} 
    \log P~=~\mathfrak{F} \left\{ \log(M/M_\odot),~\log(L/L_\odot),~\log T_{\rm eff},~\log Z \right \}.
   \end{equation} 

The most recent physical pulsation relations are given in \citet{Marconi15} and are as follows: 

  \begin{multline}\label{eq:fu} 
    \log P_F = 11.347 (\pm 0.006)~+~0.860 (\pm 0.003)~\log \left(\frac{L}{L_\odot}\right)~-~0.58 (\pm 0.02) \log \left(\frac{M}{M_\odot}\right) \\ 
-~3.43 (\pm 0.01)~\log T_{\rm eff}~+~0.024 (\pm 0.002)~\log Z 
   \end{multline} 

   \begin{multline} \label{eq:fo} 
     \log P_{FO} = 11.167 (\pm 0.006)~+~0.822 (\pm 0.004)~\log \left(\frac{L}{L_\odot}\right)~-~0.56 (\pm 0.02)~\log \left(\frac{M}{M_\odot}\right) \\ 
-~3.40 (\pm 0.03)~\log T_{\rm eff}~+~0.013 (\pm 0.002)~\log Z 
   \end{multline} 

\noindent for FU (Equation~\ref{eq:fu}) and FO pulsators (Equation~\ref{eq:fo}), respectively. 
These physical relations are the basis for all the theoretically-determined pulsation relations (e.g., those projected into observable quantities), which are discussed in the following subsections.

\begin{figure*} 
\centering 
	\includegraphics[width=\textwidth]{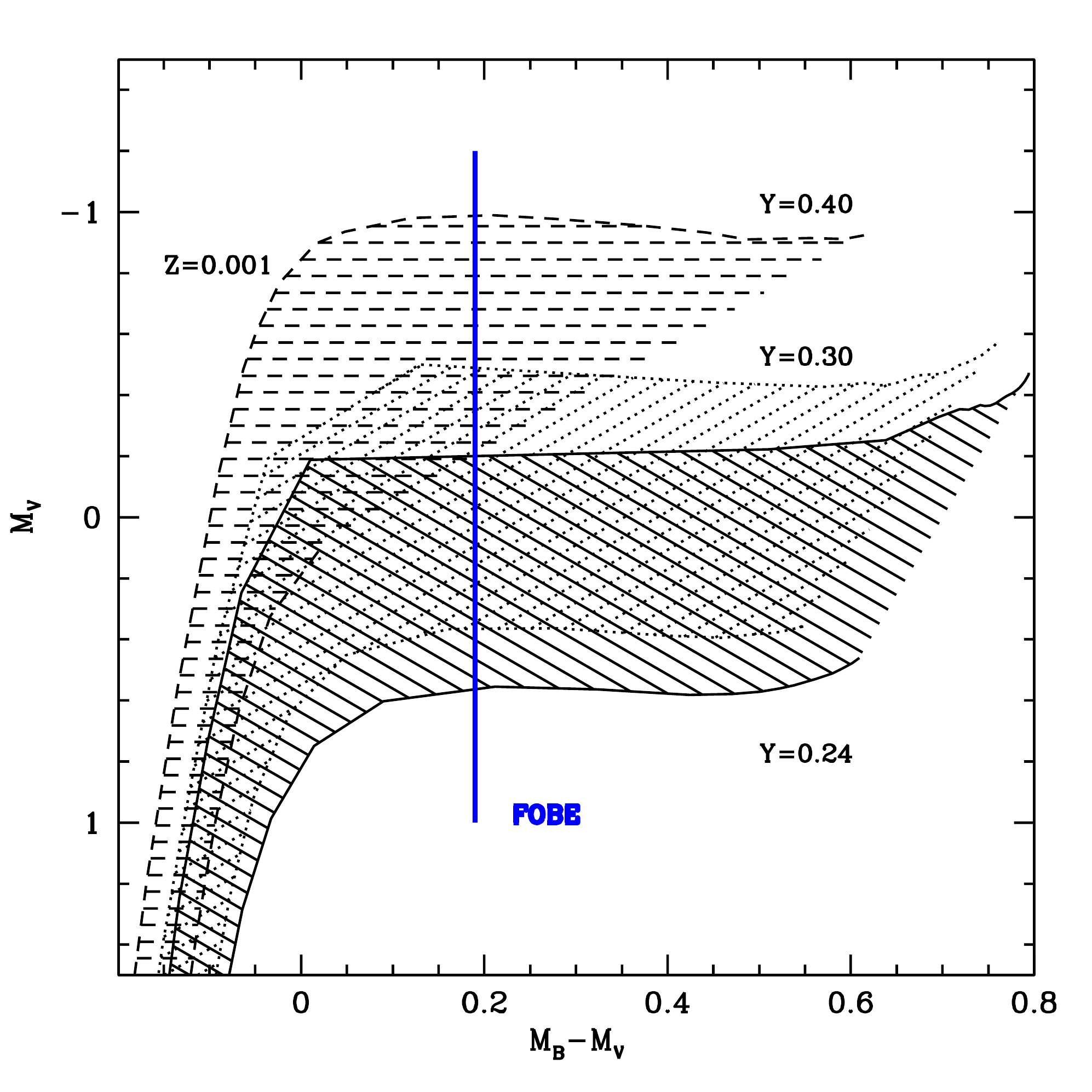}
	\caption{Theoretical HB models from in the color-magnitude diagram. Solid, dotted and dashed regions indicate different assumptions of the helium content for a fixed overall metallicity abundance. The vertical, blue line shows the color of the First Overtone Blue Edge (FOBE) that is almost constant for the metallicity range of the RRLs. All the theoretical models used here are taken from BaSTI \citep{basti}.}
	\label{fig1_rr_theory1}
\end{figure*}

We conclude here highlighting that recent theoretical investigations \cite[e.g.,][]{marconi11,marconi16} have shown that the pulsation properties of RRLs are also affected by the helium content. 
The Helium abundance impacts RRL evolutionary properties, like the total bolometric luminosity and the evolutionary timescales, which in turn impact the pulsation properties and the the interpretation of observed quantities.
Figure ~\ref{fig1_rr_theory1} shows theoretical HB models in color-magnitude space from ``a Bag of Stellar Tracks and Isochrones'' (BaSTI) \citep[most recently,][with earlier works referenced therein]{basti}\footnote{The models are publicly available: \url{http://basti.oa-teramo.inaf.it/}} for a fixed total metal content ($Z=0.001$), but with different Helium contents of $Y=0.24$ (solid hashing), $Y=0.30$ (dotted hashing), and $Y=0.40$ (dashed hashing).
The First Overtone Blue Edge (FOBE) is shown for reference.
From Figure~\ref{fig1_rr_theory1}, the impact on the mean absolute magnitude ($M_{V}$) is demonstrated, with those stars having greater Helium enrichment being systematically brighter and having a slightly different temperature distribution. 
Beyond this, at fixed metal abundance, the pulsation period is expected to increase 
and the pulsation amplitude to decrease as the helium abundance increases. 
The width of the IS, however, is minimally affected.

\subsection{RR Lyrae as Standard Candles}

This subsection focuses on the origin of various relationships that can be used to determine distances using RRL. First, optical relationships are explored, these being (i) the $M_{V}$-Metallicity relationship is explained in Section \ref{mvfeh} and (ii) the FOBE method in Section \ref{fobe}. Then the, relatively new, NIR and MIR relationships are explained in Section \ref{nirmir}. The multi-wavelength slopes are explored in Section \ref{pl_slopes}. Lastly, reddening-free relationships constructed from multi-band observations are explored in Section \ref{wes}.

\subsubsection{The visual magnitude--metallicity relation}\label{mvfeh}

\begin{figure} 
\includegraphics[width=\columnwidth]{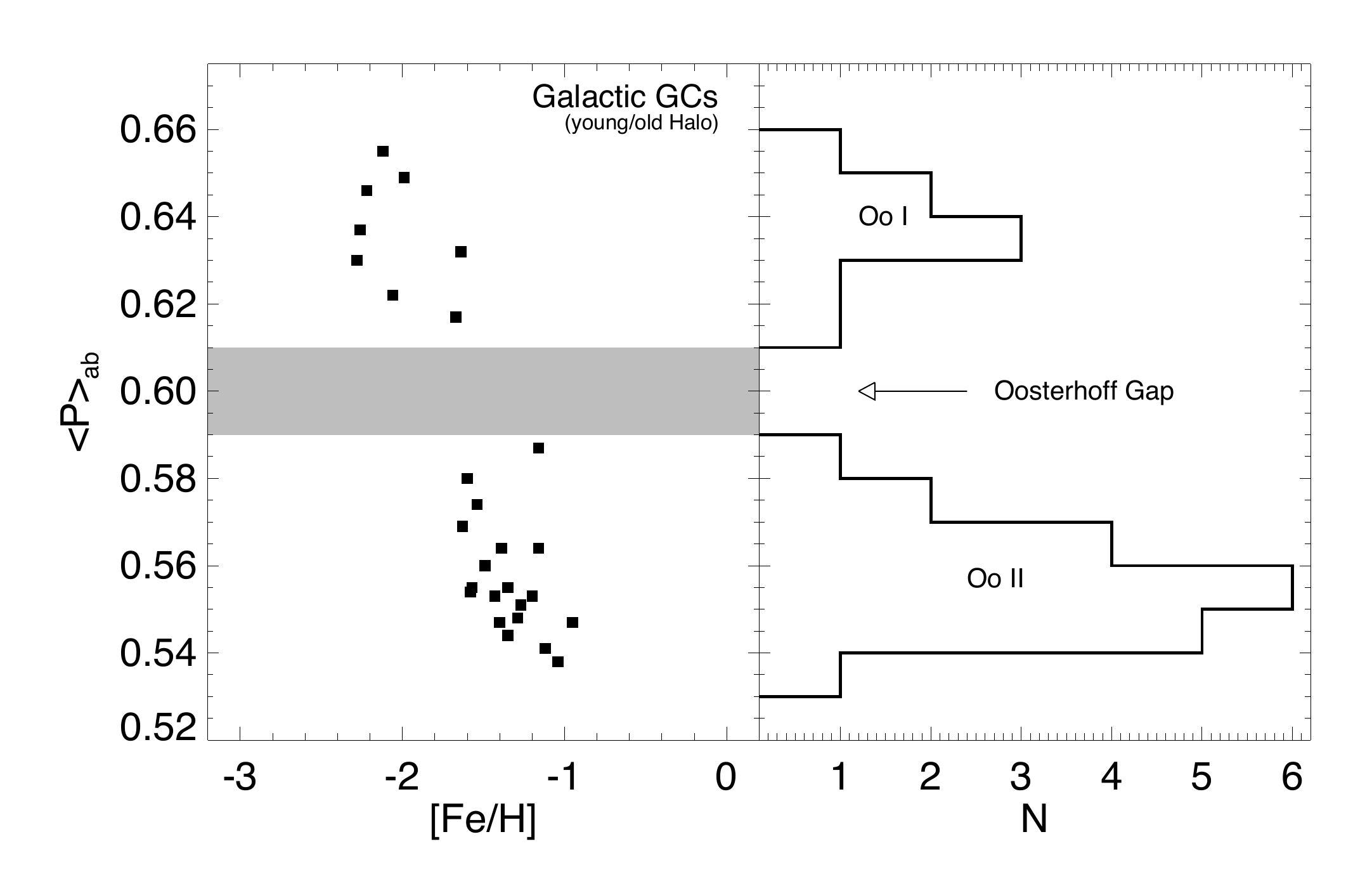}
\caption{ \label{fig:oo} 
Demonstration of the Oosterhoff classification for GGCs with more than ten RRab using the compilation of \citet[][]{catelan_2009}. Both ``young'' and ``old'' halo clusters have been included in this visualization.
\emph{Left panel:} The distribution of the mean period ($\langle$P$\rangle_{ab}$) against the metallicity ([Fe/H]) of the cluster. 
\emph{Right panel:} Histogram of of the mean periods that illustrates the distinct separation between the Oo-I and Oo-II type clusters. The lack of clusters with $\langle$P$\rangle_{ab}\sim$ 0.60 days is known as the Oosterhoff gap.}
\end{figure}

The original idea of a relation between the mean luminosity of RRLs and their metallicity dates back to \cite{baade55} and \cite{sandage58}, who discovered a correlation between the mean periods and iron abundances for cluster variables \citep[see also][]{arp1955}. 
These studies were driven by the evidence of a well defined dichotomy among the RRLs belonging to GGCs \citep{oosterhoff39}. 
\citet{oosterhoff39} found two groupings of fundamental mode RRLs, one with mean periods $P({\rm RR_{ab}}) \sim 0.55$~days and [Fe/H]$\gtrsim -1.5$ and one with $P({\rm RR_{ab}}) \sim 0.65$~days and [Fe/H]$\lesssim$-1.5. 
These two groups were later christened the Oosterhoff~I (Oo~I) and II (Oo~II) groupings, respectively. 
Subsequently, this effect was also confirmed for field stars by \citet{preston1959}. 
Figure \ref{fig:oo} demonstrates the Oosterhoff dichotomy for GGCs using data from \citet{catelan_2009}. Figure \ref{fig:oo} demonstrates two groupings of clusters in P-[Fe/H] space that separated by a region known as as the ``Oosterhoff gap.''
The dichotomy is explained as the {\it intrinsic luminosity} for the RRLs in two Oo groups being different, with the higher metallicity Oo~I clusters being fainter \citep{sandage06b}.
This physical understanding is largely supported by stellar evolutionary models \citep[e.g.,][]{cs13}, where higher metallicity stars have both a lower helium-core mass (main parameter that setting the HB luminosity) and a higher opacity in the envelope. 

\begin{figure*}
	\includegraphics[width=\textwidth]{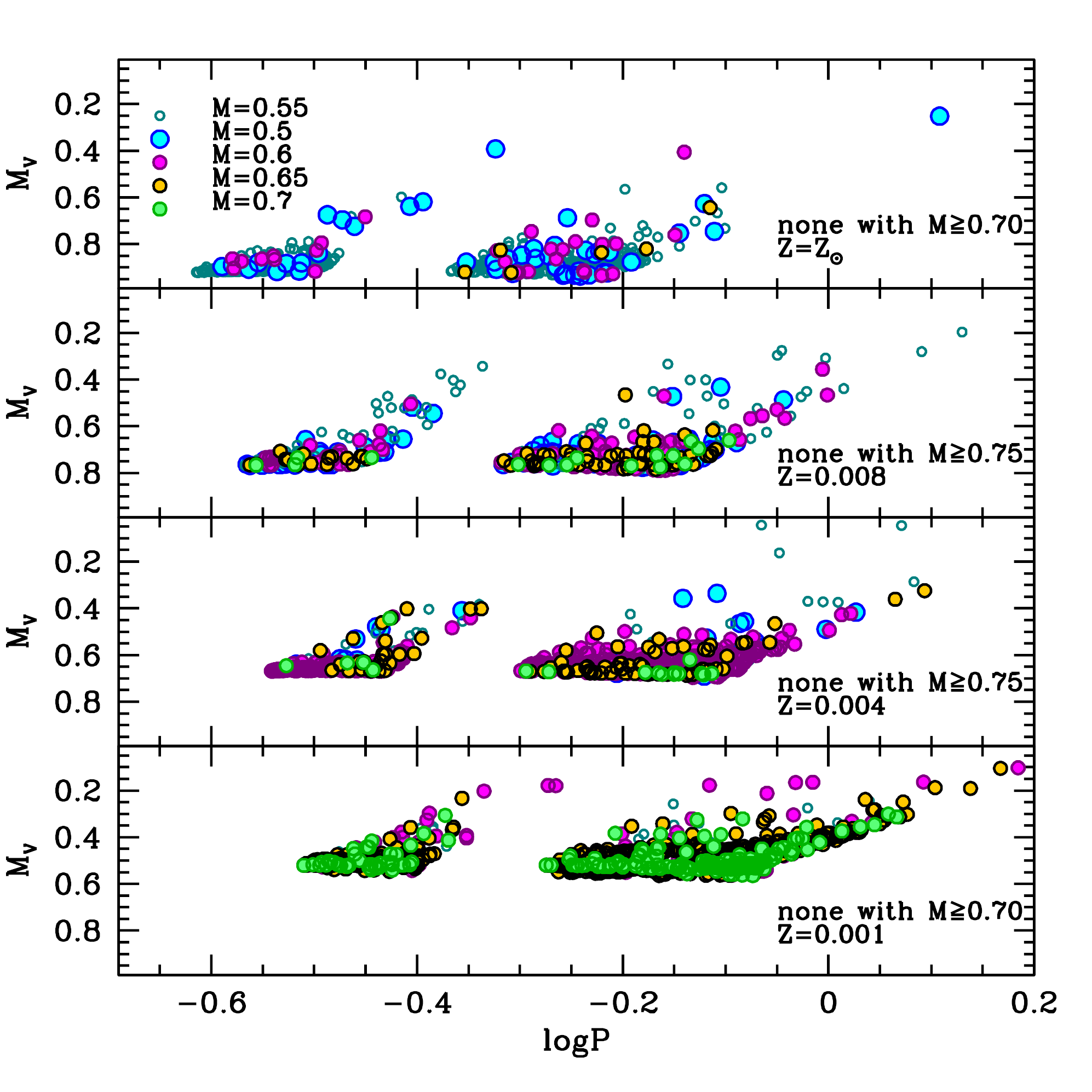}
	\caption{Synthetic period-magnitude diagrams for a population of RRLs as a function of metallicity from $Z=0.001$ (bottom panel) to Solar (top panel). A range of mean masses have been adopted for each simulation as indicated by color-coding and symbols shapes to highlight the population effects on both the mean magnitude and period.}
	\label{fig1_rr_theory2}
\end{figure*}

We demonstrate the metallicity-trend in Figure~\ref{fig1_rr_theory2} where a population of synthetic HB stars are shown for different chemical compositions ($Z$) from the RRL portion of the IS (drawn from the same models shown in Figure~\ref{fig1_rr_theory1}). 
For these stars we have computed a period using the pulsation relations given in \cite{Marconi15}. 
We show that for metal contents ranging from $Z=0.001$ ([Fe/H]$\sim -1.6$~dex) to $Z=0.02$ ([Fe/H]$\sim -0.3$~dex) and for a fixed $\alpha$-enhancement ([$\alpha$/Fe]$\sim +0.3$~dex), synthetic periods become shorter by $\Delta \log P \sim -0.1$ and luminosities become fainter by a factor of $\Delta M_V \sim +0.3$~mag (from bottom to top in Figure~\ref{fig1_rr_theory1}).
However, the offset in luminosity cannot be the only cause of the difference in the mean period of fundamental RRLs between Oo~I and Oo~II clusters. 
\citet[][and references therein]{bono97a} show that the dichotomy is largely due to the different pulsation behavior in the ``OR region,'' which is the intersection between the FU and FO IS, as a result of a hysteresis mechanism originally suggested by \cite{vanalbada73}.

Given this relation between magnitude and metallicity, it is common to assume a linear relation between the RRL $V$-band absolute magnitude, $M_V$, and the stellar metallicity, typically expressed as [Fe/H] in the following form:

  \begin{equation} \label{eq:mv}
     M_V ({\rm RR})~=~\alpha~+~\gamma~{\rm [Fe/H]}.
  \end{equation} 
  
\noindent Many different calibrations have been suggested in the literature for the slope ($\gamma$)
and the zero point \citep[$\alpha$; e.g.,][among others]{liu90b,carney92b,Mcnamara_1997,clementini03,bono03c} 

Despite the simplicity of Equation \ref{eq:mv}, the $M_V$-[Fe/H] relation as a distance diagnostic has a number of drawbacks. 
First of all, the reddening value and the extinction law have an impact on the absolute magnitude. 
The extinction coefficient is quite large in optical band, $R_V$ = 3.1 \citep{cardelli89}, 
but it decreases by one order of magnitude when approaching longer wavelengths 
(see also Section~\ref{nirmir}). 
The reddening law, itself, may not be universal \citep{kudritzki12}. 
Moreover, several complicated astrophysical effects have a strong impact on the stability of the relationship including:
 \begin{enumerate} 
   \item intrinsic deviations from the linear form of Equation \ref{eq:mv} \citep[e.g., a form non-linear with iron abundance;][]{bono03c,bono_2007},
   \item evolutionary effects within RRL populations,
   \item measurement uncertainties for the metallicity, which include both systematic differences between metallicity scales (e.g., calibrations) and methodologies (e.g., what abundance is actually being measured), 
   \item measurement uncertainties for the $\alpha$-enhancement, and
   \item deviation from static magnitudes \citep[see also][]{caputo00}. 
 \end{enumerate} 
Some additional information on these effects are given in the list below:


\begin{enumerate}
\item \textit{Linearity:}~~ The assumption of the linear $M_V$--[Fe/H] relation (Equation \ref{eq:mv}) has been challenged many times. 
There is strong empirical evidence that the $M_V$--[Fe/H] relation is not linear over the whole metallicity range covered by the observed GGCs \citep[e.g.,][]{rey00,caputo00}.
This is also supported by both pulsational \citep[e.g.,][]{bono03a,caputo00,dicriscienzo04} and evolutionary \citep[e.g.,][]{Catelan2004} models. 
The slope is observed to get steeper at approximately [Fe/H]$=-1.5$ and, as a result, a quadratic form of this relation has been proposed \citep[see e.g.,][among others]{caputo00,Catelan2004,dicriscienzo04,Sandage_2006,sandage06b,bono_2007}. 
To provide the reader with a quantitative idea of the dependence on the metallicity and of the uncertainties, we report here the calibration by \cite{clementini03} that is widely used:
  \begin{equation} \label{eq:gisella} 
    V_0({\rm RR})~=~19.064 (\pm 0.017)~+~0.214 (\pm0.047)~{\rm [Fe/H]},
  \end{equation} 
\noindent where the zero-point reflects the distance to LMC and the quoted uncertainties are only those on the coefficient of the linear regression. 
In the literature, it has been pointed out that theoretical calibrations of the $M_V$--[Fe/H] relation consider only the Zero Age HB (ZAHB) models, and hence they do not fully take into account evolutionary effects \citep[e.g.,][]{Catelan2004,Sandage_2006}. 
However, empirical calibrations such as \cite{caputo00} are built on observed data, with RRL variables at different evolutionary stages, and therefore they represent a ``mean'' evolutionary state/status for the HB. 
The quadratic form of the calibration, suitable for -2.4 \textless [Fe/H] \textless 0.0, determined in \citet{bono_2007} and applied to the Galactic Bulge and Sagittarius dwarf in \citet{kunder_2009} is as follows:
  \begin{equation} \label{eq:quad} 
    M_{V,{\rm RR}}~=~1.19 + 0.5{\rm [Fe/H]} +0.09{\rm [Fe/H]}^2.
  \end{equation} 
\noindent However, it is worth noting that no deviation from linearity has been observed within the LMC over a broad metallicity range \citep{clementini03}. Thus, the interplay between evolutionary effects (discussed below) and metallicity is difficult to decouple in observational work.
A canonical example of this difficulty is $\omega$~Centauri, which is home multiple stellar populations, and has been deemed a ``red herring'' in the review of \citet{smith_1995} due to the complexity of disentangling these effects \citep[a more recent study is][with additional references therein]{braga_2016b,braga_2018}.

\item {\it Evolutionary Effects:}~ The evolutionary effects on the $M_V$--[Fe/H] relation are shown in Figures~\ref{fig1_rr_theory1}~and ~\ref{fig1_rr_theory2}. 
We have plotted for each selected metallicity several realizations of the HB morphology coming from different assumptions on the mean mass distribution on the HB. 
This is mimicking different possible values of the mass loss efficiency during the Red Giant Branch (RGB) phase, which has the result of changing the HB morphology.
A comprehensive discussion of the role of mass loss on the HB morphology can be found in \citet[][and references therein]{scw02}. 
By an inspection of Figure~\ref{fig1_rr_theory2}, it is clear that the HB width is not narrow due to the overlap in color-space of stars on the ZAHB and stars evolving off the ZAHB on their path to the Asymptotic Giant Branch (AGB) phase. The evolved-RRL have higher luminosities. The presence of both ZAHB-RRL and evolved-RRL on the HB defines the ``evolutionary effect''
As a result of the time spent on the HB, we observe RRLs until they have burned most of the He in their cores (up to {$\sim$}90~\%). 
The region where a RRL spend most of its life time is also shown in Figure~\ref{fig1_rr_theory1} for different He values and a given mean metallicity. 
This effect implies a broadening of the optical magnitude distribution across the HB that is also correlated with the metal content, which is supported by both theory \citep{bono_1995} and observations \citep{sandage_katem_1982}. 

\item {\it Size of the He Core:}~ There are other ``ingredients'' that affect the theoretical calibration of the $M_V$--[Fe/H] relation, which are related to the size of the He core reached before the He ignition. 
These include:
\begin{enumerate} 
\item {\it The initial He value:} This can be summarized as the primordial He content plus a $\Delta$Y/$\Delta$Z, which is commonly measured from H~II regions. 
There is evidence that $\Delta$Y/$\Delta$Z may depend on the environment \citep[e.g.,][]{peimbert_20120}.
An increase in total He abundance of $\Delta {\rm He}\sim +0.2$~dex implies $\Delta M_V \sim 0.5$~mag as is demonstrated in Figure~\ref{fig1_rr_theory1}.
\item {\it Details of Red Giant Evolution:} More specifically, the processes occurring along the RGB phase that delay the ignition of the 3-$\alpha$ reaction in the electron degenerate core. 
These include atomic diffusion, electron conductive opacity, loss of energy via neutrinos, and core rotation. 
Their impact on the HB luminosity can be up to $\Delta \log (L/L_{\odot}) \sim 0.1$~dex or  $\Delta M_V \sim 0.2$~mag \citep{scw02,serenelli_2017}.
\end{enumerate} 
\end{enumerate}

\subsubsection{The FOBE method} \label{fobe}

The location in the period--magnitude diagram of the FOBE can be used as a distance indicator for a population of RRL independent of the $M_V$-[Fe/H] relationship.
The FOBE method was extensively described for the first time by \cite{caputo97}. 
This is a graphical, or topological, method that produces accurate distances for those stellar systems with sizable samples of FO RRLs \citep{fiorentino10b}. 
However, it can be reliably used only when the blue part of the IS is well populated \citep{Stetson2014,martinez-vazquez_2017}. 
Once the metallicity is known, preferably from spectroscopic measurements, and a reliable mass for the RRLs can be estimated \citep[typically in the range 0.5~$M_{\odot} \lesssim M \lesssim 0.7~M_{\odot}$;][]{bono03c}, the following theoretical relation from \cite{caputo00} can be used to fit the blue edge defined by FO RRLs to predict the absolute magnitude:

  \begin{equation} 
     M_{V}({\rm FOBE})=-0.68(\pm 0.03)-2.25\log P_{\rm FOBE}-1.26\log \left(\frac{M}{M_{\odot}}\right)+0.06\log Z,
  \end{equation} 
  
\noindent where $P_{\rm FOBE}$ is the period of the bluest FO pulsators and $M_V ({\rm FOBE})$ is the magnitude of the bluest FO pulsators for a given composition and mass.
Although, an assumption on the RRL mass has to be made \citep{bono03c}, the possibility to use the FOBE method relies on the well understood negligible dependence of the color of the FOBE on the metallicity \citep{bono97a}, which is demonstrated in Figure~\ref{fig1_rr_theory1}. 
For the bluest FO RRLs the period is most strongly dependent on luminosity and mass.

The sharp, blue edge of the IS occurs because for a given mass and luminosity, as the surface temperature increases there is less mass above the ionization zones, and in turn, their contribution to the work integral ($p\delta V$) decreases. 
Then, the FOBE color is essentially fixed by the minimum difference between the temperature of the stellar surface and that of the Hydrogen ionization region, within which pulsation is efficient.
The minimum difference is almost constant and does not depend on metallicity, at least for the metallicity range typical of RRLs. 
On the contrary, the red edge of the IS strongly depends on metallicity, because the quenching mechanism for the radial pulsations is related to convection that increases with the stellar opacity, and thus by increasing the metal content \citep{deupree77a,stellingwerf82}. 
The above theoretical prediction has been observed in GGCs; more specifically, the $B-V$ FOBE color is always {$\sim$}0.2~mag. 
Thus, the FOBE method can also be used to estimate the mean reddening of a stellar system \citep[see][for an example]{walker98}.

\subsubsection{The NIR period--luminosity relations}\label{nirmir}

\begin{figure} 
 \centering
  \includegraphics[width=\columnwidth]{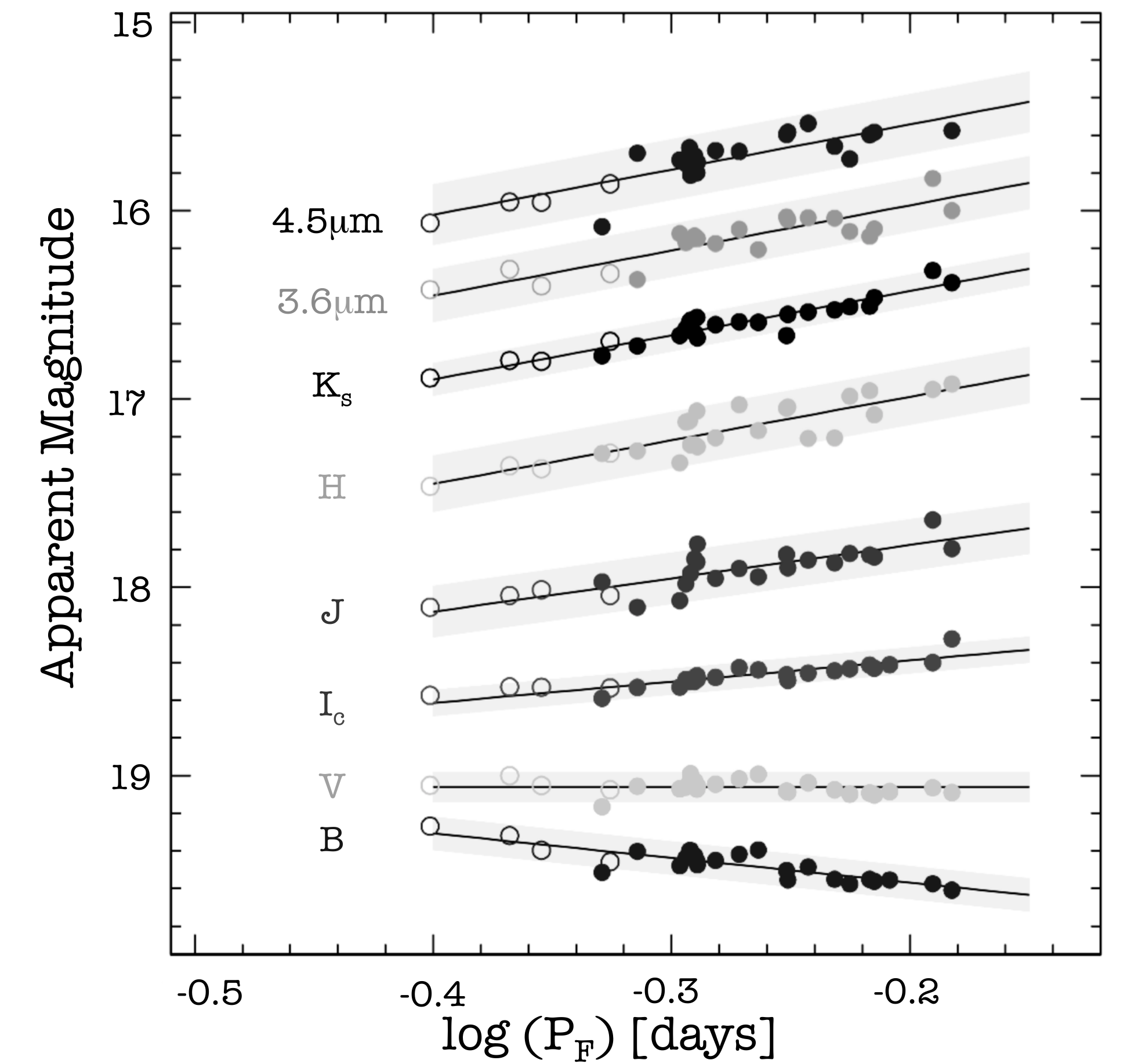}
  \caption{ \label{fig:pl_slope_lambda1} 
   Mean apparent magnitudes for RRL in the Reticulum star cluster in the optical ($B, V, I$), NIR ($J, H, K$), and MIR ([3.6],[4.5]) versus the fundamentalized period (Equation~\ref{eq:fotofu}).
   PL relationships are overplotted (solid lines) for each of the bands \citep[adapted from][Monson et al.,~in prep.]{catelan_2015}.
   The change in the slope from the optical into the mid-infrared is noticeable, with $V$ being a nearly flat slope to the mid-infrared being quite steep.}
 \end{figure}

Unlike in the optical, the observations in the IR show a true PL relationship \citep[first identified by][]{Longmore_1986}. 
In Figure \ref{fig:pl_slope_lambda1}, the mean magnitudes for RRL in the star cluster, Reticulum, are shown for eight photometric bands ($B, V, I, J, H, K_s, [3.6],[4.5]$) against the fundamentalized period, which is defined as follows \citep{bono01}:
\begin{equation} \label{eq:fotofu}
\log P_{\rm FU}~=~\log P_{\rm FO}~+~0.127.
\end{equation}
Reticulum is an ideal system to visualize the behavior of the PL slope with wavelength, because it has no appreciable metallicity spread on the RGB \citep{grocholski_2006} and, thus, scatter due to metallicity differences are minimized.
In Figure \ref{fig:pl_slope_lambda1}, preliminary PL fits are shown for each band as solid lines with the scatter about the PL shown as the shaded regions \citep[][Monson et al.,~in prep.]{catelan_2015}. 
As previously discussed, the $V$ band shows no PL slope, whereas a slope appears and becomes progressively steeper at the longest wavelengths.
In the next sections, we discuss empirical and theoretical results on the RRL PL. 

\subsubsection*{The $K$-band period--luminosity relationship}

Period--luminosity--color (PLC) relations have been predicted for RRL by even the earliest theoretical models \cite[e.g.,][]{VanAlbada1971}, but their use has always been hampered by several uncertainties on the structural parameters of the stars, in particular mass, effective temperature and metallicity. 
Evolutionary effects also play a role, because during their evolution (the He burning phase lasts at most for 100 Myr), RRLs change their luminosities and effective temperatures.
Moreover, observational uncertainties, such as those associated with reddening and extinction, have to be decoupled from true astrophysical differences.
Thus, empirical confirmation of the PLC has only come of late.

\begin{figure} 
	\includegraphics[width=\textwidth]{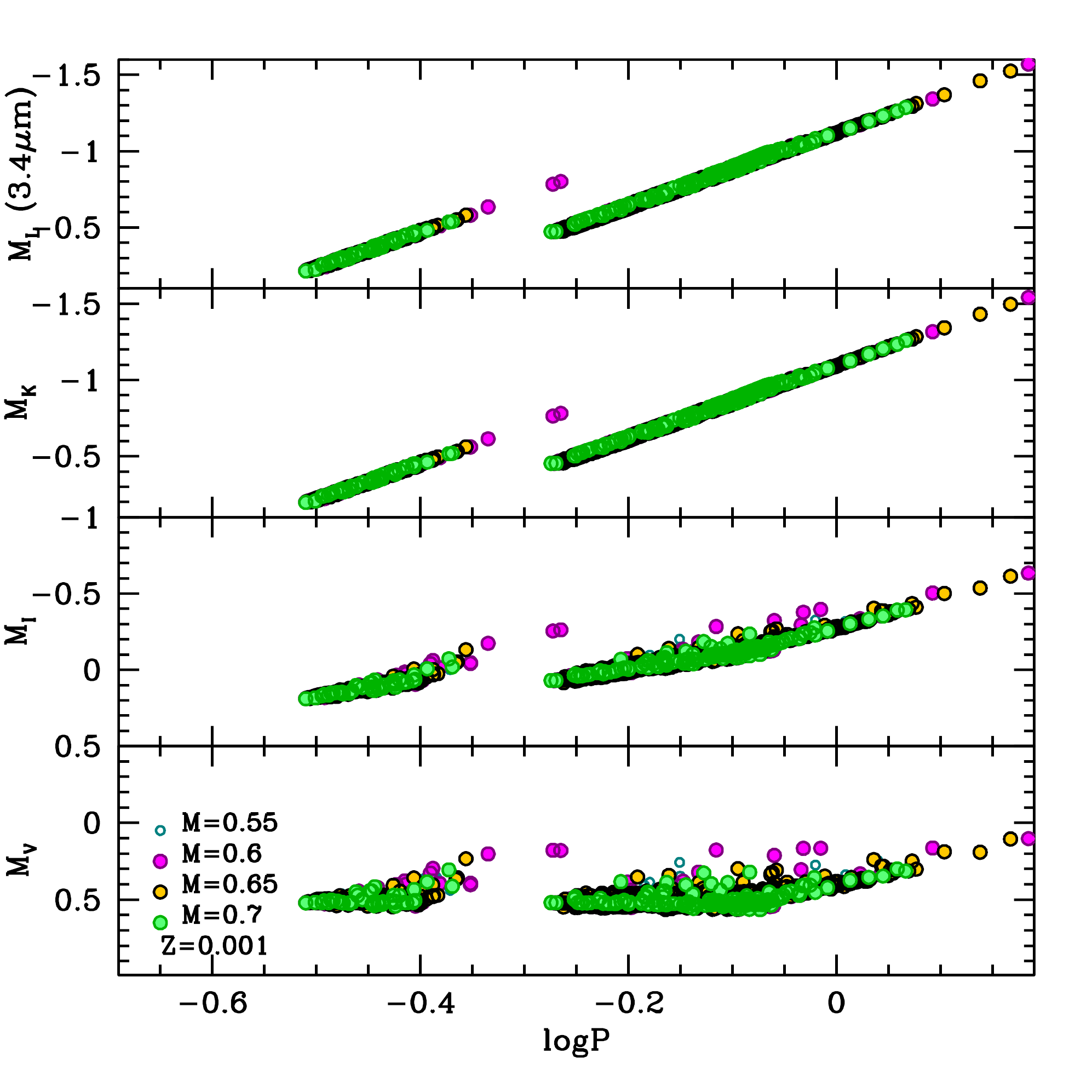}
	\caption{PL distribution from optical (bottom panel) to mid-infrared (top panel). We have used synthetic HB models as in Figures~\ref{fig1_rr_theory1}~and~\ref{fig1_rr_theory2} for fixed abundance ($Z=0.001$) and a canonical He abundance ($Y=0.24$) with different mean mass for the population distribution.} 
	\label{fig2_rr_theory}   
\end{figure} 

\cite{Longmore_1986} demonstrated, on pure empirical basis, that RRLs follow a tight linear PL relation in the $K$-band (PL$_{K}$). 
They also demonstrated that the PL$_{K}$ relation can be derived from the general equation of the pulsation, since the period dependence on the $K$-band can be derived from the $V-K$ proxy of the effective temperature (as a direct consequence of the fact that RRLs follow PLC relations). 
In other words, a PL$_{K}$ arises because of the increasing bolometric correction as a function of the effective temperature, with the redder RRLs having $K$-band magnitudes brighter than the bluer RRL (see Figure~\ref{fig2_rr_theory}, for the K band). 
Moreover, since for a given cluster, the effective temperature distribution inside the IS is related to the period distribution, such that the stars with longer periods are also cooler, the $K$ magnitude-effective temperature relation becomes a $K$ magnitude-period relation. 
A more general empirical calibration of the PL$_{K}$ was given by \citet{janes92} and these authors also considered the dependence on the metallicity, which was shown to have little impact on the slopes.

A sound and complete theoretical derivation of the $K$-band period--luminosity--metallicity (PLZ$_{K}$) relation was subsequently explored by \citet{bono01} and \citet{bono03c}, finding the following:
\begin{enumerate} 
\item the uncertainties on the mass and luminosity (e.g., on the evolutionary effects) have only a mild impact on the PLZ$_{K}$.
\item the overall dependence on the metallicity is lower, at $\gamma\sim 0.17$~mag~dex$^{-1}$ in [Fe/H], than that in the optical bands ($\gamma$ \textgreater 0.2 mag~dex$^{-1}$, see Equation \ref{eq:gisella}). 
\end{enumerate} 

On the latter aspect, we remark that \cite{sollima06} found an even lower metallicity term ($\gamma$=0.08~mag~dex$^{-1}$ in [Fe/H]), on the basis of an empirical calibration.
\cite{Muraveva_2015} combined low-dispersion spectroscopy of 70 RRLs in the LMC, with $-2.0 \le {\rm [Fe/H]} \le -0.6$~dex, with NIR photometry collected with VISTA 
finding a very low dependence on the metallicity 
($\gamma = 0.03 \pm 0.07$~mag~dex$^{-1}$ in {\rm [Fe/H]}), and a quite steep dependence on the period ($-2.73 \pm 0.25$). 
All these aspects reflect on a very low dispersion of the PLZ$_{K}$ relation, being $\sigma = 0.035$~mag rms or \textless 2\% in distance \citep{bono01}. 

PL$_{K}$ and PLZ$_K$ observational relations have been successfully tested on several stellar systems, among the others we mention the LMC cluster Reticulum \citep{dallora04} and the Galactic GGCs IC\,4499 \citep{storm04}, M\,92 \citep{delprincipe05}, Omega Centauri \citep{delprincipe06},  M\,5 \citep{coppola11}, and M\,4 \citep{Braga2015}. 
The observed scatter about the PL ranges from 0.03~mag (Reticulum) to 0.09~mag (IC\,4499), which suggests a single-star distance uncertainty ranging from 1.5\% to 4.5\% is feasible. 
Additional studies have occurred in dwarf galaxies such as the LMC \citep{ripepi_2012_lmc_rrl,moretti_2014}, SMC \citep{Szewczyk_2009,muraveva_2018_smc}, Fornax \citep{Karczmarek_2017}, Sculptor \citep{Pietrzynski_2008}, Carina \citep{Karczmarek_2015}, and IC\,1613 \citep{hatt_2017}.

Compared to the $M_{V}$--[Fe/H] relationship, the key advantages distance determination at longer wavelengths are:
  \begin{enumerate} 
    \item the impact of extinction is less by a factor of ten or more and, as a result, the impact of differential extinction is also greatly reduced; 
    \item the impact of the uncertainties the reddening measurement, itself, are minimized (this being what is determined to estimate extinction); 
    \item the pulsation amplitude is lower than in the optical bands (Figure \ref{fig:lc}) and the light curves are more symmetrical, accurate mean magnitudes can be measured even with a few data points;
    \item the availability of light curve templates \citep[e.g.,][]{jones96,monson2017,Hajdu_2018}, permits the estimation of the mean magnitudes even with a single data point, if the ephemerides of the variable are known from the optical photometry \citep[see applications in][Rich et al.~submitted]{beaton2016}.
  \end{enumerate} 
On the other hand, finding the variables, measuring their periods, and determining their pulsation mode (FU or FO) is more difficult due to smaller amplitudes, sinusoidal shape, and the more uniform light curve shapes between the sub-classes. 
Moreover, templates is primarily only available for for the RRab type stars. 

\subsubsection*{Mid-IR period--luminosity relations}\label{sec:mir}

The advantages of using PL relations at mid-infrared wavelengths (MIR, from 3.4 to 22~{$\mu$}m) with Cepheids have been well-recognized in the literature since \cite{mcgonegal82}, but only very recently it has been possible to explore this wavelength regime with RRLs.  
Indisputable advantages to the MIR PLs is that at 3.4~{$\mu$}m the impact of interstellar absorption is more than one order of magnitude lower than in the optical bands, which permits study of heavily obscured and differentially obscured regions like the Galactic bulge that contain a large population of RRL \citep[see discussion in][]{kunder_2018}.

Indeed, on the basis of the Wide Field Infrared Survey Explorer (WISE) measurements of field RRL variables, \cite{klein11} were able to produce the first calibration of the PL relations at the W1 (3.4~{$\mu$}m), W2 (4.6~{$\mu$}m), and W3 (12~{$\mu$}m) bands. 
They subsequently refined their calibrations in \cite{klein12} and \cite{Klein14}, with no metallicity term. 
Another absolute calibration of the MIR PL relations was published by \cite{Madore2013}, using four field RRLs for which individual trigonometric parallaxes, via the {\it HST} Fine Guidance Sensors ({\it HST}-FGS), were available \citep{Benedict_2011}. 
Even if with only four objects, the observed dispersion obtained by \cite{Madore2013} is a very promising 0.1~mag ({$\sim$}5\% in distance). 
\cite{Madore2013} also found, as expected, only a marginal dependence on the metallicity, because at the MIR wavelengths stars with the temperature of RRLs have few molecular bands and metallic lines. 
Based on a much larger sample of 129 RRLs with \emph{Hipparcos}-based statistical-parallaxes distances, \cite{Klein14} obtained very nice results, with observed spreads ranging from 0.01~mag to 0.05~mag, depending on the filter, on the sample (FU or FO pulsators), and assumptions on the intrinsic width of the PL.
\citet{Neeley_2017} expanded the \emph{Spitzer}-analysis to a larger sample of RRL with parallaxes from the the TGAS catalog in \emph{Gaia}-DR1 \citep[e.g.,][]{michalik_2015}, finding consistent solutions within the observational uncertainties. A complementary analysis from \citet{Sesar_2017} used \emph{WISE} photometry with the TGAS parallaxes employing a Bayesian algorithm, which \citet{Muraveva_2018_pl} has expanded upon recently using \emph{Gaia}-DR2.

The first calibration of the MIR PL using cluster variables was published by \cite{dambis_2014}, which presented an analysis of \emph{WISE} data for 15 GGCs. 
\cite{Neeley2015,Neeley_2017} published data of the GGC M\,4, collected with the Infrared Array Camera (IRAC) on-board the \emph{Spitzer} Space Telescope, obtaining PL relations in the IRAC $[3.6]$ and $[4.5]$ bands with a scatter of {$\sim$}0.05~mag (2.5\% in distance).
\citet{muraveva_2018} has recently published a calibration in the LMC star cluster Reticulum. 

\subsubsection{The RR Lyrae period--luminosity slopes}\label{pl_slopes}

Because the bolometric correction sensitivity to effective temperature is already at work starting from the $R$-band \citep[see][]{cassisi04,Catelan2004,Marconi15}, PL relations have also now been derived theoretically for the photometric bands $R$, $I$, $J$ and $H$, which are shown in Figure~\ref{fig2_rr_theory}.  
At shorter wavelengths, the effects related to the finite width of the IS and evolution of RRLs are larger and produce PL relations with larger intrinsic scatter (compare the panels of Figure~\ref{fig2_rr_theory}). 
The relationships in $R$, $I$, $J$ and $H$ are still of great interest, because these wavelengths can be observed easily from the ground.

 \begin{figure} 
 \centering
  \includegraphics[width=\columnwidth]{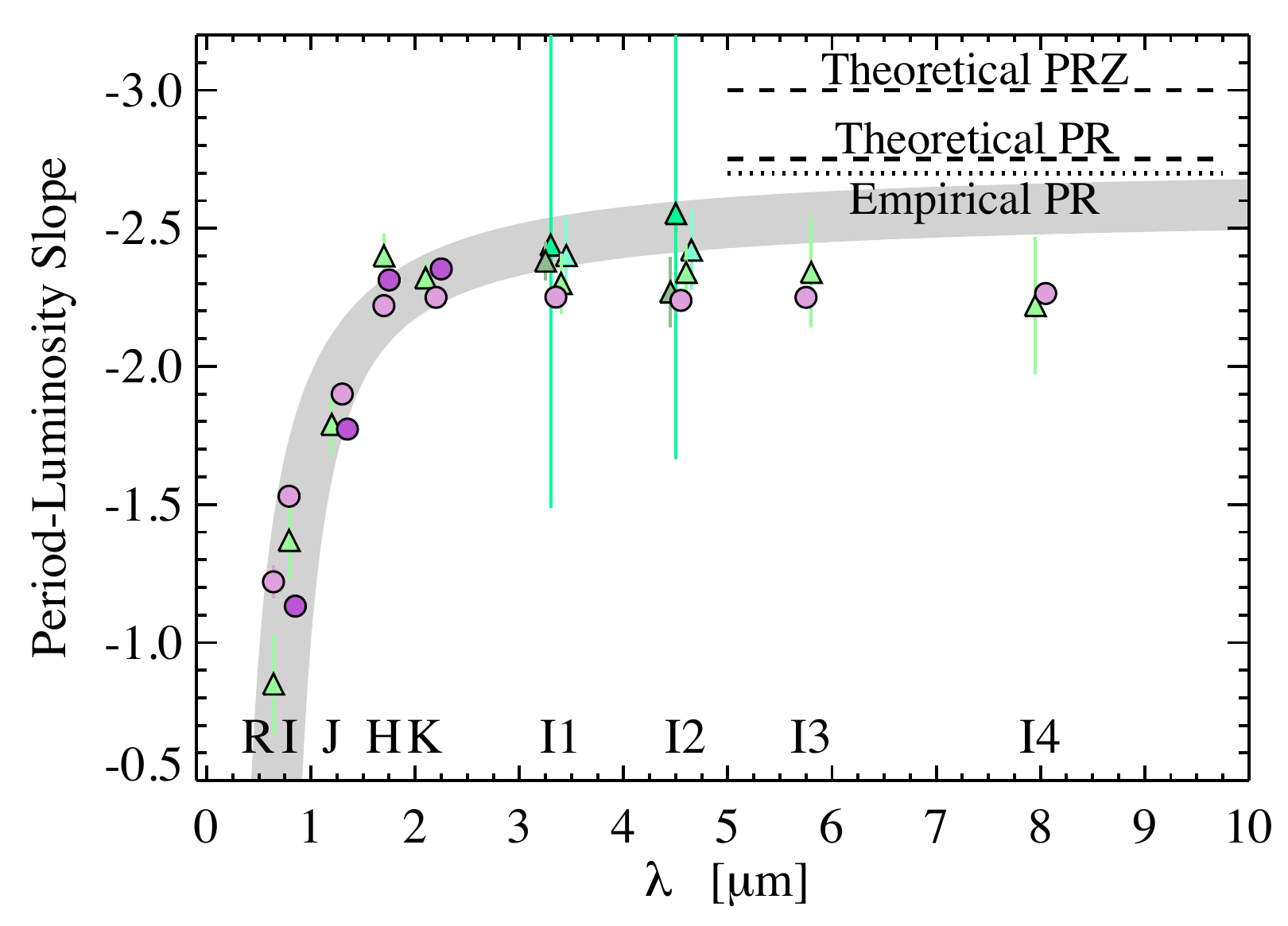}
  \caption{ \label{fig:pl_slope_lambda2} 
   Following \citet[][their figure 4]{Madore2013}, the PL slopes are plotted a function of wavelength from both theoretical \citep[circles;][]{Catelan2004,marconi05,Marconi15,Neeley_2017} and empirical \citep[triangles;][]{klein11,Madore2013,Braga2015,dambis_2014,Neeley2015,Neeley_2017} studies. 
  We note that studies in the same photometric band have been offset slightly in their effective wavelength for visualization purposes.
   The shaded region is drawn to guide the eye following \citet[][]{Madore2013}.
   The slopes for the long-wavelength filters asymptotically approach that predicted by period-radius relationship for both theory \citep[dashed line][]{Marconi15} and empirical determinations \citep[dotted line][]{burki_1986}.}
 \end{figure} 

In Figure~\ref{fig:pl_slope_lambda2}, the PL slopes determined from a subset of both empirical (triangles) and theoretical (circles) investigations are plotted versus wavelength following the example of \citet{Madore2013}.
These slopes are for the ``global'' population, e.g., the first-overtone periods have been fundamentalized following Equation~\ref{eq:fotofu}.
The shaded region in Figure~\ref{fig:pl_slope_lambda2} is meant to guide the eye to the behavior with wavelength and this is adapted from the visualization in \citet[][their figure 4]{Madore2013}.
Generally, the slopes between the different studies, both theoretical and empirical, agree well. 
Still, there are comparatively few studies on the PL for wavelengths longer than $V$ and further investigations, in particular empirical studies in star clusters with differing chemical compositions and evolutionary states, are highly relevant to determining a ``consensus'' PL slope for a given filter.

In Figure~\ref{fig:pl_slope_lambda2}, the ``predicted'' slopes from the period-radius (PR) or period-radius-metallicity (PRZ) relationships are also shown following \citet[][]{Madore2013}. 
NIR and MIR wavelengths sample the Rayleigh-Jeans tail for RRLs and, as a result, the magnitudes are much less sensitive to temperature than to the radial variations. 
This behavior is readily seen in the light curves shown and, indeed, the light curves beyond $H$ are nearly indistinguishable. 
Thus, predicting the PL slope from the period-radius relationship places an asymptotic limit on the PL slopes.

Starting from $L~=~4 \pi R^2 \sigma T_{\rm eff}^4$ and converting into a logarithmic form yields the following, 

  \begin{equation}\label{eq:logeq} 
    \log L~=~\log (4\pi\sigma)~+~2\log R ~+~4\log T_{\rm eff},
  \end{equation} 

\noindent which can be converted into magnitudes as follows, 

  \begin{equation} \label{eq:to_radius} 
    \log M ~=~-5\log R~-~20\log T_{\rm eff}~+~{\rm constant}.
  \end{equation} 

\noindent From here, we can substitute either the period-radius (PR) or period-radius-metallicity (PRZ) relationship to remove the radius term.
The PR has been determined in several ways, 
 (i)~empirically by \citet{burki_1986} using the Baade-Wesselink technique (BW),
 (ii)~stellar evolutionary models by \citet{marconi_2005} (evol), and 
 (iii)~pulsation models by \citet{Marconi15} (puls). 
The PR/PRZ takes the form, $\log R ~=~a~+~b\log P~+~c\log Z$, and the three studies find period slopes of $b_{\rm BW}=0.54$, $b_{\rm evol}=0.65$, and $b_{\rm puls}=0.55$, respectively.
Substituting, these into Equation \ref{eq:to_radius}, we estimate determine period slopes of $\beta_{\rm BW}=-2.7$, $\beta_{\rm evol}=-3.25$, and $\beta_{\rm puls}=-2.75$.
 
If we assume that the temperature term contributes very little at long wavelengths then we can ignore that term to obtain an estimate for the slope of the PL relationship.
The estimated slopes from the theoretical PR and PRZ \citep[dashed lines;][]{Marconi15} and the empirical PR \citep[dotted lines][]{burki_1986} are shown in Figure \ref{fig:pl_slope_lambda2}. 
The overall agreement between these asymptotic values and the slopes in Figure \ref{fig:pl_slope_lambda2} serves as a good cross-check on our theoretical and empirical measurements. 

\subsubsection{Reddening-free relations} \label{wes}

Even if PLZ relations have advantages for estimating individual distances, they are still affected by the uncertainties in the reddening corrections, and the problem can be severe in regions affected by strong differential reddening \cite[e.g., as in M4][or in the Galactic Bulge]{Braga2015}.

To circumvent this problem, \cite{vandenbergh75} and \cite{madore_1982} developed the Wesenheit pseudo-magnitudes.
They are, by construction, ``reddening free'' functions that redefine the observed magnitude.
The fiducial form of the transformation is defined as:

   \begin{equation} \label{eq:Wmags} 
     W(A,B)~ =~M_A~+~R_{A,B} (M_A - M_B)
   \end{equation} 

\noindent where $R_{A,B}$ is the selective-to-total extinction ratio. 
In the common practice, ``Cardelli's law'' is adopted \citep{cardelli89}.

The Wesenheit period--luminosity (PW) and period--luminosity--metallicity (PWZ) relations have the following form: 

   \begin{equation} 
     W~=~\alpha~+~\beta~\log P~+~\gamma~{\rm [Fe/H]},
   \end{equation} 
   
\noindent where each combination of filters will have its own set of PWZ parameters. 
The advantages and disadvantages of PW and PWZ relations have been discussed several times in the literature, in particular for classical Cepheids \citep[e.g.,][]{inno13,fiorentino07,ngeow05,storm11,ripepi12} 
and, more recently, with RRLs \citep[e.g.,][and references therein]{Braga2015,Marconi15}. 

Briefly, the main features can be summarized as follows: 
 \begin{enumerate}
	\item the individual distances are independent of the uncertainties in the reddening or in differential reddening within an RRL population; 
    \item a universal reddening law is assumed, but there are indications of deviations from ``Cardelli's law'' in regions of high obscuration \citep{cardelli89};
	\item PW and PWZ relations mimic PLC relations, since they host the color term: this means that the effect of the width of the IS is reduced, and individual distances are much more precise than those obtained with simple PL relations.
 \end{enumerate} 
Thus, in regions where the assumption of universal reddening law is reasonable, the PW/PWZ is a powerful tool.

On the practical side, however, use of the PW/PWZ are more observationally expensive than common PL and PLZ relations, because they need accurate mean magnitudes determined in two photometric bands. 
According to the models by \citet{Marconi15}, the dispersion around the PW and PWZ relations range from $\sim 0.01$~mag (in the NIR bands) to {$\sim$}0.07--0.08~mag (in the optical bands). 
\citet{Braga2015} provide complementary empirical PW/PWZ relations determined from M\,4 in the optical and NIR.
The PW$_{B,V}$ relation is almost independent of metallicity and are particularly useful. 

We also note that, on empirical basis, \citet{riess11a} proposed to use three-band PW relations, where the $W$ magnitude is reconstructed from the following form, 

  \begin{equation} \label{eq:3bandW} 
    W(A,B,C)~=~M_A~+~R_{A,B,C} (M_B - M_C), 
  \end{equation} 

\noindent where $R_{A,B,C}$ is the ratio between the selective absorption in the $A$ band, and the color excess for the $B-C$ color $E(B-C)$. 
The advantages of triple band PW/PWZ relations are a smaller dispersion respect to the two band relations and lower systematics from correlated errors having used the same magnitude in the color term. 
The drawback is that they need accurate mean magnitudes determined for three bands.
The most advanced theoretical study on this topic to date is that of \citet{Marconi15}, which provides theoretical three-band relations for RRL in the optical and NIR.

%
\subsection{Case studies using RR Lyrae} \label{sec:rrl_practice} 

In the previous section, a general physical understanding of RRLs as distance indicators was developed using both theoretical and empirical investigations. 
In this section, a series of case studies are presented on the practice of using RRLs for the different wavelength regimes discussed previously.
In Section~\ref{sec:rrlcase_opt}, the $M_{I}$--[Fe/H] relationship and the optical PW relationships are applied to the Sculptor dwarf galaxy
In Sections~\ref{sec:rrlcase_nir} and \ref{sec:rrlcase_mir}, the optical, near-infrared and mid-infrared PL relations are determined and applied to the nearby GGC M\,4 that suffers from strong differential reddening in the optical. 

\subsubsection{An optical study of RR Lyrae in Sculptor} \label{sec:rrlcase_opt} 

\begin{figure} 
	\includegraphics[width=\textwidth]{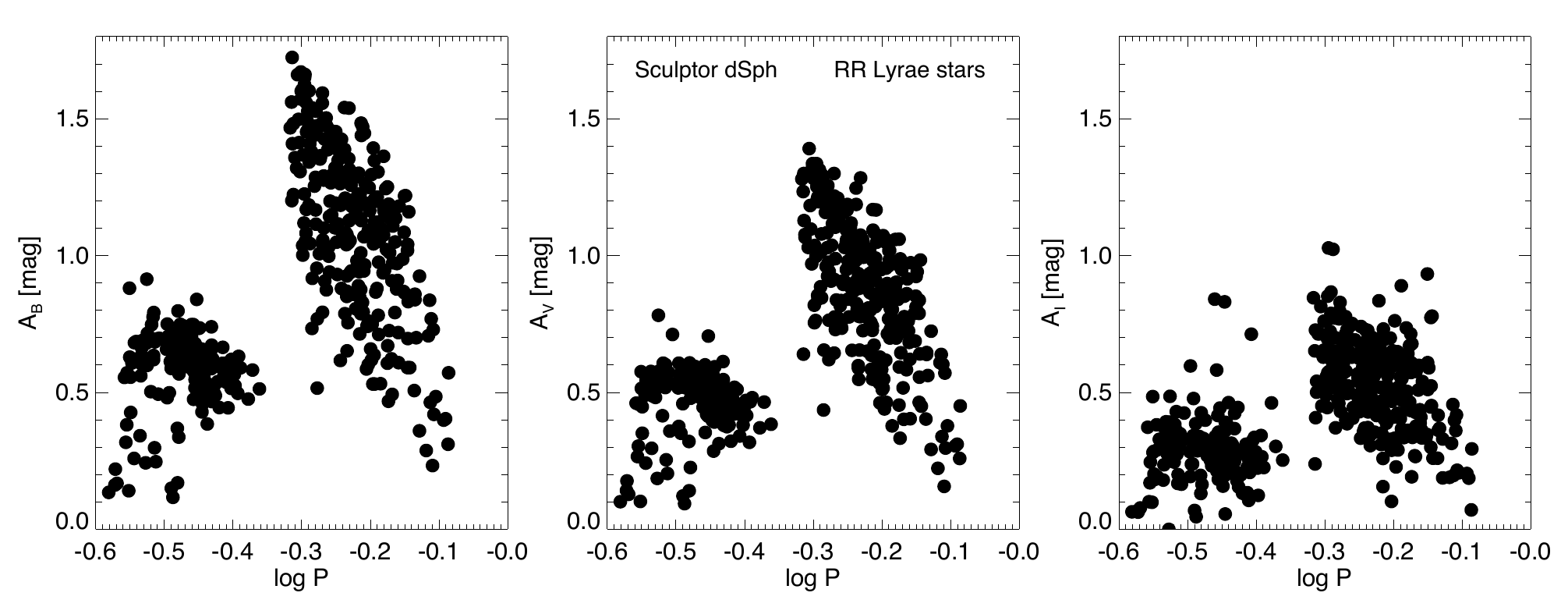}
	\caption{Period-amplitude diagrams for the RR Lyrae stars in Sculptor in three
		different optical bands: $B$ (left), $V$ (middle) and $I$ (right). RRab and RRc 
		are shown as black circles. For the sake of clarity, RRd stars are not plotted as their periods are less certain. The RRc and RRab occupy specific regions in the Bailey diagram, with the RRc having shorter periods and smaller amplitudes (i.e., in the bottom left) than the RRab's (right). The RRab's also have a broader range of amplitudes. These primary groups in the Bailey diagram can be further refined into specific sequences for the Oo groups, with two clear sequences visible for the RRab's.}
\label{fig:scl_pa}
\end{figure} 

Here we present a brief summary of a comprehensive study of the RRLs in the Sculptor dwarf galaxy \citep{MartinezVazquez2015,MartinezVazquez2016a,MartinezVazquez2016b}.
Sculptor is a Milky Way dwarf spheroidal (dSph) satellite with a complex chemical enrichment history \citep{Smith1983,DaCosta1984,Majewski1999,Hurley-Keller1999,Tolstoy2004,deBoer2012,Starkenburg2013}. 
\citet{Tolstoy2004} demonstrated that the red HB stars (RHB) are more centrally concentrated than the blue HB stars (BHB) that mirrors gradients from detailed spectrocopic studies on the RGB \citet{Battaglia2008b,Walker2007b,Walker2009b,Kirby2009,Leaman2013,Ho2015}, which suggests a that range of metallicites in also present the RRL population. 

Time series photometry was produced for stars Sculptor from a set of 5149 individual images collected over 24 years. 
Variables were identified using the Welch-Stetson variability index \citep{Welch1993} and periods obtained from a simple string-length algorithm \citep{Stetson1998} from which 536 RRLs were identified using the the shapes of the light curve. 
The final dataset contains pulsational properties and mean magnitudes for 536 RRLs in the $B$, $V$, and $I$ bands that is complete over $\sim$70\% of Sculptors area. 

\begin{figure} 
  \centering
	\includegraphics[width=\textwidth]{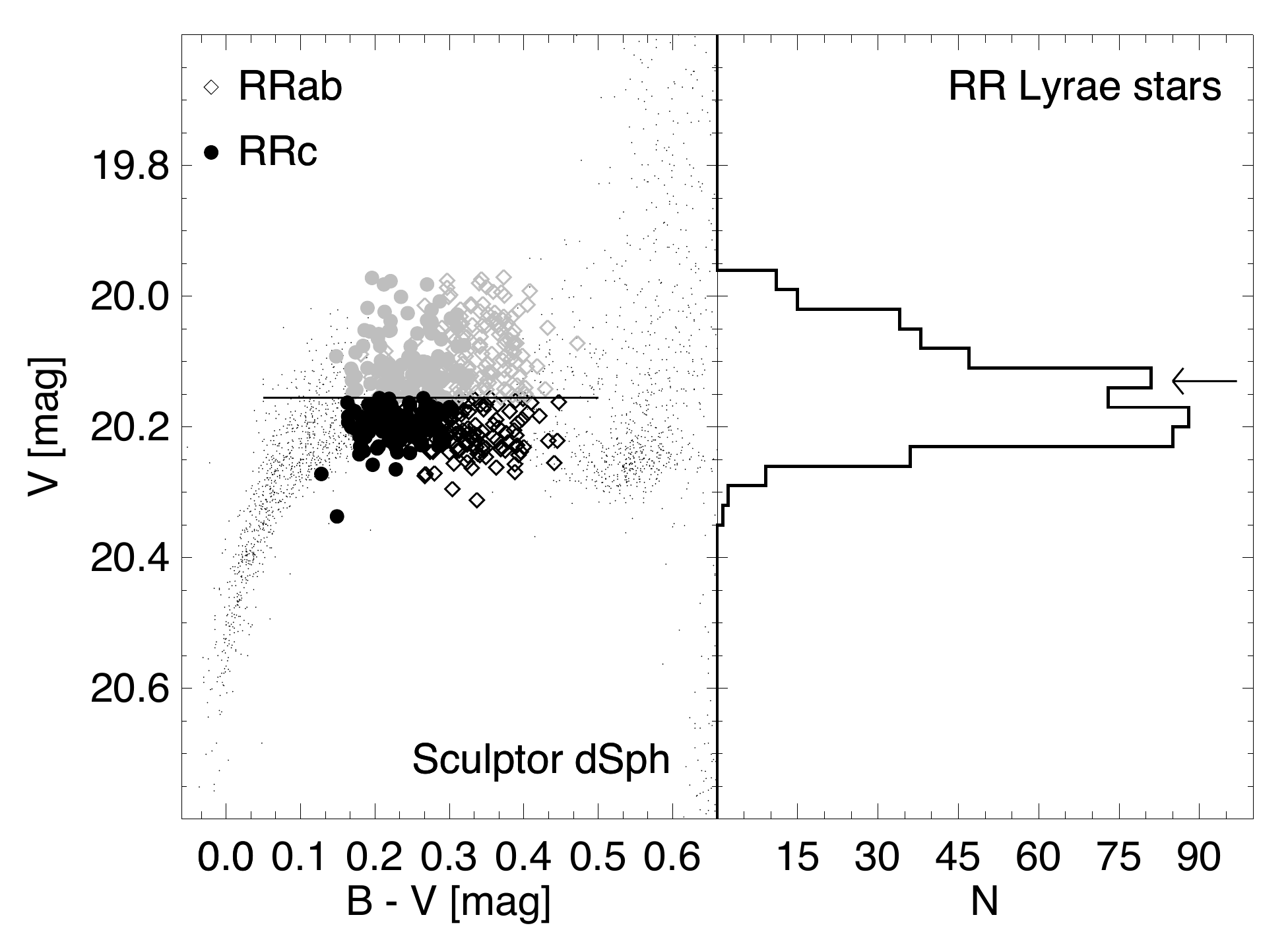}
	\caption{ \label{fig:scl_cmd} The ${V}$ distribution for Sculptor RRL. 
{\it Left panel:} Zoom-in on the HB stars in the color-magnitude diagram of Sculptor. The horizontal line (at $V=25.155$~mag) displays the split between bright (gray) and faint (black) RRL subsamples. Open diamonds and circles represent the RRab and RRc stars, respectively. 
{\it Right panel:} Luminosity distribution function in $V$ band for Sculptor RRLs. The arrow marks the magnitude adopted to split into bright and faint RRL subsamples. We note that the RRd and other more complicated RRL sub-types have been removed from this sample.}
\end{figure} 

Figure \ref{fig:scl_pa} shows the Bailey Diagrams for the $B$, $V$, and $I$ bands. 
The RRab and RRc stars separate into long- and short-period groups, that have their own amplitude behavior. 
The overall reduction of the amplitudes from the $B$ to the $I$ band is apparent. 
In the $B$ and $V$ the sub-division of the primary RRab and RRc sequences into Oo groups can also be seen, which is anticipated from the large metallicity spread. 

Figure~\ref{fig:scl_cmd} shows the $V$ magnitude distribution for the Sculptor RRLs. 
A zoom-in on the CMD is shown in the left panel and the corresponding luminosity function for the RRLs is shown in the right panel. 
The RRL stars of Sculptor show a spread in $V$ magnitude of {$\sim$}0.35~mag, which is significantly larger than the typical uncertainties in the mean magnitude ($\sigma =0.03$~mag), larger than anticipated for a mono-metallicity population showing evolutionary effects, and larger than the intrinsic spread for a mono-metallicity population.

\begin{figure} 
	\includegraphics[width=\textwidth]{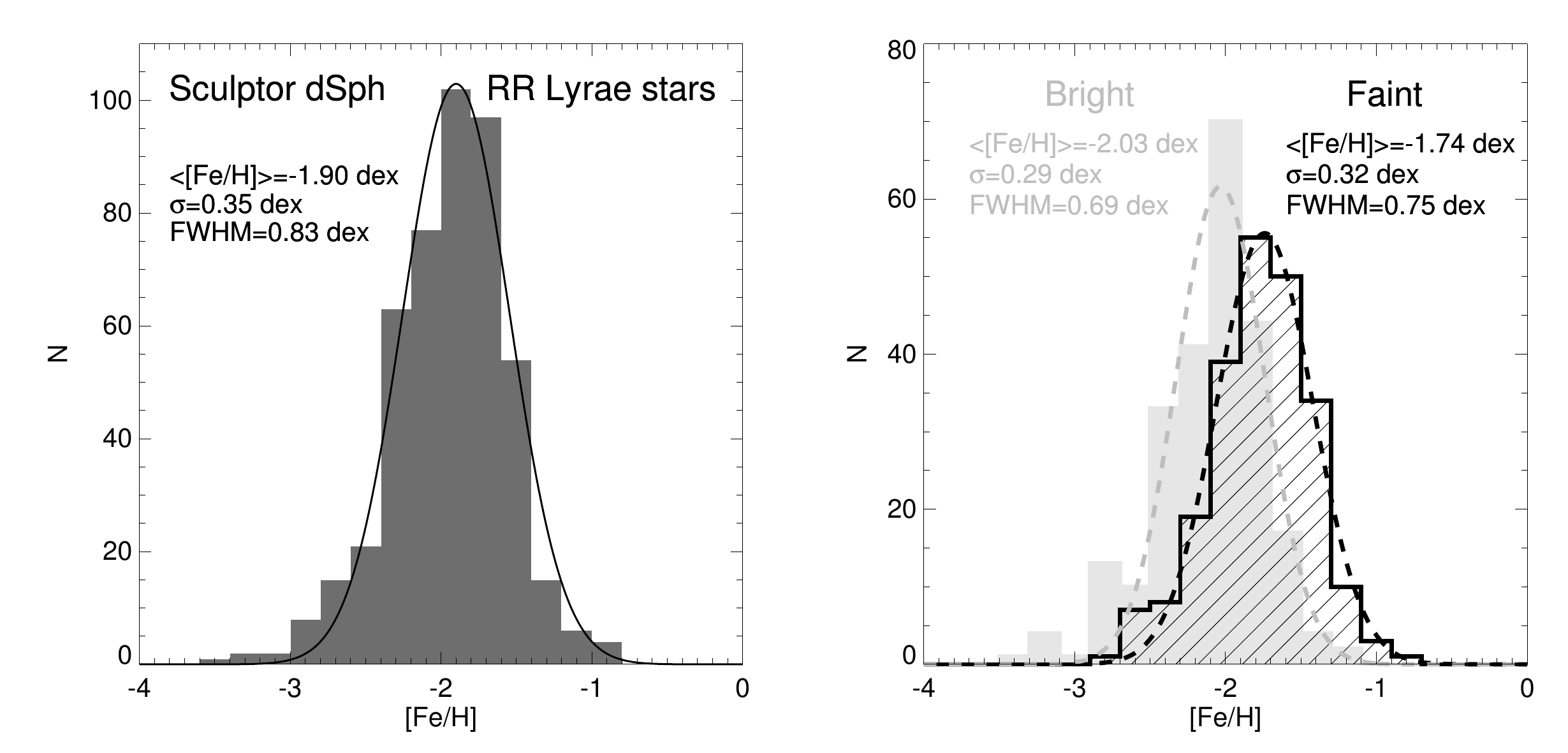}
	\caption{\emph{Left panel} Metallicity distribution obtained from the PL in $I$ band for the RRLs of Sculptor (dark grey histogram). The black solid line show the Gaussian fit performed. The parameters of the fit are labeled in the panel. 
	\emph{Right panel} Metallicity distribution for the bright (light grey histogram) and faint RRL subsample. Both distributions have been fitted to a Gaussian (dashed curves). The parameters of each fit are labeled in the panel.}
	\label{fig:scl_mdf}
\end{figure} 

The $I$-band PLZ (PLZ$_{I}$) can used to obtain the metallicity distribution for the RRLs following \citet{MartinezVazquez2016a}:  
  \begin{enumerate} 
    \item A cross comparison of the photometric metallicities to the spectroscopic measurements from  \citet[for several dozen sources;][]{Clementini2005}.
    \item Application of the technique to RRL in Reticulum, which has a well-defined spectroscopic metallicity \citep[e.g.,][]{Mackey2004} and well-sampled $I$ light curves \citep{Kuehn2013}. 
  \end{enumerate} 
Both tests validated the photometric metallicities obtained using the PL$_I$ and the results of this analysis for the Sculptor RRLs are given in Figure~\ref{fig:scl_mdf}.
In Figure~\ref{fig:scl_mdf}, the left panel shows the full distribution for Sculptor and the right panel emphasizes the difference in the population split into the Bt and Ft groups identified from the $V$-magnitudes.
 
The distribution of RRLs are centered at $-1.90$~dex, with {$\gtrsim$}90\% of the sample between $-1.2$ and $-2.6$~dex (left panel of Figure~\ref{fig:scl_mdf}). 
The Bt sample is on average more metal-poor, with mean metallicty of $-2.03$~dex, 
than the Ft sample that has a mean metallicity of  $-1.74$~dex (right panel of Figure~\ref{fig:scl_mdf}).

The distance to Sculptor was determined using the metal-independent PW relations for the $V$, $B-V$ and $V$, $B-I$ filter-color combinations (e.g., Equation \ref{eq:Wmags} and \ref{eq:3bandW}) with parameters from \citet[][their table 9]{Marconi15} and \citet[][their table 2]{MartinezVazquez2015}, respectively.
\citet{Marconi15} demonstrated that these PW relations are relatively insensitive to metallicity, which is important for the large metallicity spread in Sculptor. 
These PW relations were applied to each of three subsamples within the RRL population: (i)~RRab, (ii)~RRc, and (ii)~the global sample (both RRab and fundamentalized RRc).  

The zero-points of the PW relationships are not well constrained from observations as there are only good (\textless 10\%) trigonometric parallaxes for five RRL, even after the release of the TGAS parallaxes from \emph{Gaia} DR1 \citep[][and references therein]{lindegren_2016}
Thus, three different approaches were used to quantify the effect of both the zero-point ($\alpha$) and the slope ($\beta$) of the relation on the distances. 
These are: 
\begin{enumerate} 
  \item {\em Theoretical:} Using the zero-points and slopes of the predicted PW relations 
  \item {\em Semi-empirical:} Using the same theoretical zero-points and the empirical slopes, which were obtained from the fit performed in the plane of both observational Wesenheit magnitudes versus $\log$(P) the RRLs.
  \item {\em Empirical:} Adopting observed slopes used in the previous method, but using the zero-points 
determined from {\emph HST} trigonometric parallaxes \citep{Benedict_2011}. More specifically, RR Lyr, itself, to calibrate both the RRab and the global samples and RZ Cep to calibrate the RRc sample. 
\end{enumerate} 
The full details of these distance determinations can be found in \citet[][in particular, section~4 and their table~3]{MartinezVazquez2015}.

The zero-points are determined for each of the four RRab, global, and RRc PW relations. 
For the RRab and global relations, the zero-points agree quite well with theoretical ones. 
However, in the case of the RRc, the new zero-points are {$\sim$}30\% smaller than predicted, which is in agreement with the mean RRc magnitude determined from \citet{kollmeier_2013} from statistical parallax and highlights the need to use a large sample to determine robust zero points.

The final result is $\mu_0 = 19.62$~mag with $\sigma_{mean} = 0.04$~mag, which is in good agreement with previous determinations \citep[e.g.,][among others]{Rizzi2002,Pietrzynski_2008}.
The multiple methods required to set the zero-points and slopes of the PW relationships also emphasizes the role that high-quality trigonometric parallaxes for field RRLs, and later cluster RRLs, will play in  improving the accuracy of the distances determined via the RRL PL, PLZ, and PW relationships.

\subsubsection{A near-IR study of the RRLs in M4} \label{sec:rrlcase_nir} 

We present a case study using the GGC M\,4.
M\,4 is the closest GGC to the Sun and it hosts a sizable sample of RRLs \citep{Stetson2014}, but the cluster suffers from strong differential extinction. 
The mean reddening is $E(B-V)=0.37$~mag with a range of $\Delta E(B-V)=0.10$~mag -- or a mean $A_{B}$ ($A_{V}$) of 1.81~mag (1.37~mag), with variations at the 0.4~mag level across the cluster.
In contrast, the IR reddening is $E(J-H)=0.13$~mag with a variation of only $\Delta E(J-H)=0.04$~mag -- or a mean $A_{H}=0.22$~mag with variation at the $\sim$0.1~mag level. 

\begin{figure} 
  \centering
    \includegraphics[width=1.0\textwidth]{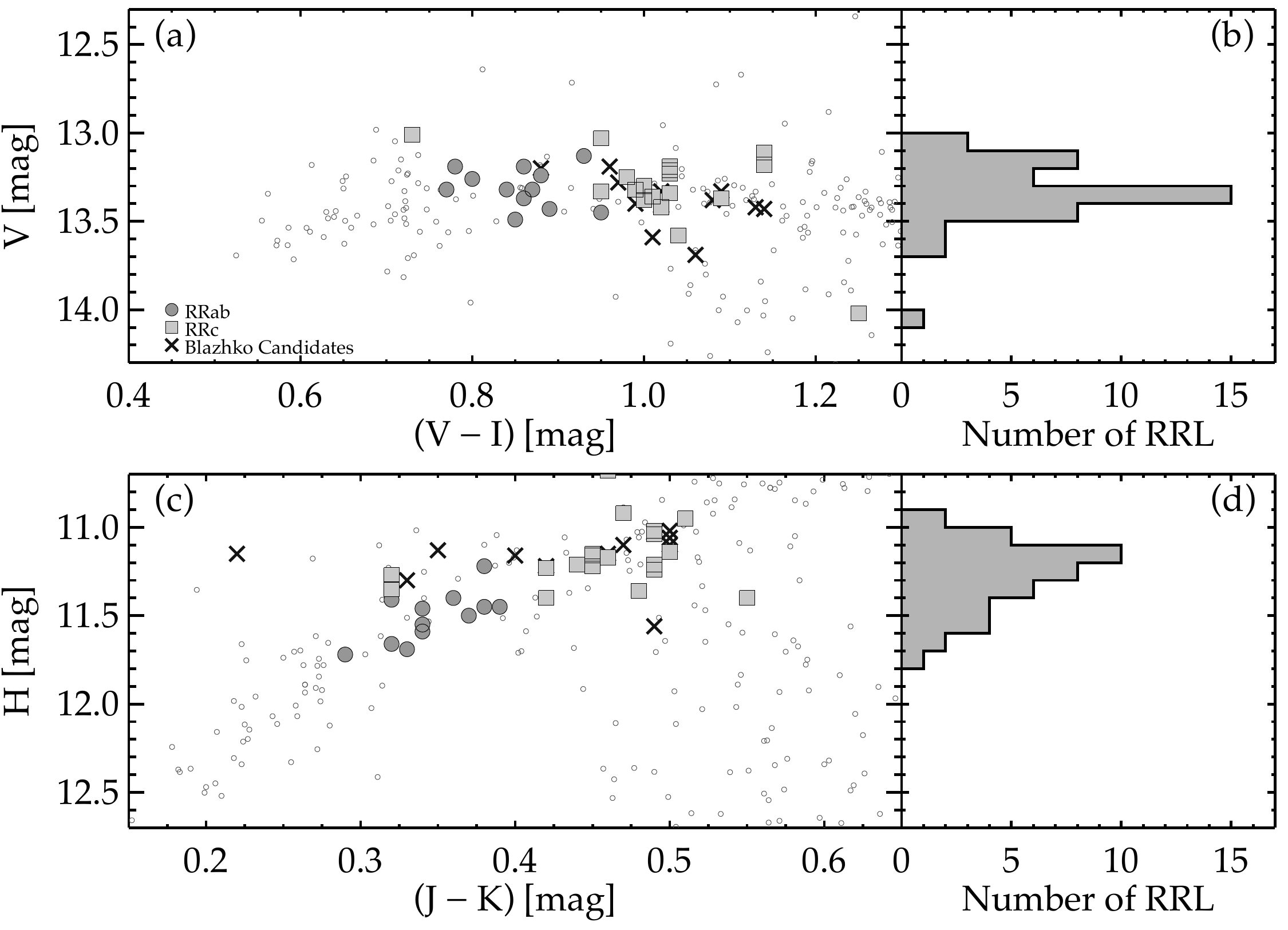}
	\caption{Optical and NIR CMDs for M\,4.
    \emph{(a)}: Optical $V-I$,$V$ CMD for the HB in M\,4 \citep[data from][]{stetson14a}. The RRab (circles), RRc (squares), and Blazhko candidate RRLs are indicated. 
    \emph{(b)}: $V$ magnitude histogram. The dispersion of the distribution is {$\sim$}0.18~mag, which, despite the differential reddening, is much narrower than that of Sculptor (Figure~\ref{fig:scl_cmd}) due to the lack of strong metallicity gradients. 
 	\emph{(c)}: NIR $J-K$,$H$ CMD for the HB in M\,4 \citep[data from][]{stetson14a}. Unlike the optical (a), the NIR HB has a noticeable slant that gives rise to the PL. 
    \emph{(d)}: $H$ magnitude histogram where the width is due to the PL.
The y-axis for all panels is 2.0~mag for ease of comparison.
	\label{fig:m4_cmds}}
\end{figure} 

The data for M\,4 is similar in its volume and time-baseline to that for the Sculptor dwarf from Section~\ref{sec:rrlcase_opt} and is described in full in \citet[][]{Stetson2014}. 
The optical data was obtained in 18 runs over 16 years for a total of over 5000 images in $U$, $B$, $V$, $R$, and $I$ filters. 
The NIR data was obtained in 18 runs over 10 years in $J$, $H$, and $K$.
The final RRL census in M\,4 contains 32 RRab and 12 RRc for a total of 45 RRL. 

Figure \ref{fig:m4_cmds}a shows the optical CMD for M\,4 and Figure \ref{fig:m4_cmds}b is a histogram of the mean magnitudes in $V$ for the RRLs, both of which can be compared directly to the corresponding panels of Figure~\ref{fig:scl_cmd} showing the same for Sculptor.
M\,4 has a dramatically smaller metallicity spread than Sculptor, but it suffers from strong differential extinction in the optical, which creates a broadened distribution of RRL magnitudes. 
Figure \ref{fig:m4_cmds}c is the NIR CMD for M\,4 and Figure \ref{fig:m4_cmds}d is a histogram of the mean magnitudes in $K$ for the RRLs. 

\begin{figure} 
	\includegraphics[width=\textwidth]{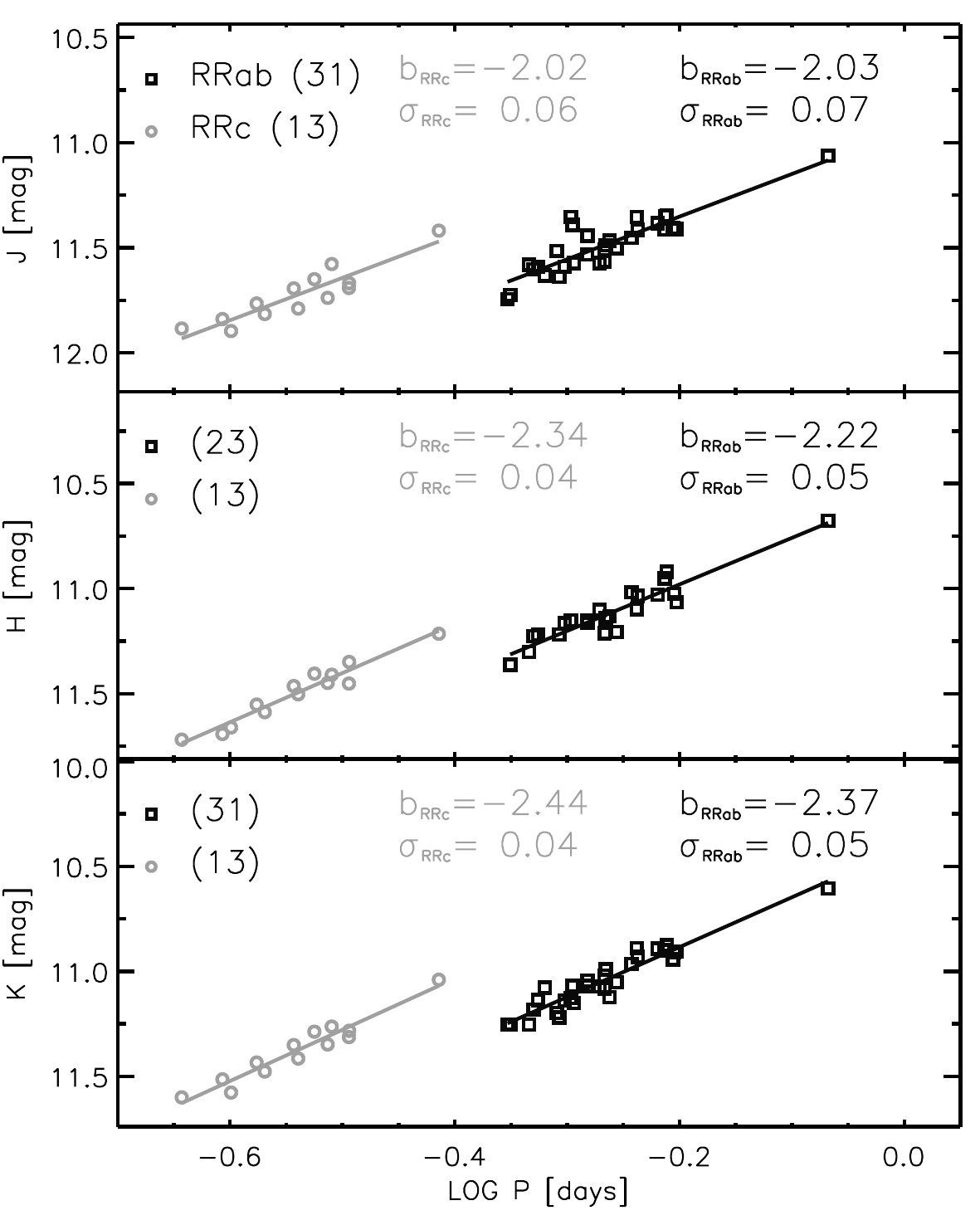}
	\caption{ NIR PL relations for RRL in M\,4.
    \emph{Top}: Empirical $J$-band PL relations of the RRLs of M\,4. 
     	Light circles represent RRcs, while dark squares mark the position of RRabs. 
    	Slopes ($b$) and dispersions ($\sigma$) of the relations are labeled.
 	\emph{Middle}: same as top, but for $H$-band PL relations. 
	\emph{Bottom}: same as top, but for $K$-band PL relations}
	\label{fig:m4_plnir}
\end{figure} 

\subsubsection*{Period--luminosity relations} 

Figure \ref{fig:m4_plnir} shows the empirical PL relations in the $J$ (top), $H$ (middle), and $K$ (bottom) bands for the RRab and RRc in M\,4. 
The RRab and RRc RRL are shown independently in Figure \ref{fig:m4_plnir} with their native periods.
The resulting PL relations are given below. 
First, for the RRab stars:
\begin{equation}
 \begin{split}
  J_{RRab} =& -2.030~(\pm~0.204)~\log P~  + 10.946~(\pm~0.056) ~\sigma=0.065 \\
  H_{RRab} =& -2.215~(\pm~0.176)~\log P~  + 10.537~(\pm~0.047) ~\sigma=0.050 \\
  K_{RRab} =& -2.372~(\pm~0.142)~\log P~  + 10.410~(\pm~0.039) ~\sigma=0.045  
 \end{split}
\end{equation}
For the RRc, 
\begin{equation}
 \begin{split}
  J_{RRc}  =& -2.020~(\pm~0.273)~\log P~ + 10.634~(\pm~0.148) ~\sigma=0.056 \\
  H_{RRc}  =& -2.340~(\pm~0.179)~\log P~ + 10.232~(\pm~0.097) ~\sigma=0.037 \\
  K_{RRc}  =& -2.440~(\pm~0.198)~\log P~ + 10.058~(\pm~0.108) ~\sigma=0.041 
 \end{split}
\end{equation}

\noindent Lastly, the global PL uses fundamentalized periods for the RRc variables (e.g, Equation~\ref{eq:fotofu}), which is not shown explicitly in Figure \ref{fig:m4_plnir}: 
\begin{equation}
 \begin{split}
  J_{Glob} =& -1.793~(\pm~0.109)~\log P_F~ + 11.002~(\pm~0.035) ~\sigma=0.064 \\
  H_{Glob} =& -2.408~(\pm~0.082)~\log P_F~ + 10.492~(\pm~0.027) ~\sigma=0.046 \\
  K_{Glob} =& -2.326~(\pm~0.074)~\log P_F~ + 10.420~(\pm~0.024) ~\sigma=0.043 
 \end{split}
\end{equation}

\noindent These slopes and their dispersions are also labeled on the top of each panel in Figure \ref{fig:m4_plnir}. 
As expected, the slope increases with wavelength as PL approaches the PR relation (e.g., Figure~\ref{fig:pl_slope_lambda2}). 
The dispersion in $K$ (0.06~mag) is half that of the $R$ for this cluster \citep{Braga2015}, further emphasizing a strength of the NIR PL. 

The distance to M\,4 can be determined once the zero-point of the PL is calibrated. 
Similar to the methods for Sculptor, this can be done either on an empirical or on a theoretical basis, as we describe below:
 \begin{enumerate} 
   \item {\it Empirical}: Zero points are determined by adopting the \emph{HST} parallaxes \citep{Benedict_2011} and the mean $JHK$ magnitudes \citep{sollima08} for RR Lyr (an RRab). Note that the \emph{HST} parallaxes have smaller uncertainties than those from TGAS in \emph{Gaia} DR1 \citep{lindegren_2016}.
   \item {\it Theoretical}: Zero points are adopted from the predictions of the theoretical PLZ relations \citep{Marconi15}. The metallicity term is taken into account by adopting a fixed [Fe/H]$=-1.1$~dex, which is the average iron abundance of M\,4 \citep{marino08,malavolta14}.
 \end{enumerate} 

\noindent The two methods provide distances of $\mu_{0}=11.35\pm 0.03$ (standard error) $\pm 0.05$ (standard deviation) mag and $\mu_{0}=11.283\pm 0.010$ (standard error) $\pm 0.018$~mag (standard deviation), respectively \citep[see][for additional details]{Braga2015}. 
Both are in agreement within 1~$\sigma$ with estimates of the distance modulus from the literature, in particular that derived from eclipsing binaries of $\mu_{0}=11.30\pm 0.05$~mag \citep[1-$\sigma$ agreement;][]{kaluzny_2013}.

\begin{figure} 
   \includegraphics[width=\textwidth]{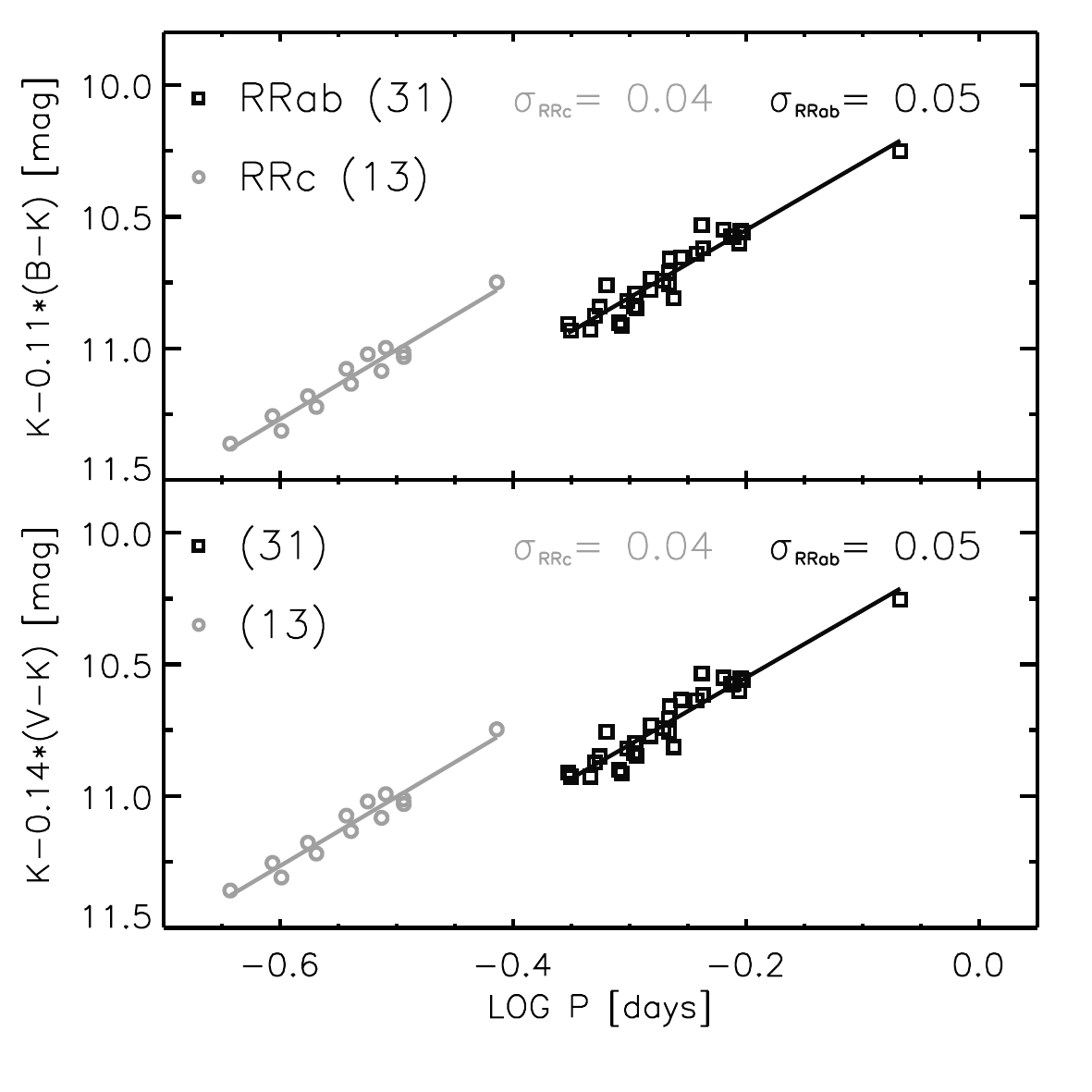}
   \caption{Three-band PW relationships for RRL in M4.
      {\emph Top:} Empirical PW($K$,$B-K$) relations of the RRLs of M4. 
       Light circles represent RRcs, while dark squares mark the position of RRabs. 
       The dispersions ($\sigma$) of the relations are labeled.
      {\emph Bottom:} same as top, but for PW($K$,$V-K$) relations}
    \label{fig:rrl_pw}
 \end{figure}

Figure \ref{fig:rrl_pw} shows two PW relationships in $B$,$B-K$ and $K$,$V-K$ for M\,4 from \citet{Braga2015} that combine both optical and NIR magnitudes. 
Using these relationships, calibrated in a similar fashion, provide $\mu_{0}=~11.272\pm 0.004 \pm 0.013$~mag, in perfect agreement with the distance moduli determined from the PLs. 
For the two relationships shown in Figure \ref{fig:rrl_pw}, the dispersions are between 0.04 and 0.05~mag.
The small dispersions and the agreement between these distance moduli and those from the NIR PL provide empirical validation that the PW relationships are indeed insensitive to reddening and that the assumptions of the PW relationship (e.g., ``Cardelli's Law'') work well for M\,4 in the NIR.

\subsubsection{A MIR study of RR Lyrae in M4} \label{sec:rrlcase_mir} 
\begin{figure} 
	\includegraphics[width=\textwidth]{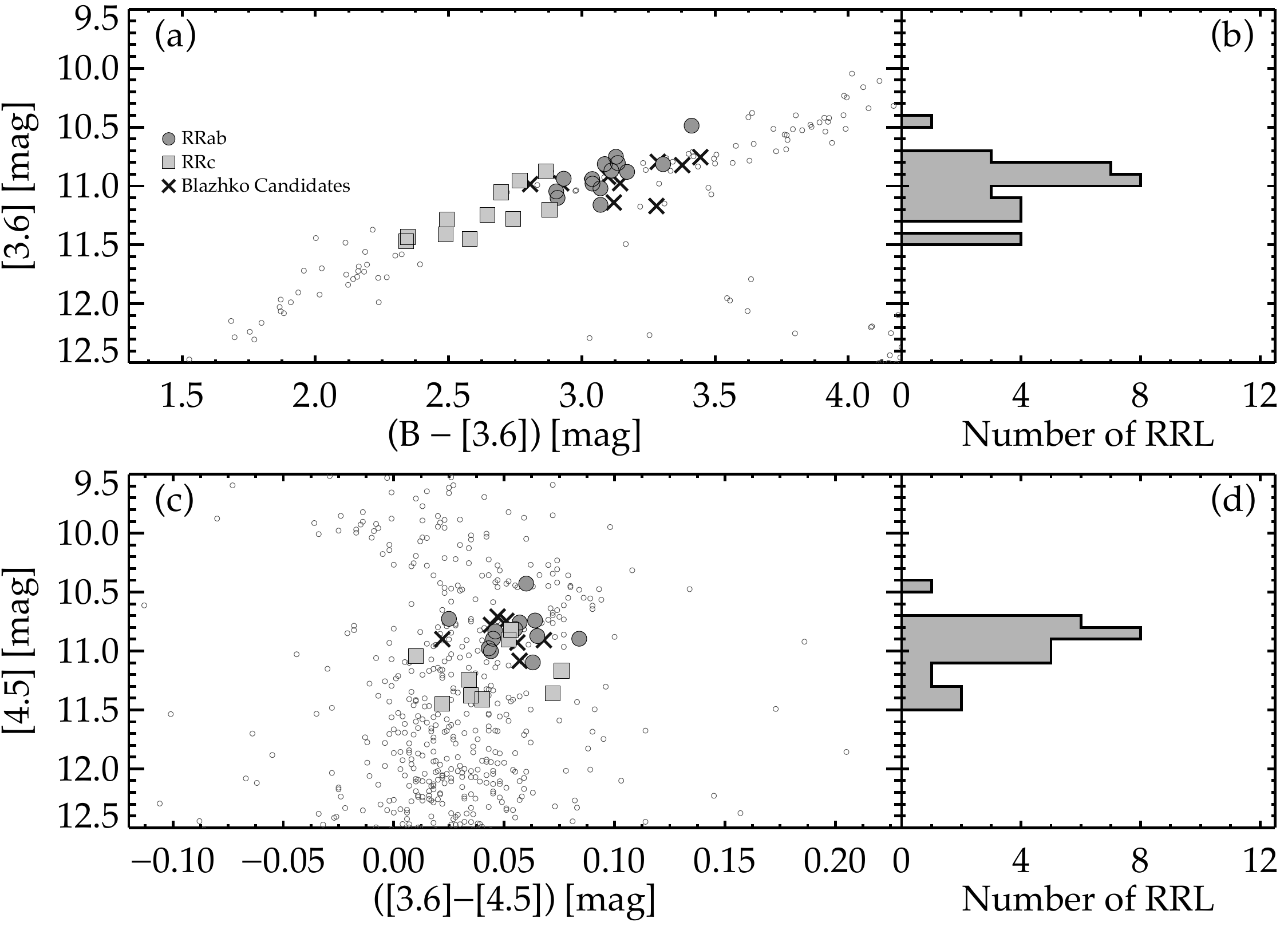}
	\caption{MIR photometry for the HB in M\,4. 
     (a) Joint opt-MIR CMD for the HB in M\,4 \citep[data from][]{stetson14a,Neeley2015} with the RRL common to both datasets indicated.
     (b) Distribution of [3.6]-mag for the RRLs in M\,4.  
     (c) MIR CMD for the HB in M\,4 \citep[data from][]{Neeley2015} that demonstrates the HB is not readily identifiable in MIR CMDs. Identification and classification of RRLs is much more readily completed in the opt or NIR, but the RRc variables are noticeably fainter. 
     (d) Histogram of [4.5] magnitudes for the RRLs in M\,4.  
	\label{fig:m4_cmdmir}}
\end{figure} 

The MIR case study also uses the closest GGC, M\,4. 
Due to the smaller amplitudes at longer wavelengths, it is not practical to identify and derive the pulsation characteristics using only MIR data (e.g., Figure \ref{fig:lc}). 
Therefore we rely on ground-based optical and NIR data described previously to provide the coordinates and periods of our target RRL, which then sets the optimal strategy to derive well sampled light curves in the MIR \citep[see also][]{hendel_2018,muraveva_2018}.

Intensity mean magnitudes are computed either from Fourier fitting techniques \citep[as in][]{dambis_2014} or the GLOESS method \citep[as in][]{monson2017}.
Using the exquisite multi-band data available for M\,4 and described in this chapter, the extinction and distance to M\,4 can be measured using a semi-empirical method, as described in \citet{Neeley_2017}.
The photometry was simultaneously fit in each band to theoretical PLZ relations, finding an extinction of $A_V = 1.45\pm0.12$ \citep[assuming $R_V=3.62$ measured directly in the foreground of M\,4 by][]{hendricks_2012} and a distance modulus of $\mu_0~=~11.257~\pm~0.035$~mag. 
Both this value and the fully empirical results above are in good agreement with other distance methods, \citep[e.g. eclipsing binaries from][]{kaluzny_2013} as well as from RRL in the NIR \citep{Braga2015}. 

\begin{figure} 
\includegraphics[width=\textwidth]{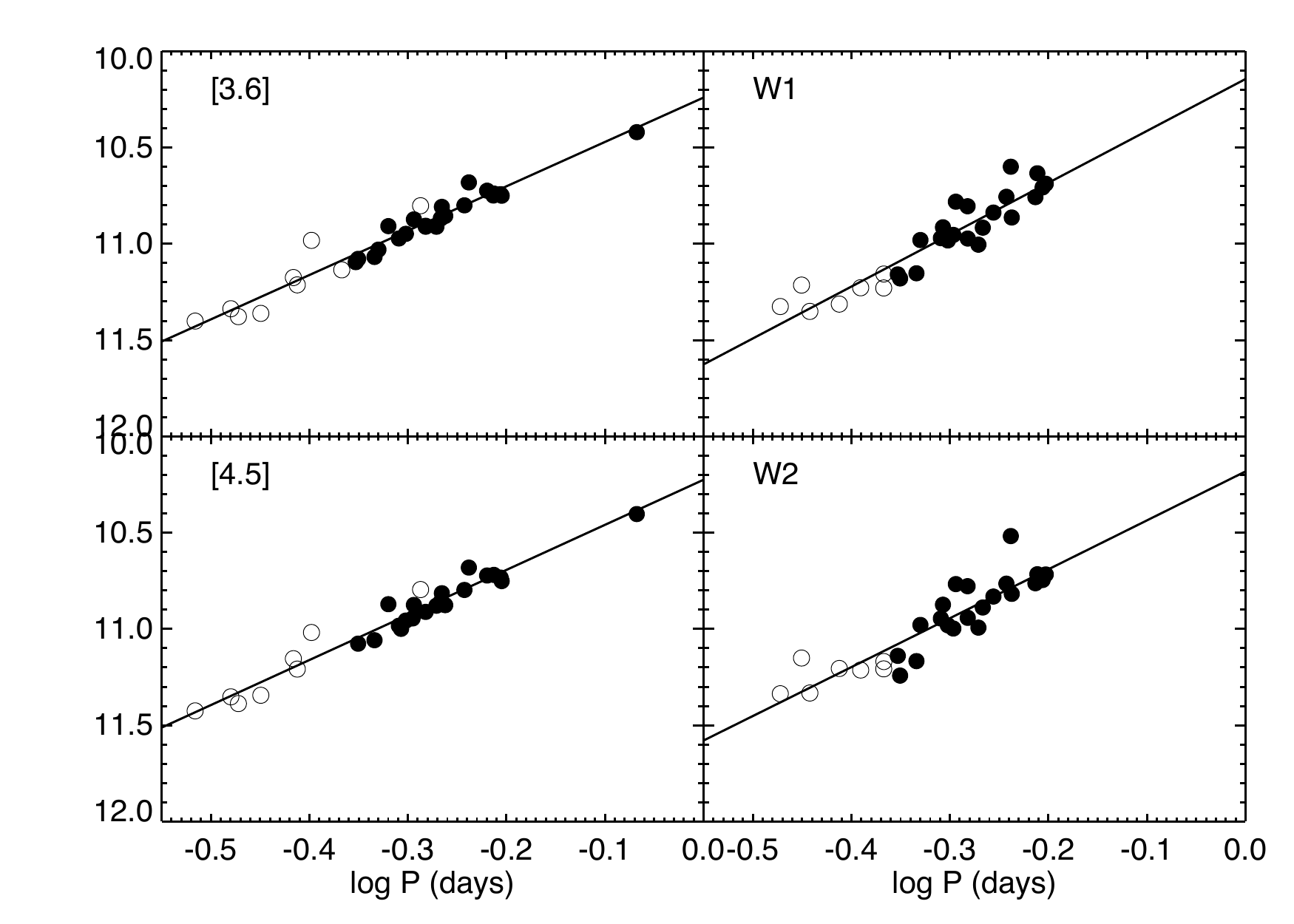}
\caption{IRAC (left) and WISE (right) PL  relations for RRL in M\,4 using data from \citet{Neeley2015} and \citet{dambis_2014}. Periods of the first overtone pulsators (open circles) have been fundamentalized (e.g., Equation~\ref{eq:fotofu}).}
\label{fig:M4-pls}
\end{figure}

The empirical PL relations using both \emph{Spitzer} and \emph{WISE} data are given in Figure~\ref{fig:M4-pls}. 
Periods of the RRab stars (shown as open circles) have been fundamentalized following Equation~\ref{eq:fotofu}. 
The empirical MIR PL relations for M\,4 for the ``global sample'' are as follows:
\begin{equation} \label{eq:m4_mir_pl} 
\begin{split} 
  {[3.6]}_{Glob}~& =~-2.304(\pm0.105)~\log P_F~+~10.241(\pm0.034) \\
  {[4.5]}_{Glob}~&=~-2.340(\pm0.104)~\log P_F~+~10.225(\pm0.034) \\
  W1_{Glob}~&=~-2.694(\pm0.213)~\log P_F~+~10.145(\pm0.213) \\
  W2_{Glob}~&=~-2.540(\pm0.248)~\log P_F~+~10.182(\pm0.249).
\end{split} 
\end{equation} 

\noindent The dispersion on the IRAC relations is 0.05~mag and is approaching the expected intrinsic scatter of the PL relation ({$\sim$}0.03~mag). 
The difficulty of deriving accurate photometry in clusters using WISE data (due to the larger PSF) results in a larger dispersion of {$\sim$}0.09~mag. 
The observed slopes are in good agreement with theoretical PL relations derived from pulsation models \citep[see discussion in][]{Neeley_2017}.

To use the PL to determine a distance, the absolute zero point must be calibrated with RRL of known distances.
Similar to the NIR study of M\,4, there are two methods:
\begin{enumerate} 
\item {\it Empirical} : Adopting observed slopes from M\,4 RRLs and zero points are determined using trigonometric parallaxes for Galactic RRLs (either from {\emph HST}-FGS or {\emph Gaia}). 
\item {\it Semi-empirical}: The distance and extinction are measured simultaneously by fitting multi-wavelength data to the theoretical PLZ relations \citep{Marconi15,Neeley_2017}.
\end{enumerate} 

The distance modulus of M\,4 can be derived by comparing the empirical and calibrated PL relations, as was done in \citet{Neeley_2017}.
This results in $\mu_{[3.6]} = 11.353~\pm ~0.095$~mag and $\mu_{[4.5]} = 11.363\pm 0.095$~mag, which agrees with the \citet{kaluzny_2013} and \citet{Braga2015} measurements.
Using the exquisite multi-band data available for M\,4 and described in this chapter, the extinction to M\,4 can be measured by fitting the distance and extinction law directly to each star as was described in \citet{Neeley2015,Neeley_2017}, finding $A_{V}=1.45\pm 0.12$~mag using $R_{V}~=~3.62$ measured directly in the foreground of M\,4 by \citet[][taking into account M\,4 is directly behind the $\rho$~Oph cloud]{hendricks_2012}.

\subsection{Summary}

RRLs have been so far excellent standard candles \citep[e.g.,][]{tam08}. 
RRL can be used in individual Galactic structures, extra-galactic systems, and are likely the only relatively-luminous standard candle that can be used to trace ultra-faint galaxies and tidal streams) that populate the Galactic halo \citep[e.g.,][]{fiorentino15a,fiorentino16b,Andrievsky_2018}. 
Although they are typically discovered and classified in the optical, photometry of RRL can be used to determine distances in a number of passbands, from the optical to the mid-infrared using theoretical and/or empirical calibrations of the RRL PLZ or the FOBE method.

On the theoretical side, current uncertainties in the PL relations come from our lack of perfect knowledge of the evolutionary effects (mixing length, turbulent convection, mass loss, helium abundance), of the input physics (atomic diffusion, extra-mixing, neutrino losses), and of the color-temperature transformations that are adopted to transform bolometric magnitudes and effective temperatures, into the observed magnitudes and colors. Given that theoretical relations are often preferred over empirical ones, these improvements are important.

As demonstrated by the case studies, the distances from the RRL are limited by the incomplete knowledge of the RRL PLZ (in all bands), in particular by the lack of a strong trigonometric parallax calibration. 
As such, single studies will often apply several calibrations to determine the distance. 
Studies using RRL are also limited by the imprecise knowledge of the metallicity for individual RRLs, although there are some forms of the RRL PLZ that have only a small dependence on metallicity.

The next generation of large telescopes will permit use of RRLs to distances of {$\sim$}5--6~Mpc \citep[see e.g.,][]{deep11}, which have previously only been accessible via Cepheids. 
The proposed adaptive optics systems for 30-meter class telescopes will be optimized for the NIR wavelengths, which necessitates sound calibration of the NIR PLZ and PWZ relations.


\section{The type II Cepheids} \label{sec:2cephs}

\defcitealias{Soszynski_2008}{S08}
\defcitealias{Soszynski_2010}{S10}

\subsection{Overview} 

Type II Cepheids (T2Cs) are evolved, low-mass stars (typically \textless~1~M$_{\odot}$) and have periods between 1 and 100~days, which roughly spans the range between RRLs and Miras \citep[for the former, see the previous section and for the latter see][]{Subramanian_2017}.
The discovery of T2Cs played a critical role in the establishment of the modern extragalactic distance scale in the 1950s \citep[formalized in the reviews of][at the seminal Vatican Conference]{baade58b,baade58c}.
Before its revolutionizing discovery, the stars adopted to build the PL relation of Cepheids were a mixture of both (i)~young, intermediate-mass stars characterized by a radial distribution typical of thin disk stars and (ii)~old, low-mass stars with a radial distribution typical of the Galactic halo or bulge. 
The former are the classical Cepheids of Population-I, while the latter are known as Type-II Cepheids (T2Cs), to further emphasize the difference in age and stellar mass of their progenitors. 
By discovering the presence of two distinct classes of variables Baade effectively doubled the size (and the age) of the Universe \citep{Baade_1956,Fernie_1969} by identifying and resolving a major issue in the determination of the Hubble constant at that time.

The discovery of T2Cs can be considered as one of three seminal discoveries that formed the modern distance ladder, joining (i)~the measurement of stellar parallax by \cite{bessell1871} and (ii)~the discovery of the Leavitt Law\footnote{The consensus of the scientific community present for the {\it Thanks to Henrietta Leavitt Symposium} held on Nov~8, 2008 was to officially adopt the term `Leavitt Law' for the Cepheid PL relationship. The Council of the American Astronomical Society provided similar advice shortly thereafter. The reader is refferred to \url{https://www.cfa.harvard.edu/events/2008/leavitt/} and \url{https://aas.org/archives/Newsletter/Newsletter_146_2009_05_May_June.pdf}, respectively.} or PL, first noted by \citet{Leavitt_1908} and then quantified by \citet{leavitt12}. 
For a review of the Leavitt Law, readers are referred to \citet{Subramanian_2017}. 
After the separation of classical Cepheids and T2Cs, the latter have received far less attention than the former.
Nevertheless, there are several good reviews on T2Cs and related objects which are recommended to interested readers:
\citet{Harris_1985, Wallerstein_1984, Wallerstein_2002, Sandage_2006, Welch_2012, Feast_2010, Feast_2013, catelan_2015}.

Dating back to almost half a century ago \citep{wallerstein70}, T2Cs have been associated with GGCs characterized by a blue (hot) HB morphology. 
This means that the HB for these clusters is well populated on the blue tail, which is comprised of hot and extreme HB stars. 
Roughly one hundred of these variables have been detected in GGCs \citep{clement01}.  
They have also been observed in stellar systems with complex star formation histories like the Magellanic Clouds
 \citep[\citetalias{Soszynski_2008,Soszynski_2010},][]{Groenewegen_2017a,Groenewegen_2017b} and there are a preliminary identifications in the extragalactic stellar systems with examples being: IC\,1613, M\,31, M\,33, M\,106 and NGC\,4603 \citep{Majaess_2009}. 
On the other hand, Fornax is the only nearby dSph galaxy for which even a handful of T2Cs have been detected \citep{Bersier_2002}; if these stars were a signpost for old metal-poor populations, then dSphs, which are home to many RRLs, would be expected to be hosts as well.
The paucity of these objects in dSphs has been explained by their lack of blue HB morphologies. 
However, an observational bias cannot be excluded, since we still lack long-term photometric monitoring for the bulk of distant and diffuse stellar systems, although recent efforts have dramatically improved our variable star census \citep[e.g.,][among others]{Stetson2014}. 

\subsection{Observed pulsational properties}\label{obs}
 
\begin{figure} 
   \centering
   \includegraphics[width=0.9\columnwidth]{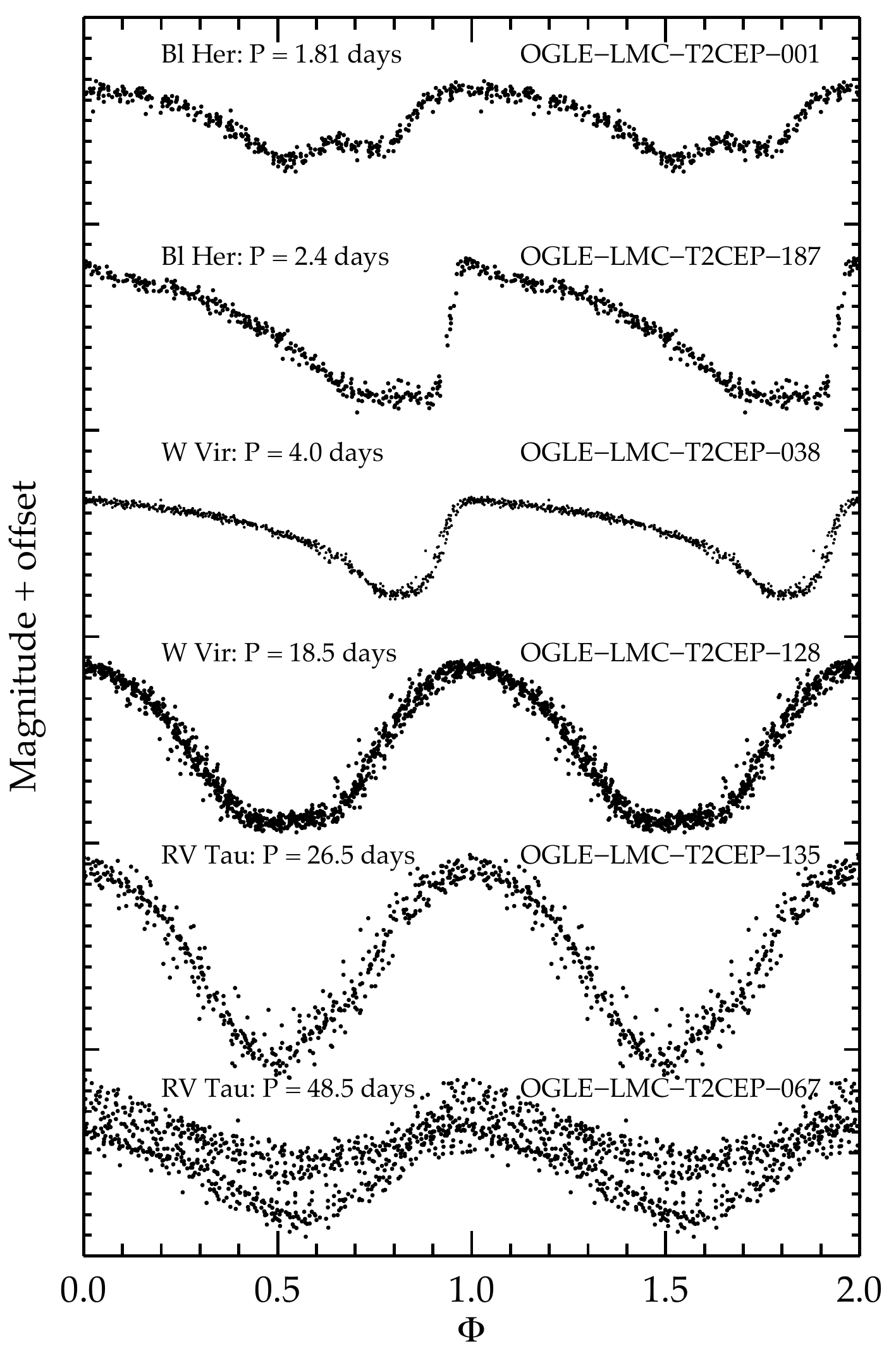}
   \caption{ \label{fig:t2c_lc} 
     Example optical, $I$-band, light curves for T2C in the LMC from the OGLE survey \citepalias{Soszynski_2008}. 
     From top to bottom there are two representative light curves for stars of each of the BL~Her, W~Vir, and RV~Tau classes. 
     The light curves have been normalized to amplitudes of 1~mag to allow comparison of the shapes and structures, with the minor-tick marks being 0.1~mag. 
     The light curve shapes are easily distinguished from those of RRL with similar periods (e.g., Figure~\ref{fig:lc}).}
\end{figure}

\begin{figure*} 
	\centering
	\includegraphics[width=\textwidth]{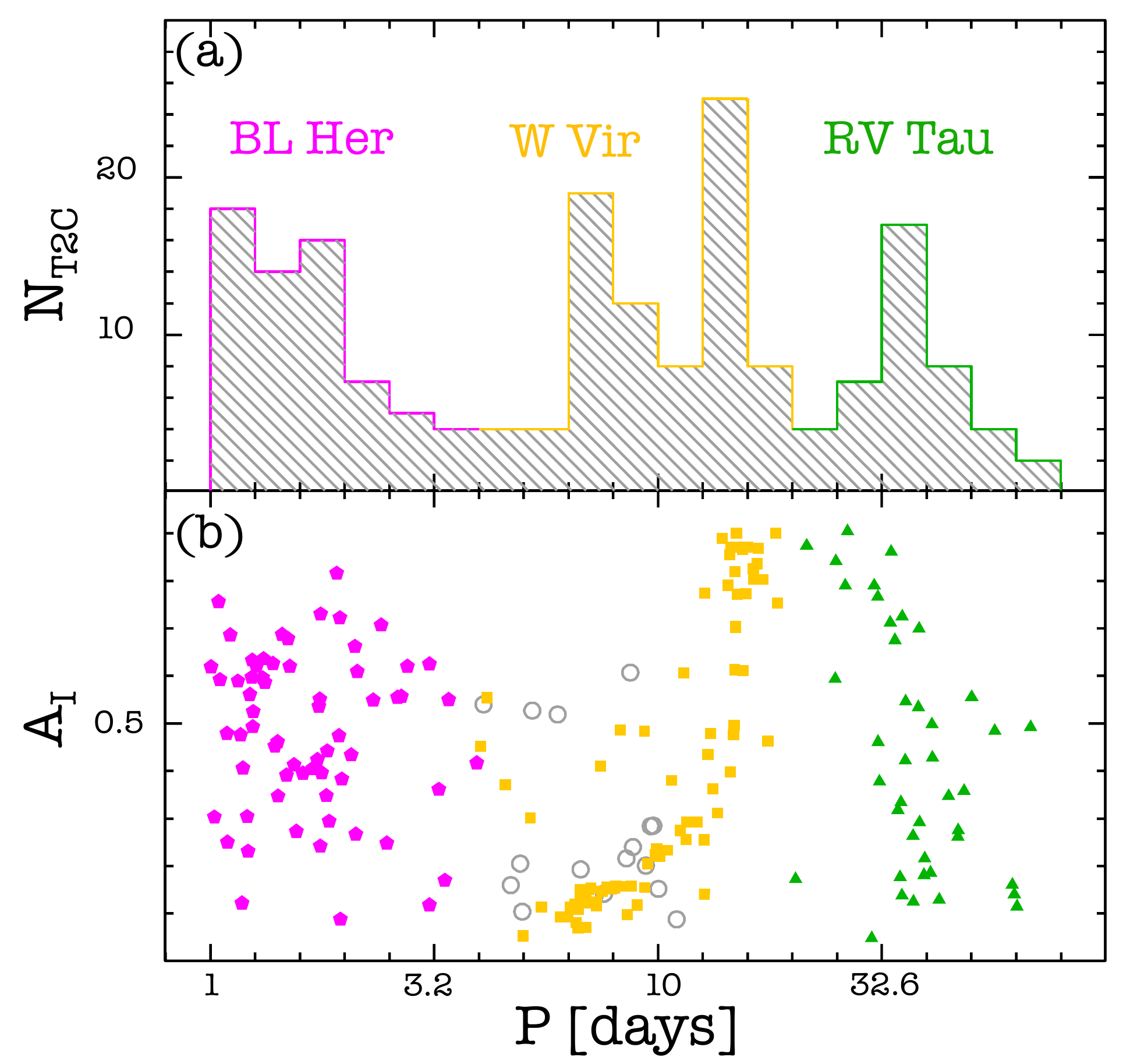}
	\caption{Basic properties of the T2Cs as defined in the LMC from the OGLE survey \citepalias{Soszynski_2008}. 
   {\emph (a)}~Period distribution of the T2Cs as divided into the BL~Her, W~Vir, and RV~Tau objects following the \citetalias{Soszynski_2008} classification scheme. 
   {\emph (b)}~$I$-band amplitude ($A_{I}$) versus period diagram (known as the Bailey diagram for the RRLs). The pentagons correspond to BL~Her stars, the squares to W~Vir stars, the open circles to peculiar W~Vir objects, and triangles to RV~Tau stars.} 
	\label{periods}
\end{figure*}

Example light curves for T2Cs for a wide span of periods are given in Figure~\ref{fig:t2c_lc} using data from OGLE \citepalias{Soszynski_2008}. 
The T2Cs show a striking degree of variation in comparison to the RRLs (Figure~\ref{fig:lc}), because T2Cs are often divided into subpopulations.
The period distribution from \citetalias{Soszynski_2008} is given in Figure~\ref{periods}a and shows three groupings by period.  
Considering this period distribution, \citetalias{Soszynski_2008} presented a classification scheme for the T2Cs in the LMC using the pulsation properties of the stars, including their amplitudes, light curve shapes, and periods. 
These pulsation properties are both distance and reddening independent and, thus, are good means for typing the stars.
The three groups are divided into the classical sub-types with the following criteria:
\begin{enumerate}
  \item BL Herculis (or BL~Her) with 1$\le$P$\le$4 days, 
  \item W Virginis (or W~Vir) with 4$<$P$\le$20 days, 
  \item RV Tauri (or RV~Tau) with $P$$>$20 days. 
\end{enumerate}

The period-amplitude diagram in the $I$-band ($A_{I}$) for the LMC T2Cs is given in Figure~\ref{periods}b with the amplitudes ranging from {$\sim$}0.1 to {$\sim$}1~mag. 
Figure~\ref{periods}b further demonstrates that there is a set of structure in the period-amplitude diagram similar to what is seen in the Bailey diagram for RRLs (see e.g., Figure~\ref{fig:scl_pa}), albeit the T2C groups are less well defined than the RRLs. 
In addition to these classical groupings, \citetalias{Soszynski_2008} have also proposed a new group of T2Cs, the peculiar W Vir (or pW) stars known as pW variables, that display peculiar light curves and are brighter than typical T2Cs at a fixed period; these stars are indicated by open circles in Figure~\ref{periods}b. 
This classification scheme, however, has not been well settled, because the two minima of the period distribution that drive the definition of the sub-groups depends on the metal abundance and, likely, also on the environment. 
This means that the quoted period ranges will not universally define the three sub-groups and, thus, further investigations in a more diverse set of objects may yield further sophistication to our sub-classifications. 

It is also important to note that different classification schemes have been adopted in different studies throughout the years. 
For example, General Catalog of Variable Stars\footnote{Available: \url{http://www.sai.msu.su/gcvs/gcvs/}} \citep[GCVS, ][]{samus_2017} uses a boundary of 8 days to separate the T2Cs, `CWA' for $P>8$~days and `CWB' for $P<8$~days. 
In addition, GCVS uses the label `RV' for RV~Tau stars for radially pulsating supergiants characterized by the presence of alternating primary and secondary minima.
Different schemes were proposed by \citet{Joy_1949} and later by \citet{Diethelm_1990} that take into account factors beyond the period, such as the light curve shape and spectroscopic features \citep[see also][]{Sandage_2006}.
Several authors have further suggested that the RV~Tau stars are also heterogeneous as a class.
While six objects in GGCs have been claimed to be RV Tau stars, some authors have doubted this classification from both the photometric \citep{Zsoldos_1998} and the spectroscopic point of view \citep{Russell_1998}. 
The reader should be cautious, when exploring this topic, due to the complexities of these classifications. 

\begin{figure} 
	\centering
	\includegraphics[width=\textwidth]{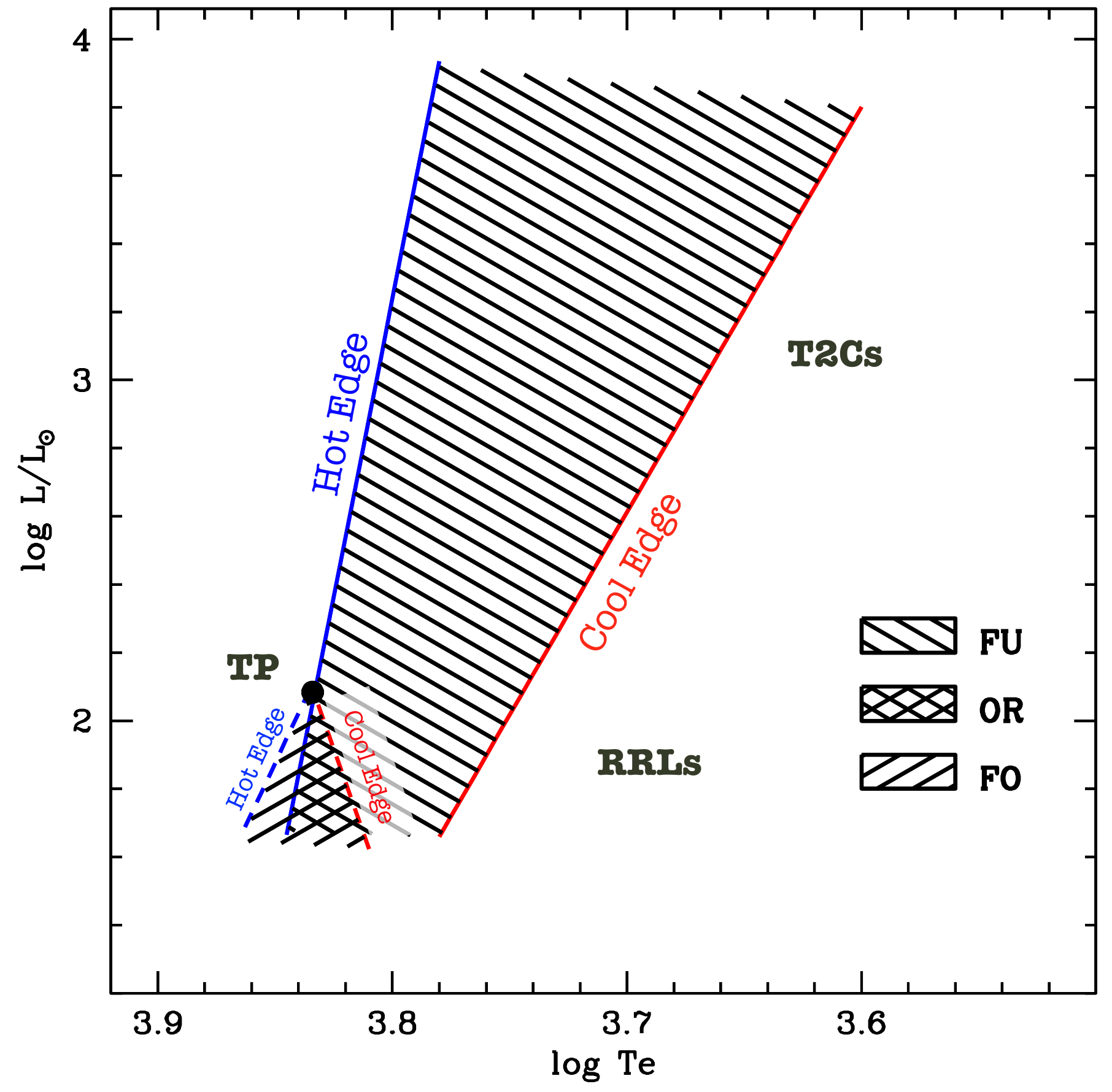}
	\caption{Predicted topology of the instability strip in the luminosity and 
		effective temperature range typical of RRLs and T2Cs \citep{Marconi15}. 
        The the hot and cool edges of the strip are indicated for the FU mode (solid edges), FO mode (dashed edges), and their overlap region (``OR''; double hatched).  
		Stars more luminous than the transition point (``TP''; see the text for definition) do not pulsate in the FO. 
		Only in the region demarcated by the dashed edges can FO pulsators (RRc) attain a stable limit cycle.  
        In the region  labeled ``OR'', variables can pulsate simultaneously in the FO and in the FU mode (RRd, or mixed-mode pulsators). 
        The region bounded by the solid edges represent the regions where FU pulsators can attain a stable limit cycle. 
        The luminosity for the bulk of the T2Cs excludes them from the FO region. }
	\label{figinsta}
\end{figure} 

\subsection{Theoretical pulsation predictions}\label{theory}

There is a general consensus that T2Cs are in a post-HB phase with a C+O degenerate core with H-shell burning as the main energy source.  
As is well known, Zero-Age HB stars have a wide range of effective temperatures depending on the total mass (or the envelope mass). 
After the He exhaustion in the core, they start to climb up in the HR diagram evolving mainly into AGB, but this climb can begin at a range of effective temperatures on the HB.
The evolutionary paths of the post-HB phase are predicted to be highly dependent on the ZAHB position. 
Stars coming from the blue HB tail that go through the IS at luminosities brighter than those of RRL, are found to be pulsating stars and commonly called T2Cs. 
Calculated evolutionary tracks were pioneered by \citet{Gingold_1976,Gingold_1977} with \citet{Smolec_2016} and \citet{Bono_2016} providing up-to-date theoretical calculations. 
However, we still lack a firm understanding of the production routes for T2Cs, in particular specific observational evidence for or against the postulated paths \citep[though, we note that detailed studies such as][provide a solid starting point]{Groenewegen_2017a,Groenewegen_2017b}. 
An interesting question is, for example, the possibility of excursions, first suggested by \citet{Schwarzschild_1970}, from the AGB to the hotter side reaching the instability strip. 
The early investigations \citet{Gingold_1976,Gingold_1977} suggested such routes ({\it Gingold nose}) play an important role on explaining a fraction of T2Cs, but such excursions are not expected to take place commonly if at all according to more recent calculations \citep[see][]{Smolec_2016,Bono_2016}. 

In contrast to the RRLs, theoretical investigations of the pulsation properties of T2Cs are quite limited. 
Early studies based on linear models were discussed by \citet{Wallerstein_1984} together with their period distribution, light- and velocity curves, and their location in the color magnitude diagram. 
Subsequently, a detailed analysis of T2Cs pulsation properties was performed by \citet{Fadeev_1985} using non linear radiative models. 
In particular, \citet{Fadeev_1985} investigated the modal stability and provided theoretical constraints on the PL relation for T2Cs, but their approach neglected the convective transport and, in turn, the physical mechanism that causes the quenching of radial oscillations. 
This means that this approach cannot predict the location of the cool (red) boundary of the IS. 
Using a similar approach, a thorough analysis of the limit cycle stability of T2Cs was presented by \cite{Kovacs_1988}. 
They have shown that the pulsation behavior changes from single periodic, to period doubling, and eventually to chaotic when increasing the pulsation period.

A more comprehensive theoretical scenario including a time-dependent convective transport was presented for BL~Her stars by \cite{bono97e,marconi07,dicriscienzo07}.
In \cite{bono97e}, a systematic theoretical investigation of T2Cs has shown that fundamental pulsators are primarily expected among this class of variable stars.
Moreover, they derived a period--luminosity--amplitude relation and found that it is independent of metallicity for the metal poor regime of $-2.3\le {\rm [Fe/H]}\lesssim -1.5$ that was explored in the study. 
These theoretical predictions match quite well the pulsation properties observed for GGCs.

\citet{marconi07} presented the full morphology of the theoretical IS for both fundamental and first overtone pulsators and a detailed atlas of both the light and radial velocity curves. 
Their models confirm that the FO IS is both narrow and limited to faint luminosities,
providing new constraints on the {\sl transition point}. 
The topology of the IS for the luminosity and effective temperature range of the RRLs and T2Cs is given in Figure~\ref{figinsta}. 
T2Cs populate the IS at luminosities brighter than the so-called {\sl transition point} \citep{Stellingwerf_1979}.
The {\sl transition point} corresponds to the luminosity at which the cool (red) edge of the FO IS matches the luminosity and the effective temperature of the hot (blue) edge of the FU. 
This means that FO pulsators do not attain a stable limit cycle for luminosities that are brighter and effective temperatures that are cooler than the {\sl transition point}. 
In consequence, most T2Cs are fundamental pulsators. 
The first overtones, if any, are only permitted for the low luminosity or short period T2Cs. 
The empirical evidence, originally brought forward by \cite{Mcnamara_1995} and more recently by \citetalias{Soszynski_2008} and \citetalias{Soszynski_2010} for Galactic and Magellanic T2Cs, confirms that they primarily pulsate in the fundamental mode.  
There is theoretical evidence that some stability islands for the FO can appear at luminosities brighter than the {\sl transition point}, but their true nature needs to be further clarified \citep{Marconi15}. 
Furthermore, the FO IS tends to vanish in the metal-rich regime (e.g., [Fe/H]$\sim -1.0$). 
Finally, using these same pulsation models and the evolutionary tracks from \cite{basti}, \cite{dicriscienzo07} derived analytical relations for the boundaries of the IS as a function of the adopted stellar parameters, as well as the PL and PW relations for these objects.
They derived the T2C-based distances to GGCs and concluded that there is good agreement with the distance scale set by the RRLs.

Similarly to what happens for classical Cepheids, the PL relation for T2Cs can be derived using the fundamental pulsation relation given in Equation~\ref{eq:pulsationeq} \citep{VanAlbada1971} valid for radial pulsators, together with the mass--luminosity relation typical of low-mass AGB evolutionary models \citep[e.g., for BL~Her stars, ][]{dicriscienzo07}. 
Indeed, these AGB pulsators follow an inverse relation between stellar mass and luminosity, since lower mass pulsators are crossing the IS at a luminosity brighter than those with larger masses \citep{Bono_2016}. 
The pulsation relation can then be averaged in temperature because the width of the IS is quite narrow ({$\sim$}1500~K), originating a PL relation \citep[see also][]{Fadeev_1985,Matsunaga_2006}.
Interestingly enough, the slope of the overall PL for T2Cs is less steep than that of classical Cepheids, which is also predicted by theoretical models \citep[e.g.,][]{bono99a,dicriscienzo07}.

\subsection{Optical and near-IR period--luminosity and period--Wesenheit relations}\label{sec:t2cpl}

It has long been recognized that T2Cs follow, at least, a loose PL relation.
From previous studies on T2Cs in GGCs, it is known that the T2Cs do follow a PL in $BVI$.
However, the relation has not been well established until recent work.
While \citet{Harris_1985} and \citet{Mcnamara_1995} claimed that the optical PL slopes become steeper at around $\log P=1$, \citet{Pritzl_2003} did not find such a feature for the T2Cs in NGC\,6388 and NGC\,6441 based on \emph{HST} photometric data.
As \citet{Pritzl_2003} claimed, many earlier studies for the T2Cs were based on magnitudes determined from photographic plates and, from these earlier works, it had been unclear if this class were useful as a distance indicator. 

\begin{figure*}
	\centering
	\includegraphics[width=\textwidth]{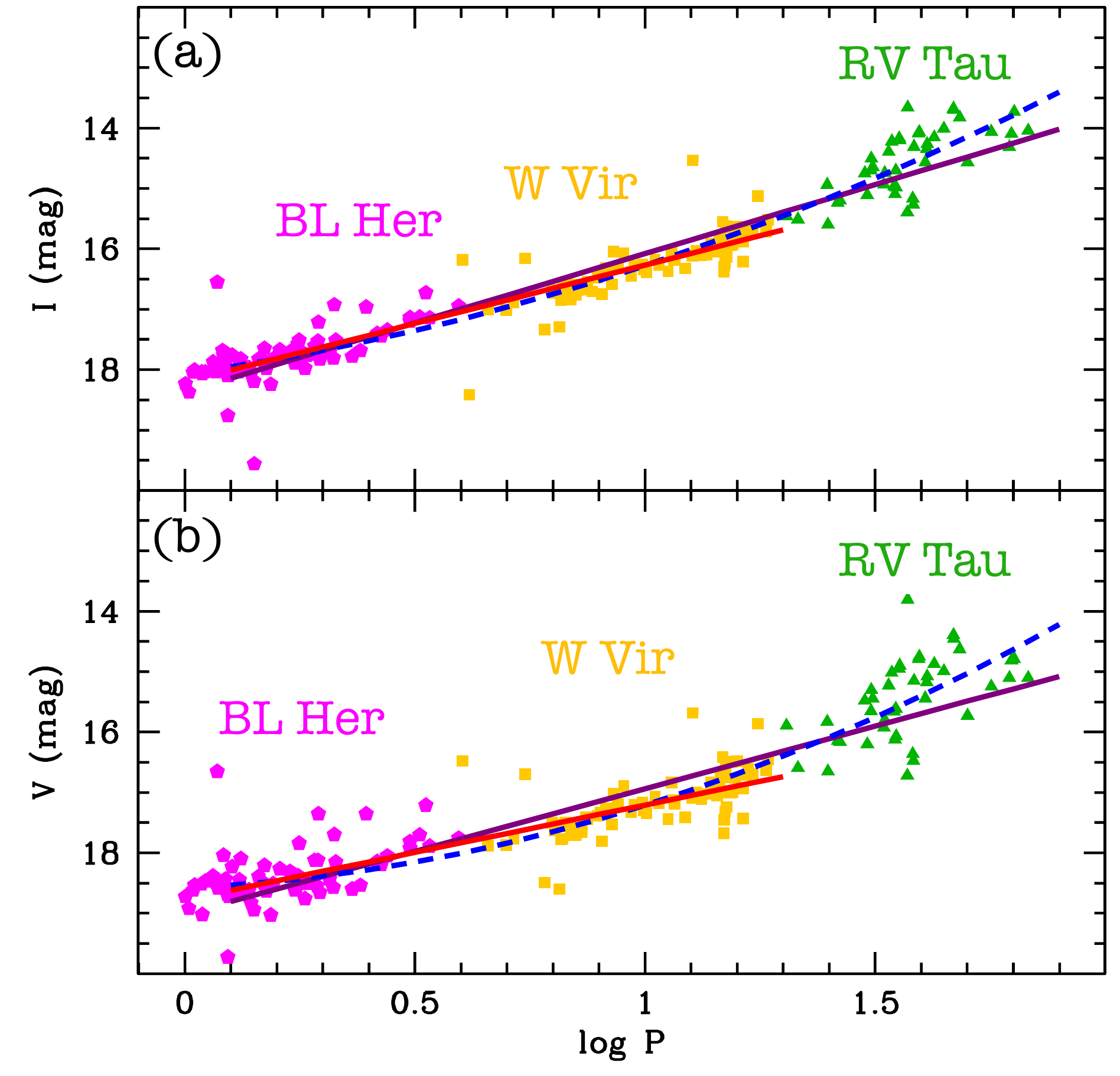}
	\caption{The PL relations for T2Cs in the LMC: 
     (a)~$I$-band and (b)~$V$-band PL relations. 
        The \citetalias{Soszynski_2008} T2C classification is maintained, but the peculiar W Vir stars are removed.
		The three PLs discussed in the text are shown for both panels: 
        (i)~a linear global fit for all three sub-types (solid line; Equation \ref{eq:t2c_linfull}), 
        (ii)~a quadratic global fit that better accounts for the structure of the RV~Tau types (dashed line; Equation \ref{eq:t2c_quadfull}), and 
        (iii)~a linear fit just to the BL~Her and W~Vir type stars (solid line; Equation \ref{eq:t2c_linshort}). } 
	\label{pl}
\end{figure*}

Breakthroughs were brought by modern photometric surveys in both the optical and the infrared.
In particular, large-scale microlensing surveys that started in the late 1990s have systematically discovered T2Cs in the Magellanic Clouds.
In the MACHO microlensing survey, \citet{alcock_1998} determined that W~Vir and RV~Tau sub-types follow a PLC relation in the form of 
\begin{equation} \label{eq:alcock}
 M_{V}~=~a~+~b \log P+ c (V-R) 
\end{equation}
\noindent with a relatively small scatter of 0.15~mag, which is identical (c.f.~\citealt{Bono_1999}) to a PW relation of the form, 
\begin{equation} \label{eq:alcock2} 
V~-~c(V-R)~=~a~+~b\log P.
\end{equation}
The photometric analysis performed by \citetalias{Soszynski_2008} using $VI$ optical data from the OGLE survey provided solid, empirical optical PL relations that support the theoretical predictions.
The LMC sample is larger than the samples in GGCs and, because they are in the same system, can be studied as an ensemble without uncertainties from distances to the individual clusters. 
Figures~\ref{pl}a and \ref{pl}b show the period-luminosity distribution for the T2Cs from \citetalias{Soszynski_2008} for the $I$ and $V$ bands, respectively; the BL~Her stars are shown as filled hexagons, the W~Vir as filled squares, and the RV~Tau stars as filled triangles, while the pW stars are removed (following the classification by \citetalias{Soszynski_2008}). 
The PL relations can be fit in three ways:
\begin{enumerate}
\item 
First, a linear regression can be performed for the entire sample, which results in the following relations,
 \begin{equation} \label{eq:t2c_linfull}
    \begin{split}
	   M_V~&=~19.01~-~2.07~\log P~~(\sigma_V~=~0.6), \\
	   M_I~&=~18.37~-~2.29~\log P~~(\sigma_I~=~0.4).
    \end{split}
 \end{equation} 
A linear regression over the entire period range shows a decrease of the dispersion from {$\sim$}0.6 to {$\sim$}0.4~mag when moving from the $V$ to the $I$ band, indicating that the scatter decreases for longer wavelengths. 
\item 
As is evident in Figures~\ref{pl}, the PL relations for T2Cs in these photometric bands are far from being linear and they could be better represented by a higher order fit, more specifically a polynomial quadratic in $\log P$.
Fitting this form of the PL to the \citetalias{Soszynski_2008} data results in the following fits,
 \begin{equation} \label{eq:t2c_quadfull}
   \begin{split}
       M_V~&=~18.58~-~0.34~\log P~-1.03~(\log P)^2 ~~ (\sigma_V~=~0.5), \\
	   M_I~&=~18.06~-~1.04~\log P~-0.74~(\log P)^2 ~~ (\sigma_I~=~0.3).
  \end{split}
 \end{equation}
This reduces the scatter by 0.1~mag, corresponding to improvements in distance precision by 5~\%. 
\item 
As has been done in previous works, we also consider excluding RV~Tau stars from the sample, because they show a larger spread in luminosity at fixed period for all wavelengths. 
Effectively, this is a cut in period for the \citetalias{Soszynski_2008} sample to those stars with $P\le 20$~days.
Fitting a linear relation over this period range results in the following,
 \begin{equation} \label{eq:t2c_linshort}
  \begin{split}
	 M_V~&=~18.78~-~1.57~\log P~~ (\sigma_V~=~0.5), \\
     M_I~&=~18.20~-~1.94~\log P~~ (\sigma_I~=~0.3). 
  \end{split}
 \end{equation} 
However, it is noteworthy that the scatter of these relations and those of the quadratic fits for all three sub-types are similar. 
Exploration of additional systems with different metallicities and in different environments may help distinguish which of the relations is more representative for a larger range of properties.
\end{enumerate}

One has to keep in mind the presence of peculiar T2Cs that are scattered above the PL relation for BL~Her and W~Vir type T2Cs.
They have some distinctive characteristics compared to W~Vir stars (e.g., light curve shapes) but otherwise may be difficult to distinguish without long term monitoring. 
\citet{Soszynski_2017} further investigated the light curves of this pW group. 
A significant fraction of the pW stars identified by \citetalias{Soszynski_2008} and \citetalias{Soszynski_2010} were found to be in eclipsing binary systems.
These objects are interesting in terms of evolutionary paths to produce T2Cs (and related objects), but to obtain accurate distances to stellar systems with T2Cs these atypical objects need to be identified and then excluded.

There is also evidence that T2Cs display well defined NIR PL relations that are linear over the entire period range. 
\citet{Matsunaga_2006} reported relations with a scatter of 0.15~mag in $JHK_s$ for BL~Her to RV~Tau stars in GGCs.
Because they used RRL-based distances to individual GGCs to combine T2Cs onto the single relation, the well-defined relations suggest some consistency with the RRL distance scale.
\citet{Ripepi_2015} found smaller dispersions in both the $J$ (0.13~mag) and $K$ (0.09~mag) for T2Cs in the LMC, further supporting the evidence that they are good distance indicators. 
Recent updates on the NIR relations based on a larger photometric sample can be found in \citet{Bhardwaj_2017a,Bhardwaj_2017b}.

\begin{figure*} 
	\centering
	\includegraphics[width=\textwidth]{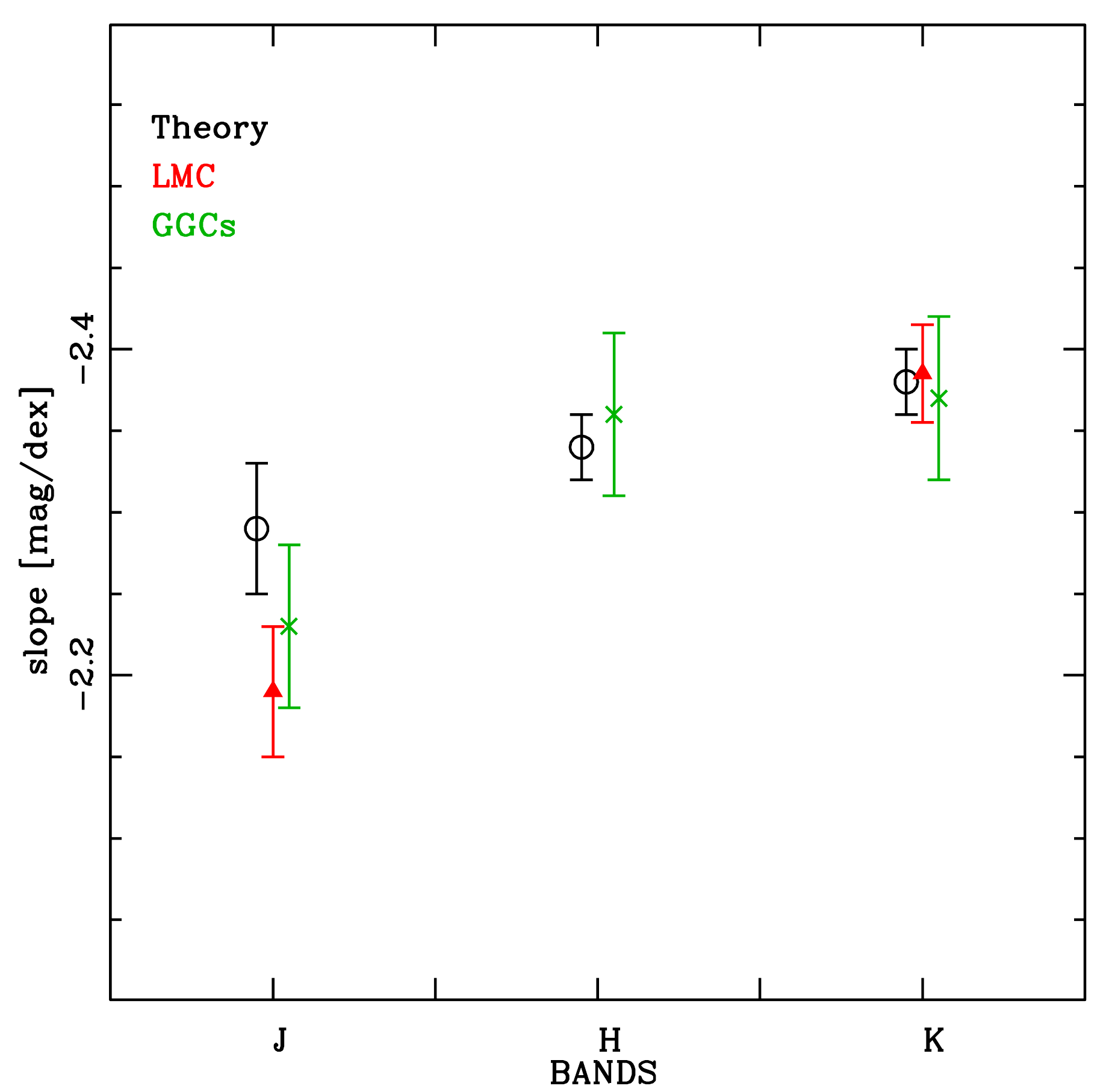}
	\caption{Observed slopes of NIR ($JHK$) PL relations for T2Cs in GGCs \citep[xes,][]{Matsunaga_2006} and in the LMC (triangles, \citetalias{Soszynski_2008}). The slopes predicted by \citep{dicriscienzo07} are plotted as open circle. 
 The vertical bars display the uncertainties on the slopes.}
	\label{t2c_pl2}
\end{figure*}

The slopes of these NIR PL relations do not change dramatically over the metallicity range spanned by these observations, as shown by Figure~\ref{t2c_pl2} that displays the values for $JHK$ PL relations obtained from GGCs \citep{Matsunaga_2006} and the LMC \citep{Ripepi_2015}.
Spectroscopic measurements are still too limited to draw firm conclusions (only a few dozen have measurements); those stars with metallicity measurements, however, span a range of metallicities similar to that of RRLs. 
The agreement becomes more stringent if we account for the theoretical  NIR PL relations provided by \citet{dicriscienzo07}, which are shown in Figure~\ref{t2c_pl2} as the open circles. 
Note that a similar agreement is also found in a detailed comparison of the zero-points. 
Theoretical work, however, suggested a dependence of the zero-point on the metal content at the level of 0.04 to 0.06~mag~dex$^{-1}$. 
Observations, on the other hand, suggest a minimal, if any, dependence \citep{Matsunaga_2006}.
A different result is found when considering the T2Cs in the SMC \citep{Ciechanowska_2010}; more specifically, the slopes are shallower with $-1.95$ in the $J$-band ($\sigma_J=0.24$~mag) and $-2.15$ in the $K$-band ($\sigma_K =0.2$~mag). 
However, the difference with the slopes shown in Figure~\ref{t2c_pl2} is within the associated 1~$\sigma$ errors, and it is probably due to the high scatter of the mean NIR magnitudes derived from single epoch observations in \citet{Ciechanowska_2010}. 
In addition, the NIR sample of T2Cs in \citet{Ciechanowska_2010} includes only 50\% of the total optical sample for the SMC \citepalias[40 objects,][]{Soszynski_2010} and thus a more complete sample with more well-sampled light curves may resolve the discrepancy.

The use of empirical relations to determine the LMC distance employed by \citet{Ripepi_2015} returned a distance modulus in excellent agreement with RRLs, but with a difference of {$\sim$}0.1~mag with respect to the distance modulus based on classical Cepheids ({\bf most likely due to parallax sample used, see Section \ref{sec:primary_absscale}}). 
Even though this is not a statistical significant difference (e.g., within the formal 1~$\sigma$ uncertainty), it suggests a possible population bias in the LMC distances \citep{Inno_2016}. 
Theoretical models for the BL~Her period range are in extremely good agreement with the above evidence, since they predict dispersions in these passbands equal to $\sigma_J=0.13$~mag and $\sigma_K=0.06$~mag \citep{dicriscienzo07}.

\citet{Matsunaga_2006} also pointed out that the T2Cs and RRLs seem to follow the same, or continuous, PL relations in the infrared. 
The fact that RRLs and T2Cs follow very similar NIR PL relations \citep[see also][]{Feast_2011} 
seems to be supported by the smooth and natural transition of the evolutionary properties of evolved RRLs and BL~Her type stars. 
Although it was also discussed that BL~Her and W~Vir stars could have distinct NIR PL relations, recent findings based on accurate NIR mean magnitudes for LMC T2Cs suggest they are, within the errors, very similar \citep[][and references therein]{Ripepi_2015}. 
This does not apply to RV~Tau variables, because they typically display a larger spread compared to W~Vir and BL~Her stars. 
The increase in the spread might also mask a possible change in the slopes of the NIR PL relations 
\citep{Ripepi_2015}. 
This occurrence appears to be associated with the properties of the circumstellar envelopes typical of these variables, which may also contribute to the larger scatter for this sub-type observed in the optical bands (see Figure~\ref{pl}).

\begin{figure} 
	\centering
	\includegraphics[width=\textwidth]{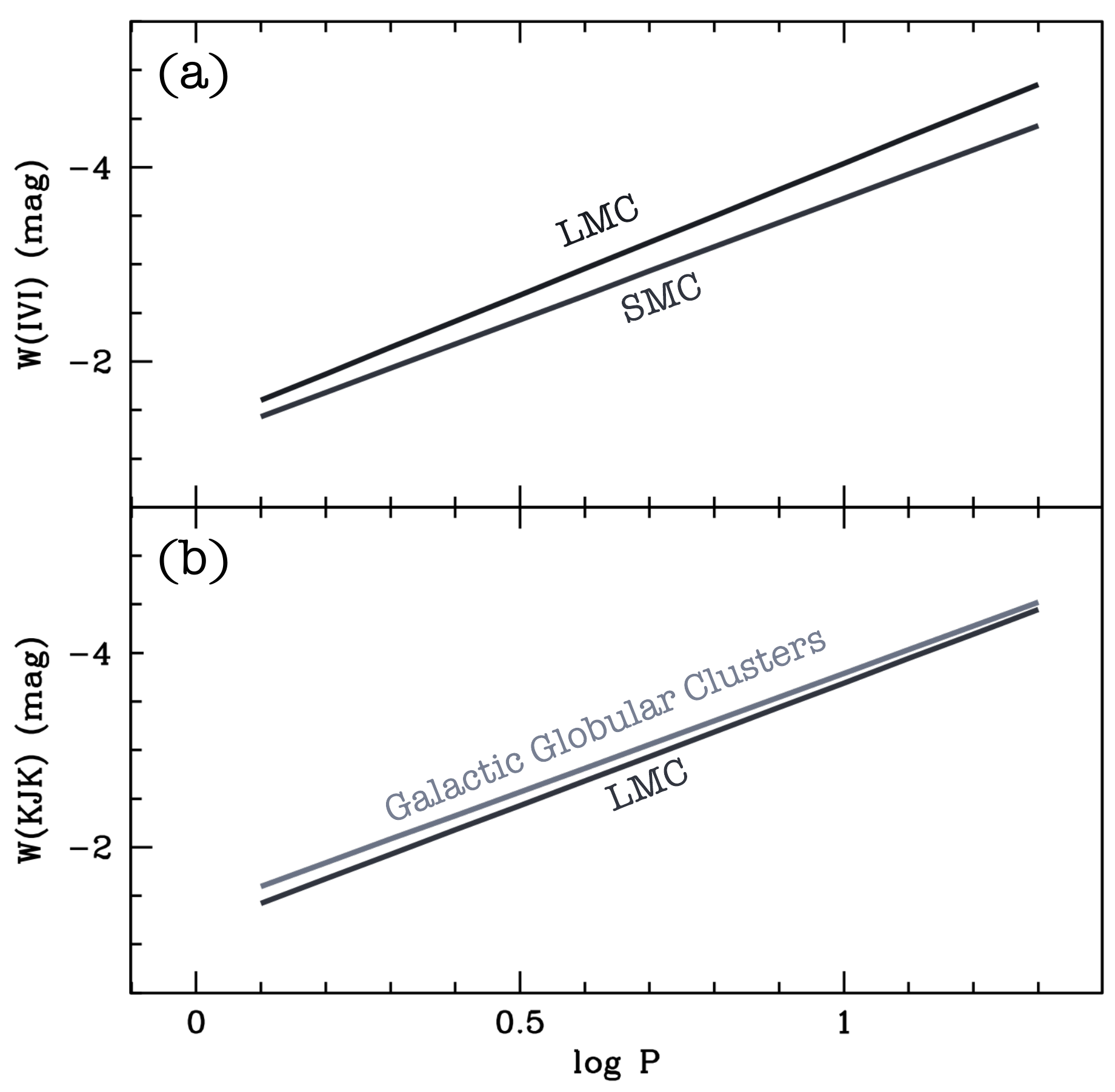}
	\caption{
      \emph{(a)} Optical ($I$,$V-I$) PW relations for T2Cs in the LMC and SMC derived as described in the text using data are taken from \citepalias{Soszynski_2008} and \citepalias{Soszynski_2010}, respectively. 
      \emph{(b)} NIR ($K$,$J-K$) PW relations for T2Cs. 
        The LMC data is adopted from \cite{Ripepi_2015}, whereas the for GGCs it has been derived using NIR magnitudes from \cite{Matsunaga_2006}. 
       The precise form of the relations is given in Equations \ref{eq:t2c_pw_opt} and \ref{eq:t2c_pw_nir} and described in the text.}
	\label{t2c_pw}
\end{figure}

PW relations are also commonly used to reduce the impact of extinction in distance determination and here we derive these for T2Cs based on the most conspicuous samples available in literature \citep[see also][]{Bhardwaj_2017a}. 
\begin{enumerate}
\item
The most common Wesenheit magnitude in the optical takes the form of $W(I,V-I)=I-1.55(V-I)$. 
Performing a linear regression to the BL~Her and W~Vir stars we find:
  \begin{equation} 
    \begin{split}  \label{eq:t2c_pw_opt}
	 W(I,V-I)_{\rm LMC}~&=17.31-2.50 \log P + \mu_0({\rm LMC})~~(\sigma = 0.2), \\
	 W(I,V-I)_{\rm SMC}~&=17.63-2.71 \log P + \mu_0({\rm SMC})~~(\sigma = 0.4),  
   \end{split}
  \end{equation} 
for the LMC \citepalias{Soszynski_2008} and SMC \citepalias{Soszynski_2010} data, respectively.
To compare the relations on the absolute scale, we assume the consensus distance moduli, $\mu_0({\rm LMC}) =18.49$~mag \citep[][and references therein]{degrijs14} and $\mu_0({\rm SMC})=18.96$~mag \citep[][and references therein]{degrijs15}, respectively \citep[see also][in this series]{degrijs_2017}. 
Figure~\ref{t2c_pw}a compares these $W(I,V-I)$ relations.
It is worth noticing that when decreasing the host galaxy metallicity (from that of the LMC to that of the SMC), the slopes get steeper and the dispersion around the mean relation increases, although the difference of slopes is still within 1~$\sigma$.
\item
The most common NIR Wesenheit magnitude takes the form of $W(K,J-K) = K - 0.69 (J-K)$.
Performing a linear regression to the BL~Her and W~Vir type stars for the LMC \citep[as given by][]{Ripepi_2015}, SMC \citep[IR magnitudes and periods from][]{Ciechanowska_2010} and GGCs \citep[data from][]{Matsunaga_2006}, we find the following PW relations:
  \begin{equation} 
  \begin{split} \label{eq:t2c_pw_nir}
	W(K,J-K)_{\rm LMC}~ &=~17.32~-2.52~\log P~+\mu_0({\rm LMC})~~(\sigma = 0.08),  \\
	W(K,J-K)_{\rm SMC}~ &=~17.41~-2.17~\log P~+\mu_0({\rm SMC})~~(\sigma = 0.3),  \\
	W(K,J-K)_{\rm GGCs}~&=~-1.35~-2.44~\log P~~(\sigma = 0.2). 
    \end{split}
  \end{equation} 
Figure~\ref{t2c_pw}b compares NIR PW relations;
 the SMC relation is not included because the mean magnitudes have a large dispersion about the relation due to sparse sampling of the NIR light curve. 
As shown in Figure~\ref{t2c_pw}b, the slopes for LMC and GGCs are very similar, suggestive of only a mild dependence on the metal content. 
However, for the SMC we note that the slope decreases down to $-2.17$~mag~dex$^{-1}$, which is opposite to what happens in the optical regime.
There are no doubts that a more detailed investigation of T2Cs in the SMC will be crucial to understand the universality of the NIR PW relations.
\end{enumerate}

\subsection{T2Cs in context with RRLs and CCs} \label{comp}

In this section we discuss the pulsation and evolutionary properties of T2Cs as compared
to the most popular distance indicators, namely RRLs and classical Cepheids. 
The aim is to highlight their possible new role for these stars in the \emph{Gaia} era that is expected to refine the extragalactic distance scale. 
In the following, we list their properties and assess their advantages and disadvantages compared to classical Cepheids and RRLs:   

\begin{enumerate}
\item {\it NIR PL relation}:~~As discussed in the previous section, T2Cs follow well defined optical and NIR PL relations \citep{Matsunaga_2006,matsunaga13}. Moreover, there is evidence that their pulsation properties are either independent of, or minimally affected by, the metal abundance \citep{bono97e,dicriscienzo07,lemasle15}. RRLs also follow linear NIR PL relations, however their luminosity (and luminosity range) is much smaller, which largely limits their application to within the Local Group (e.g., distances up to {$\sim$}1~Mpc).
T2Cs are roughly 1.5~mag fainter than classical Cepheids at fixed period with the precise difference being period dependent. Moreover, the long period tail of classical Cepheids exceeds 100 days, while the long period tail of T2Cs approaches {$\sim$}70~days.
\item {\it Linearity}:~~The current empirical evidence indicates that BL~Her and W~Vir stars follow NIR PL relations that are linear over the entire period range. With current data, however, it is unclear if RV~Tau objects follow a long period extension of the same PL relations, or whether their PL relation is distinct. Thus, RV~Tau subgroup is typically neglected when fitting PL relations and in determining distances. 
\item {\it Brightness}:~~T2Cs are brighter than RRLs and cover a similar magnitude range as Classical Cepheids. The difference with RRLs ranges from at least half magnitude ($M_V\sim -0.5/-1$~mag) to roughly three 
magnitudes ($M_V\sim-3$~mag). This means that they can be employed to measure distances and trace old (blue HB) stellar populations beyond the Local Group.
\item {\it (Short) Evolutionary Lifetime}:~~T2Cs are typically approaching the AGB phase. This means that their evolutionary time scale is roughly two orders of magnitude faster than RRLs \citep{Marconi15,Bono_2016}. RRLs are low-mass stars during their core He burning phase, they have been identified in all stellar systems hosting stellar populations older than $\sim$10 Gyr, and from a theoretical point of view stellar mass and chemical abundance ranges that can give origin to RRLs are well known \citep{bono96,bono97a,bono97d}. The same is true for classical Cepheids, that are young core 
helium burning stars. The ratio of evolutionary times of classical Cepheids to the entire lifetime of their progenitors is longer than for T2Cs \citep{marconi05}. This has a strong impact on the occurrence of T2Cs in a given system such that T2Cs are more rare than both RRLs and classical Cepheids.
As shown in Figure~\ref{fig:cmd}a, the LMC contains {$\sim$}3,000 classical Cepheids, {$\sim$}20,000 RRLs, but only {$\sim$}200 T2Cs \citep{fiorentino12c}.
\item {\it Binary or Evolutionary Channel}:~~The fraction of T2Cs in binary systems is not well constrained. 
Very recently \citet{Soszynski_2017} and \citet{Soszynski_2018} presented additional monitoring of T2Cs in the LMC and in the Bulge that find new irregularities in the light curves consistent with binary companions. 
\citet{maas02} present evidence that several RV~Tau stars may be in binary systems. More importantly, there is photometric evidence that the pW stars may also be binaries \citep[most recently][]{Pilecki_2017}. 
This working hypothesis is also supported by spectroscopic evidence of binary companions for some T2Cs \citep{maas07,Soszynski_2010,jurkovic16} and, in turn, casts some doubts on the proposed small initial mass of their progenitor stars.
These kind of objects have been dubbed {\em Binary Evolution Pulsators} \citep{karczmarek16,Karczmarek_2017_bep}. 
Long term monitoring of T2Cs can help identify their evolutionary path and determine if binary formation is the dominant formation path. Period changes can tell us the directions of evolution within the instability strip and this, in turn, provides additional evidence for the evolutionary paths of T2Cs \citep[e.g,][]{Diethelm_1996,Wallerstein_2002,Rabidoux_2010}.
\item {\it Shape of the light curve}:~~The coupling between shape of the light curve, e.g., the Fourier parameters and luminosity amplitudes, and the pulsation period is less distinctive for T2Cs than for RRLs (e.g., compare Figure~\ref{fig:t2c_lc} to Figure~\ref{fig:lc} for light curves and Figure~\ref{periods} and Figure~\ref{fig:scl_pa} for the amplitude-magnitude diagrams). They partially overlap with classical Cepheids \citepalias{Soszynski_2008}.  
\item {\it Host stellar systems}:~~T2Cs have been identified in a variety of stellar systems hosting an old ($t\geq 10$~Gyr) stellar population, e.g., both early and late type stellar systems. Although this appears as a solid empirical evidence, we still lack firm identification of T2Cs in dSphs. The lack of T2Cs in dSphs can hardly be explained as an observational bias, since they are systematically brighter than RRLs and some of the classical dSphs have been surveyed for very long time scales \citep[e.g.,][among others]{stetson14a,Coppola2015,MartinezVazquez2016b}. The working hypothesis suggested by \cite{Bono_2011,Bono_2016} to explain the desert of T2Cs in dSphs is that they typically lack hot and extreme HB stars \citep{salaris13a}. This means that the fraction of HB stars evolving from the blue (hot) to the red (cool) side of the CMD is small. This is relevant, given that the stellar populations of dSphs do cover the same metallicity range of GGCs that do have well extended blue tails on the HB. The lack of hot/extreme HB stars, and in turn of T2Cs, may be due to an environmental effect between dSphs and GGCs, that otherwise are composed of similar stellar populations.  
\end{enumerate}
On the whole, T2Cs have many enticing properties and show potential as distance indicators, but also still have some 
more shadowy areas left to be resolved before their application en masse. 

\subsection{Summary and final remarks}\label{conc}

In this section, the observational and theoretical properties of T2Cs have been described. 
Although there is no doubt that they are radially pulsating stars crossing the IS, their evolutionary origin has not yet been completely addressed \textbf{\citep[e.g., the Binary Evolution Pulsator scenario, see][]{karczmarek16,Karczmarek_2017}.} 
Both theoretical and observational studies demonstrate that T2Cs mainly pulsate in the fundamental mode, given that they reach luminosities brighter than the {\sl transition point}. 
The possible occurrence of short-period, fainter T2Cs evolving along the so-called {\sl blue hanger} \citep{Bono_2011} cannot be excluded. 
However, they are expected to add little contamination to the fundamental mode NIR PL and PW relations.     

T2Cs are {\sl largo sensu} solid distance indicators, for they follow well defined optical and NIR PL relations, and both theoretical and empirical results suggest a minimal dependence on metal-abundance. 
Furthermore, they also follow similar optical and NIR PW relations. 
The key advantages of these diagnostics is to be independent of reddening uncertainties, but rely on the assumption that the reddening law is universal.   
This further suggests that T2Cs can be considered {\it stricto sensu} ideal distance indicators, given that they can provide very accurate relative distances, independent of uncertainties in the zero-point of the PL relations.
The small metallicity dependence of the slopes of the NIR PW relations further supports the use of these variables as distance indicators for stellar systems affected by differential reddening \citep[e.g., as with M\,4 in][]{Braga2015}.
To obtain the tighter PL relations, only BL~Her and W~Vir stars (i.e., those T2Cs with periods less than 20 days and $M_V$ fainter than $-2$~mag) are considered, excluding the brighter RV~Tau objects.
This means that one can detect and employ T2Cs as distance indicators out to {$\sim$}10~Mpc with \emph{HST} (limiting magnitude $V \sim 30$~mag).
However, because their periods have a wide distribution ranging from a few days to 20~days, complete photometric surveys with long temporal baseline have to be conducted in order to collect complete T2C samples.

The absolute distance scale of T2Cs, i.e., the zero point of the PL relation, is calibrated using the populations in the Magellanic Cloud and GGCs.
There are only a handful of trigonometric parallax measurements for T2Cs and the uncertainties for the measurements that do exist make it difficult to establish a distance scale directly tied to parallax. 
The uncertainties affecting the zero-points of these relationships, and in turn, the absolute distances based on T2Cs, are going to be largely reduced during the next few years. This is thanks to the very accurate trigonometric parallaxes that will be provided by \emph{Gaia}, not only for the nearest T2Cs, but for more than 100 field T2Cs and at least a dozen in nearby GGCs using end-of-mission predictions \citep[][]{Harris_1985,Matsunaga_2006}.

The observational outlook concerning T2Cs appears even more promising if we take account 
of the fact that \emph{JWST} will be capable of producing a complete census of T2Cs for Local Group 
and Local Volume galaxies. The halos of large galaxies are, indeed, marginally affected 
by crowding problems. Moreover, 30 meter-class telescopes will provide a unique opportunity to trace old stellar populations in the bulges and the innermost regions of galaxies in the Local Volume due to their unprecedented spatial resolution \citep{Bono_2017}.


\section{The Tip of the Red Giant Branch} \label{sec:trgb}

The ultimate origin of the tip of the red giant branch (TRGB) as a standard candle is the canonical work of \citet{baade_1944}, who used red-sensitive plates to first resolve the stellar populations in early-type stellar systems, M\,32, NGC\,205 and the central region of M\,31. 
From these observations developed the concept of the stellar classes of Pop~I and Pop~II. 
\citeauthor{baade_1944} also concluded that the brightest stars in these objects had similar magnitudes and colors.
\citet{sandage} later pointed out that galaxies in the Local Group invariably contained a background sheet of red stars. 
Comparing the Cepheid distances to three Local Group galaxies, M\,31, M\,32, and IC\,1613, he determined a mean absolute magnitude of the brightest red stars to be $M_V$=-3.0~$\pm$0.2~mag.
This work led to the early applications using RGB stars for distance determination in the early eighties \citep[see a summary in][]{lee93}. 

A formal method to detect the TRGB in color-magnitude-diagrams (CMDs) of galaxies, a calibration and a thorough analysis of its uncertainties were first published by \citet{lee93} and \citet{mf95}, respectively. 
More specifically, \citet{lee93} compared empirical RGB loci for 6 GGCs \citep{da90} with theoretical stellar evolutionary tracks \citep[Yale Models,][]{green_1987}, and concluded that the $I$-band absolute magnitude of the TRGB is accurate to $\pm 0.1$~mag for resolved stellar systems of low metallicity (e.g., [Fe/H]\textless $-0.7$).
\citet{lee93} introduced Sobel ``edge detection'' kernels for detecting the discontinuity caused by the TRGB in $I$-band luminosity functions and this methodology was rigorously tested by \citet{mf95}.
Since those works, the TRGB has been used extensively in Pop~II dominated systems. 
With current observational capabilities the TRGB can be used to determine distances out to {$\sim$}20~Mpc \citep[see e.g.,][and references therein]{jang_2015,jang_2017b,jang_2018}.
Indeed, several authors have discussed the TRGB as an alternative to CCs for the basis of the extragalactic distance scale \citep[e.g.,][]{mould_2008,beaton2016}. 
So far, more than 400 galaxies have been measured their distances with the TRGB \citep{tul16}.

The TRGB as a precise distance indicator for resolved stellar systems has held against several observational tests \citep{lee93, fer00, bfp, sak04}. 
Several studies have shown that the TRGB distance estimates to nearby galaxies agree well with those based on other precise distance indicators, such as CCs, RRLs, and Megamasers \citep{fre88, lee93, sak04, fer00, tam08, mag08,freedman_2010,jang_2017a}. 
It indicates that the precision of the TRGB is comparable to those of other primary distance indicators. 

In this section, physical understanding of the TRGB is given in Section \ref{sec:trgb_phys}. 
Section \ref{ssec:trgb_detect} discusses empirical methods to measure the TRGB.
A case study of using TRGB in the optical is described in Section \ref{ssec:trgb_practice}.
In Section \ref{ssec:trgb_practice_ir}, the IR properties of the TRGB are discussed.
We conclude looking to future improvements of the method in Section \ref{sec:trgb_future}.

\subsection{Physical Description} \label{sec:trgb_phys}
The TRGB corresponds to the termination of the RGB evolution in old stellar populations, due to the onset of the He-flash in the electron degenerate cores of low-mass stars. 
For a given  initial chemical composition, the bolometric luminosity of the TRGB is determined by the He-core mass at the He-flash.  This mass is basically constant, e.g., within a few 0.001$M_{\odot}$, for stellar masses that reach He-ignition at ages from 1.5-3~Gyr to the age of the Universe (the starting age is a function of the initial metallicity). 

The use of the TRGB for distance estimates has been traditionally based on $I$-band photometry, in the implicit assumption that the observed RGB of a galaxy has an age comparable to the age of GGCs, or in any case larger than 4-5~Gyr and metallicity [M/H] below $\sim-$0.7 \citep[see, e.g.,][]{da90, lee93, scw02}.
In this age range $M_{bol}^{TRGB}$ is roughly constant for a given [M/H], but becomes brighter with increasing [M/H]. 
Two effects influence $M_{bol}^{TRGB}$ in more metal rich stars: 
(i) increased efficiency of the hydrogen burning shell, which makes the stars brighter, and 
(ii) decreasing He-core mass, which makes the stars fainter; overall, $M_{bol}^{TRGB}$ is more luminous in metal-rich stars from these two effects.
At the same time the effective temperature of the TRGB decreases with increasing [M/H] and the bolometric corrections, $BC_I$, decrease with decreasing $T_{eff}$. 
The resulting rate of decrease of $BC_I$ with increasing metallicity very nearly matches the rate of increase of $M_{bol}^{TRGB}$ with [M/H]. 
Given that $M_{I}^{TRGB}$=$M_{bol}^{TRGB} - BC_I$, this implies that $M_{I}^{TRGB}$ stays almost constant with [M/H] for the range between $\sim -$2.0 and $\sim -$0.7 and for ages older than 4-5~Gyr.

   \begin{figure}
   \centering
   \includegraphics[width=\textwidth]{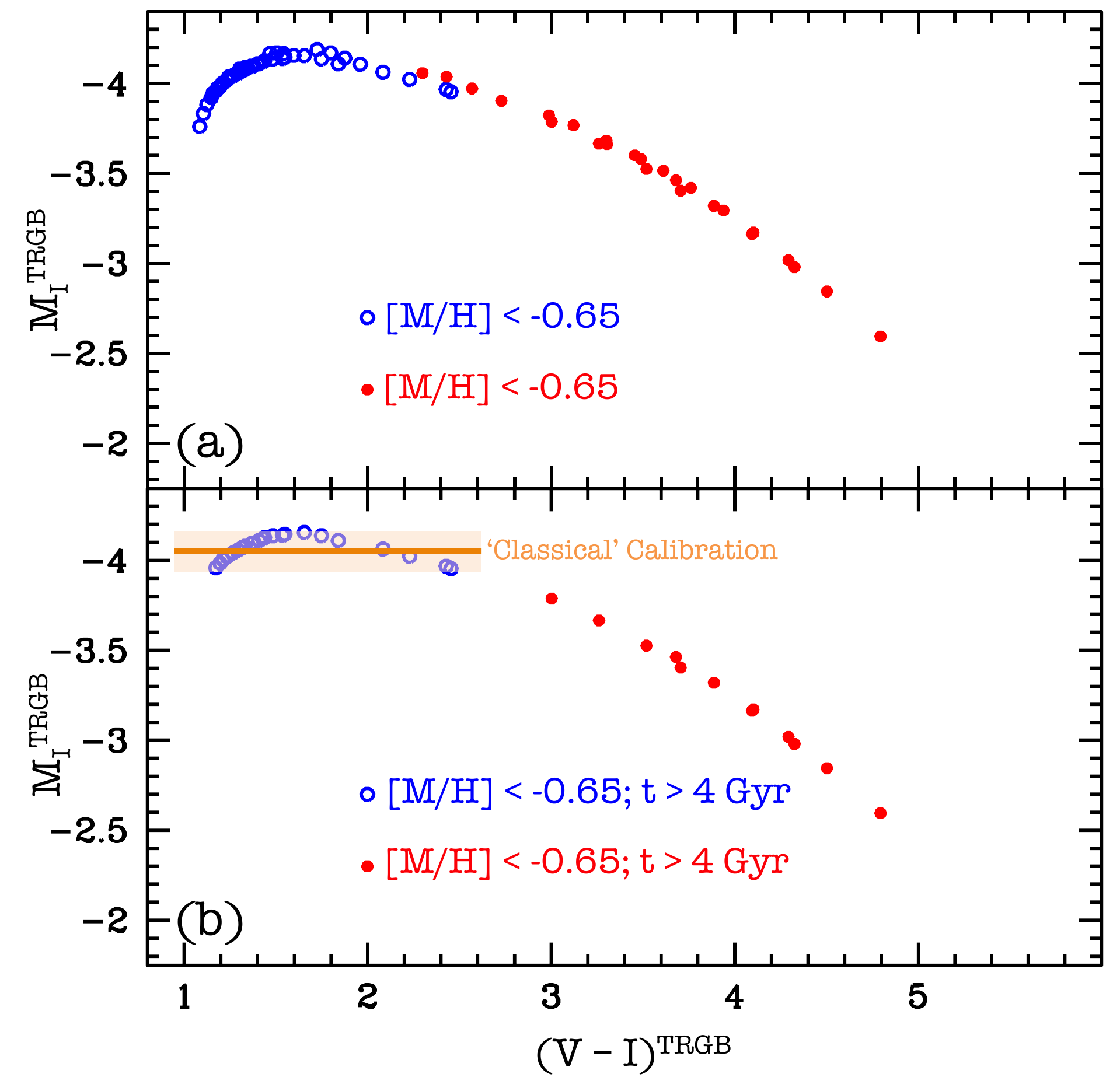}
    \caption{$M_{I}^{TRGB}-(V-I)^{TRGB}$ distribution for the TRGB from theoretical models from BaSTI \citep{basti}. (a) The full range of ages (1.5\textless~t~\textless~14 Gyr) and metallicity ($-$2.2 \textless~[M/H]\textless0.3 dex) for the models and (b) only models with ages older than 4~Gyr. In both panels, the blue open circles are for [M/H]~\textless~0.65 dex and the red filled circles are for [M/H]~\textgreater~0.65 dex. The solid orange line represents the empirical calibration of $M_{I}^{TRGB}$=-4.05 mag \citep{lee93,sc97,tr}, with the shading giving the typical uncertainty of $\pm$0.10 mag. The `classical' calibration is a reasonable approximation to the theory for stars in this age and metallicity range (corresponding to V-I from 1.0 to 2.5 mag).}
         \label{theory_1}
   \end{figure}
   
From the variation of the absolute magnitude from theoretical models, it follows that applying a single-value for the absolute magnitude of the TRGB to galaxies with an extended star formation history can cause large systematic errors. 
Adopting a single value in a system assumes that the RGB stars in these galaxies are {\it all} older than 4-5~Gyr with [M/H]$< -$0.7, which can only be assumed safely in specific conditions.
This effect has been studied theoretically by \citet{bsh04}, \citet{sg05}, \citet{cs13}, and also verified empirically by \citet{gorski}.
A recent, detailed study is given in \citet{serenelli_2017}.

Figure~\ref{theory_1} displays a theoretical $M_{I}^{TRGB}-(V-I)^{TRGB}$ calibration in Johnson-Cousins filters using the \citet[][BaSTI]{basti} stellar evolution models after correcting the TRGB bolometric magnitudes for the effect of including the updated electron conduction opacities by \citet[][]{opa} and using the empirical bolometric corrections by \citet{wl11}. 
In Figure~\ref{theory_1} [M/H] lower(larger) than $-$0.65, are displayed with open(filled) circles, respectively.
The BaSTI theoretical models cover a range of metallicity [M/H] between $-$2.2 and 0.3 and ages between 1.5 and 14~Gyr.
The full age and metallicity range of the models is shown in Figure~\ref{theory_1}a, which demonstrates that decreasing age at constant [M/H] pushes $M_{I}^{TRGB}$ to bluer colors, whereas increasing [M/H] at fixed age has the opposite effect.
The smooth $M_{I}^{TRGB}-(V-I)^{TRGB}$ relation displays a maximum around $(V-I)^{TRGB}$$\sim$1.6, with the TRGB magnitude increasing steadily for redder colors. 

Figure~\ref{theory_1}b shows TRGB magnitudes and colors for ages above 4~Gyr with [M/H]$> -$0.65 and [M/H]$< -$0.65 using the same symbols as Figure~\ref{theory_1}a compared with a `classical' calibration (orange). 
This `classical' calibration provides a median $M_{I}^{TRGB}$=$-$4.05, with typical observational uncertainties of $\pm$0.10~mag over the range of $V-I$ from 1.0 to 2.5 mag \citep[shown with the shading;][]{lee93,sc97,tr}. 
The theoretical behavior of $M_{I}^{TRGB}$ vs $(V-I)^{TRGB}$ in this age and metallicity range is approximately quadratic, but using a constant average value of $M_{I}^{TRGB}$ is  a decent approximation as is shown by the shading in Figure~\ref{theory_1}b.  

   \begin{figure}
   \centering
   \includegraphics[width=\textwidth]{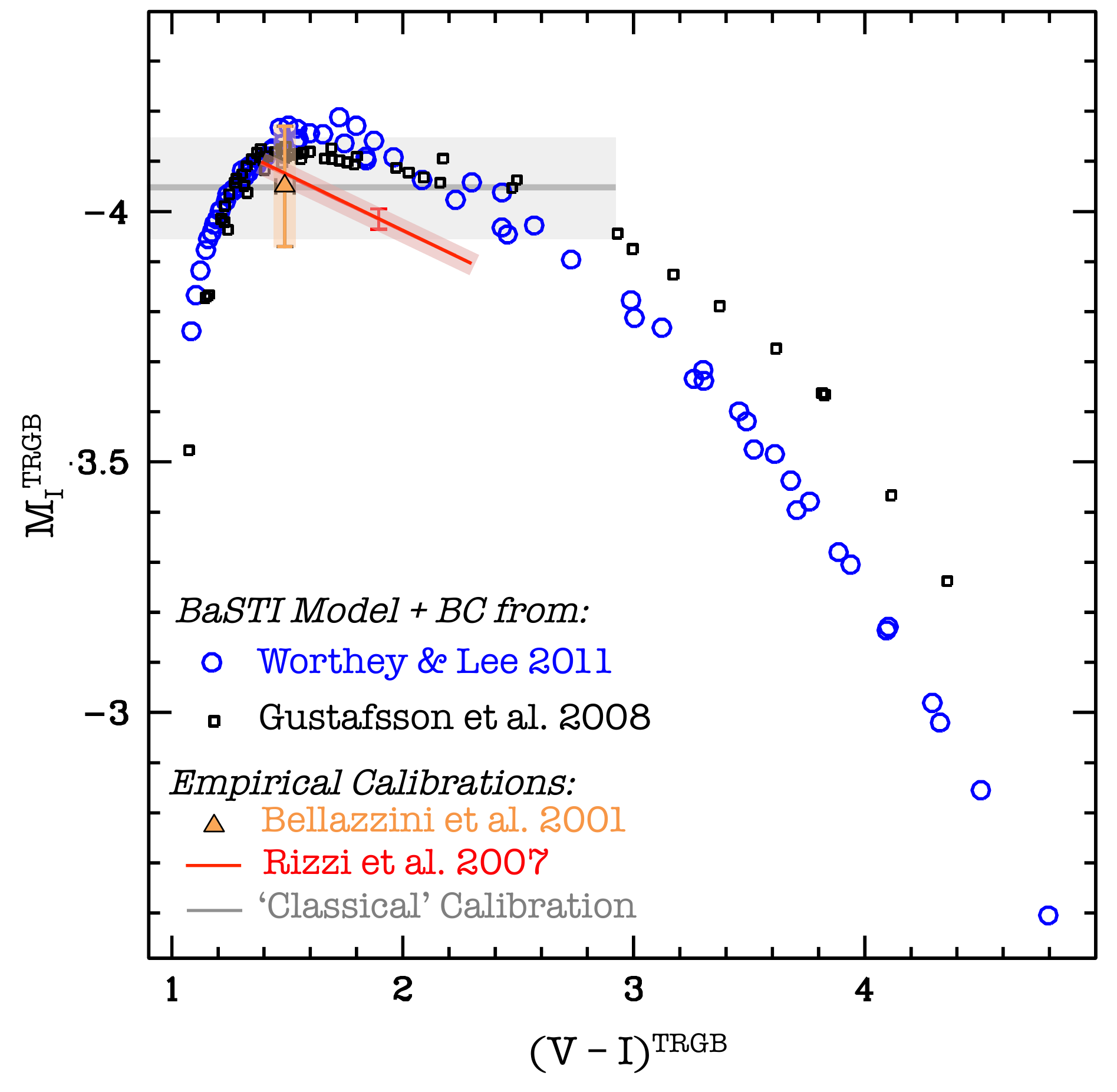}
      \caption{Theoretical $M_{I}^{TRGB}-(V-I)^{TRGB}$ distribution at the TRGB for the full age range of the BaSTI models, but limited to sub-solar [M/H]. 
      Two different sets of bolometric corrections are employed, more specifically those of \citet[][blue circles]{wl11} blue circles and of \citet[][black squares]{marcs}.
 We also show the empirical estimate \citep{bfp} for $\omega$~Centauri \citep[orange triangle;][]{bfp} and the empirical relationship by \citet{rizzi07} based on a sample of galaxies (red line). 
 The range of the `classical' calibration is shown in gray. The the zero point uncertainties are shown by the shading. See also \citet{serenelli_2017}.}
         \label{theory_2}
   \end{figure}

It is evident that the $M_{I}^{TRGB}-(V-I)^{TRGB}$ relationship is very smooth for all [M/H], and irrespective of any age sub-selection, and can be used as distance indicator for any stellar population with an RGB.
Figure~\ref{theory_2} compares the theoretical relationship of Figure~\ref{theory_1} (open circles) to one obtained using the same models, but applying the bolometric corrections by \citet[][open squares]{marcs}. 
For these bolometric corrections we include the TRGB only up to [M/H]$\sim$ solar, because at super-solar metallicities TRGB models have negative log($g$) (surface gravity) that are not covered by these bolometric corrections \citep[additional discussion is given in][]{serenelli_2017}.
We compare the theoretical relations to the empirical relationship determined by \citet{rizzi07} on a sample of galaxies (solid red line) and the determination of the TRGB absolute magnitude for $\omega$~Centauri, which has a broad range of metallicities and ages \citep{bfp} and the `classical' calibration from Figure \ref{theory_1}.

\citet{rizzi07} calibration has roughly the same slope of the theoretical ones for $(V-I)$ above $\sim$1.6-1.7, but with a magnitude offset of $\sim$0.1~mag.
The uncertainty for the $\omega$~Centauri datapoint is large and does not put very strong constraints on the TRGB absolute magnitude.
The two sets of bolometric corrections applied to the theoretical calculations provide similar results as long as $(V-I)$ is below $\sim$2.0. 
Increasingly larger discrepancies appear for redder colors.

   \begin{figure}
   \centering
   \includegraphics[width=\textwidth]{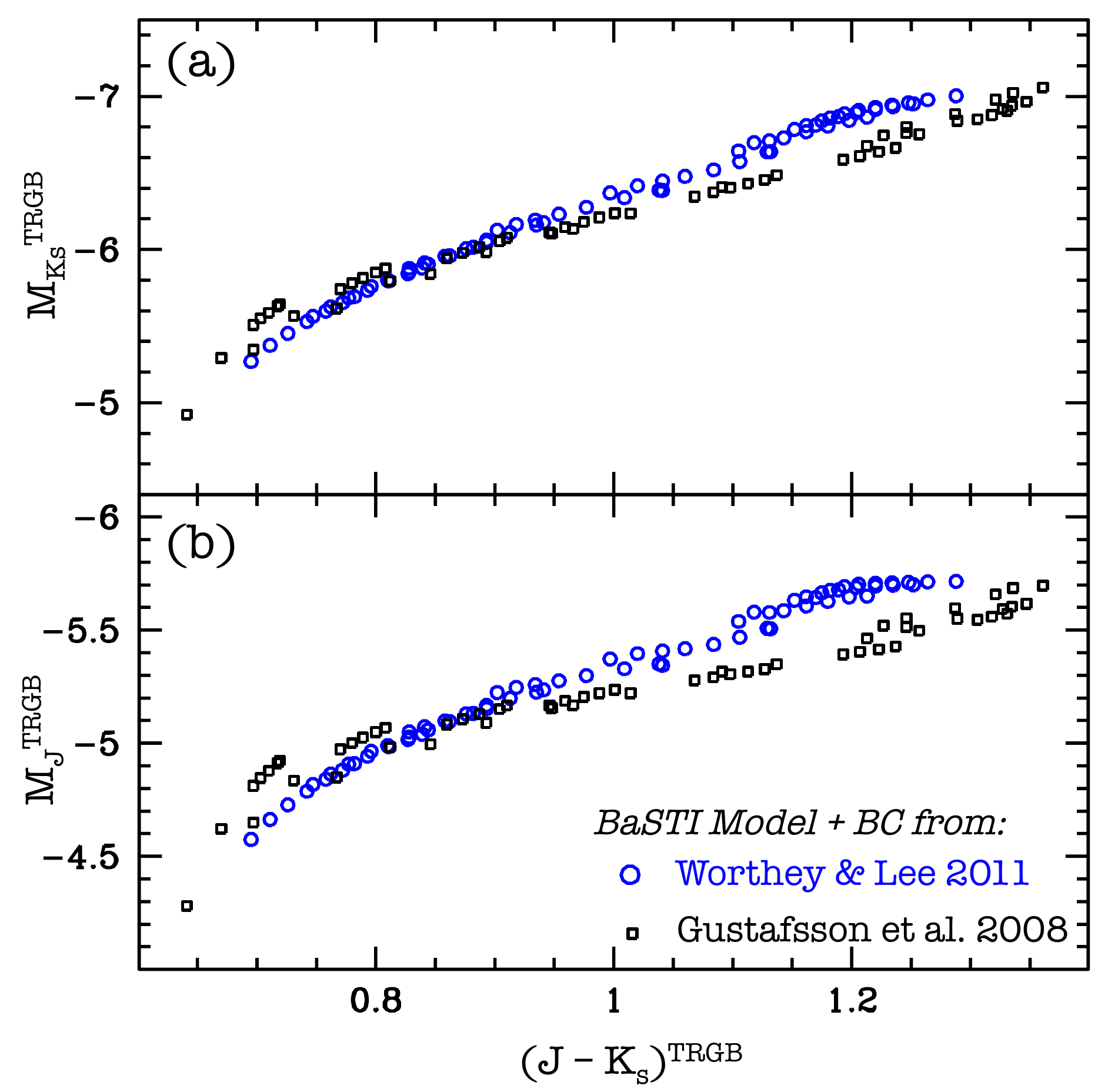}
      \caption{ Theoretical $M_{NIR}^{TRGB}-(J-K_s)^{TRGB}$ distribution at the TRGB for the full age and [M/H] range of the models, more specifically the panels are: 
      (a) $M_{K_s}^{TRGB}-(J-K_s)^{TRGB}$ and (b) $M_{J}^{TRGB}-(J-K_s)^{TRGB}$. 
      The color coding matches that of Figure \ref{theory_2}, more specifically open circles denote results obtained with the \citet{wl11} bolometric corrections, whilst open squares display results obtained with the \citet{marcs} bolometric corrections. 
      Overall, the NIR-TRGB magnitude shows an approximately linear relationship with $J-K_s$ color and, compared to the optical (Figures \ref{theory_1} and \ref{theory_2}), it more than 1~magnitude brighter, spans a range of $\sim$1~mag over this range of [M/H] and ages, but spans a much more narrow color-range. See also \citet{serenelli_2017}. } 
    \label{theory_3}
   \end{figure}

The panels of Figure~\ref{theory_3} shows theoretical TRGB absolute magnitude-color calibrations for the infrared filters $K_s$ and $J$ (Figure ~\ref{theory_3}a and ~\ref{theory_3}b, respectively). 
Notice an almost linear behavior and again a smooth and tight correlation over the whole age and metallicity range covered by the models, with the added bonus of a lower sensitivity to reddening. 
Notice also how the dynamical range of the TRGB magnitude in $J$ is reduced compared to the case of the $K_s$ and $I$ bands.

The panels of Figure~\ref{theory_3} also compare the results from the different bolometric corrections, as in Figure~\ref{theory_2}.
The uncertainty on the bolometric corrections is at the moment the major drawback for theoretical calibrations in these filters. 
Notice the different overall slope of the two sets of results, that cause systematic differences of the TRGB absolute magnitudes at fixed color up to $\sim$0.2~mag.

\begin{figure} 
   \centering
   \includegraphics[width=\textwidth]{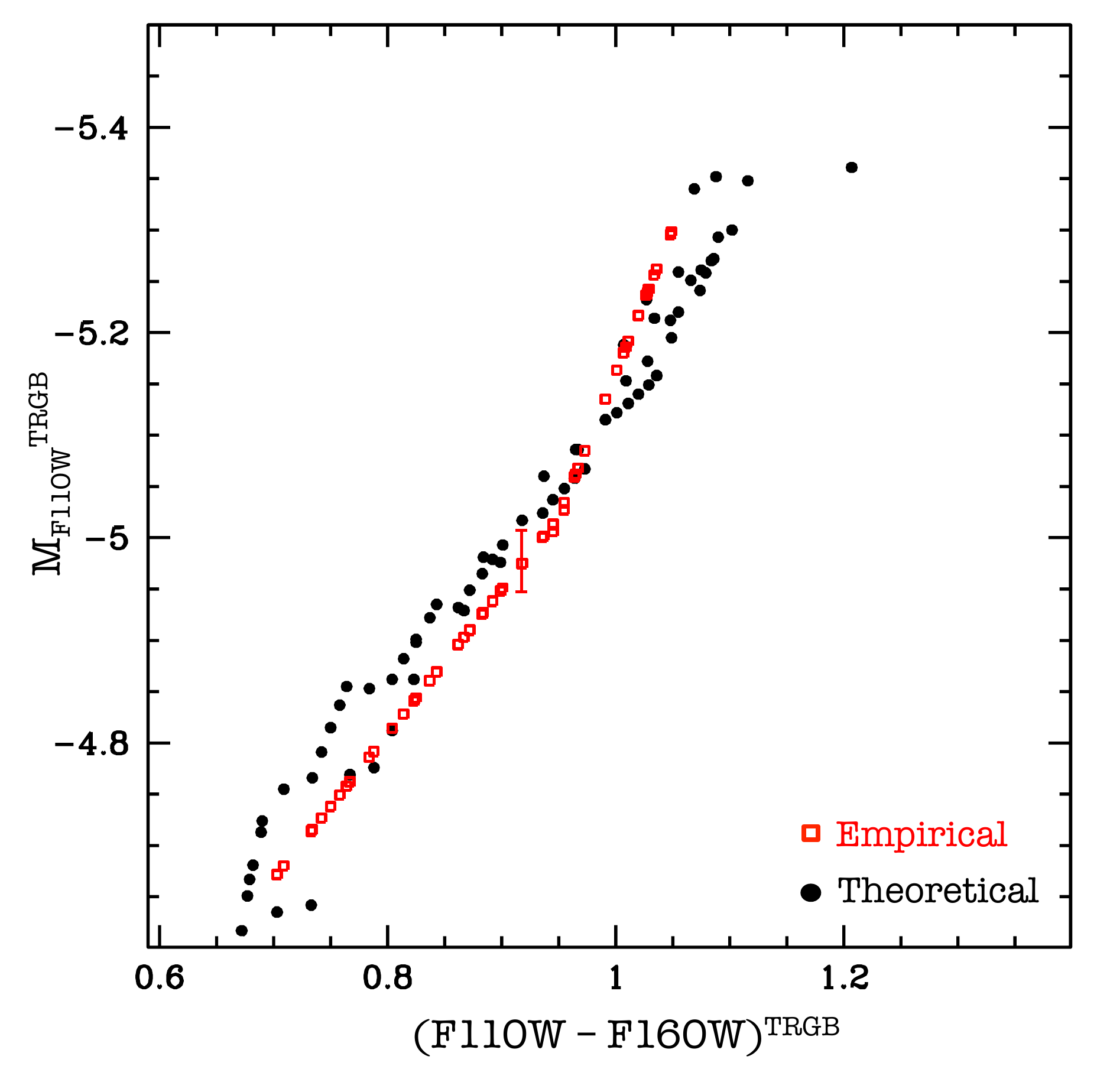}
      \caption{\label{theory_4} Comparison of theoretical and empirical calibrations of the NIR-TRGB color-magnitude distribution for the \emph{HST}+WFC3/IR filters F110W and F160W, which are similar to the 2MASS $J$ and $H$. The filled black circles are the theoretical predictions from BaSTI using the \citet{marcs} bolometric corrections and the open squares are the empirical measurements from \citet{wu14}. The error bar on the zero point of the empirical calibration is also displayed.}
\end{figure}

\citet{wu14} have also derived empirically relationships in the corresponding infrared filters of the WFC3-\emph{HST} system, namely $F110W$ and $F160W$ following the same methods as \citet{rizzi07}. 
Figure~\ref{theory_4} compares their $M_{F110W}-(F110W-F160W)$ TRGB relationship (red boxes) with the theoretical one based on \citet{marcs} bolometric corrections (filled circles). 
The overall shapes are different in the sense that the empirical calibration has a clear change of slope at $(F110W-F160W)\sim$0.95 that is not seen in the theoretical counterpart. 
Overall differences of the TRGB absolute magnitude at fixed color are however typically within $\sim\pm$0.05~mag.  

\subsection{Detecting the TRGB}\label{ssec:trgb_detect}

In this subsection, we discuss the different techniques employed to determine the apparent magnitude of the TRGB, which is a sharp discontinuity along the observed luminosity function (LF). 
In the following, we discuss the major components of the TRGB detection process: edge-detection algorithms in Section \ref{edge-detection}, strategies used to account for the TRGB shape in Section \ref{trgb_sharp}, and brief caution regarding the overlap between the TRGB and other stellar sequences in Section \ref{agb_effects}. 

\begin{figure} 
\centering
\includegraphics[width=0.8\columnwidth]{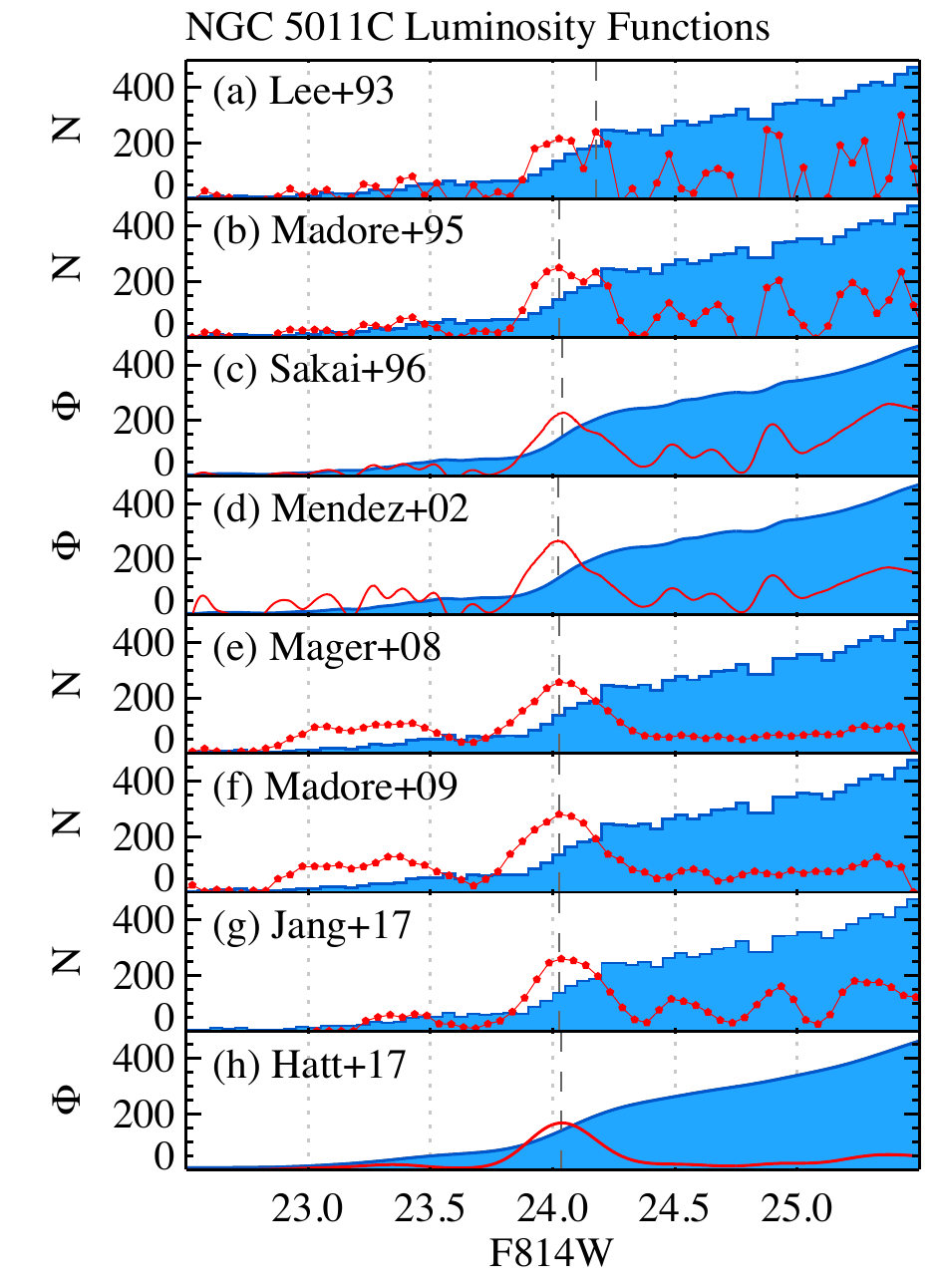} 
\caption{ \label{fig:trgb_comp} Following \citet[][their figure 9]{jang_2018}, F814W luminosity functions (blue histograms) and edge-detection responses (red lines) applied to NGC\,5011\,C. Each panel applies a specific TRGB algorithm to the same underlying data, as follows: (a) \citet{lee93}, (b) \citet{mf95}, (c) \citet{sakai_1996}, (d) \citet{mendez_2002}, (e) \citet{mag08}, (f) \citet{Madore_2009}, (g) \citet{jang_2017b}, and (h) \citet{hatt_2017}. The histograms show modifications for smoothing as employed in each of the studies with y-axis labels of $N$ for binned starcounts and $\phi$ for smoothing. The TRGB is detected in each plot as the maximum of the response function, which is indicated by the dashed vertical line. Generally the algorithms agree for this dataset, but differences can occur as in \citet{jang_2018}.}
\end{figure} 

\begin{table*} 
\centering
\caption{Discrete Approximations to the First Derivative Used for TRGB Detection} \label{tab:trgbdetect}
\begin{tabular}{ll}
\hline \hline 
Reference & Derivative Approximation \\
\hline
\citet{lee93} & $[-2, 0, +2]$ \\
\citet{mf95} & $[-1, -2, 0, +2, +1]$ \\
\citet{sakai_1996} & $\Phi(I+\sigma_m)-\Phi(I-\sigma_m)$\\
\citet{mendez_2002} & $\sqrt \Phi[log(\Phi(I+\sigma_m))-log(\Phi(I-\sigma_m))]$\\
\citet{mag08} & $\sqrt{\sum N_m}[log(\Sigma N_m)-log(\Sigma N_m)]$ \\
\citet{Madore_2009} & $[-1, -1, -1, 0, 0, 0, +1, +1, +1]$ \\
\citet{jang_2017b} & [--1, --2, --1, 0, +1, +2, +1] \\
\citet{hatt_2017} & [--1, 0, +1] \\
\hline \hline
\end{tabular}
\end{table*} 

\subsubsection{Edge-Detection Techniques} \label{edge-detection}

Practically, the LF is constructed from binning the magnitudes of RGB stars from selection over an appropriate color range. 
The TRGB is detected by finding the point of greatest change in the LF, either using tools that approximate the first derivative or tools that model the components of the LF itself (e.g., individual stellar sequences). 
Poisson noise occurs in the LF, especially in smaller samples (like in GGCs), which in turn create spurious spikes in the edge-detection response. 
Thus, smoothing is generally used to mitigate this noise, which can either be (i) incorporated into the functional form of the edge-detection kernel, (ii) applied to the LF itself, or (iii) folded into the model.
Edge detection techniques for the TRGB come in the following forms:
\begin{enumerate}
\item discrete approximations to the derivative (e.g., the Sobel kernal), 
\item discrete approximations to the derivative that incorporate smoothing (e.g., a Gaussian formulation of the Sobel kernel), 
\item maximum-likelihood fitting techniques (e.g., fitting a functional form to the LF with the TRGB as a parameter).
\end{enumerate}
\noindent These methods can applied LF that are either discrete (e.g., with large bins) or a ``continuous'' (e.g., with bins over very small intervals). Each method has advantages and disadvantages that should be weighed for the science application in question.  
As an example, simulations by \citet{mf95} demonstrated that at least 100 stars within one magnitude below the TRGB to detect unambiguously the TRGB for their algorithm and similar tests should be performed for any algorithmic approach to provide guidance for its application.
Popular forms of the discrete derivative approximations are given in Table \ref{tab:trgbdetect}.
Alternative parametric and non-parametric methods have been also applied to determine the TRGB magnitude in a number of stellar systems; the studies of \citet[][]{cioni, conn, makarov} are representative examples.\footnote{Additional non-parametric studies are listed in \citet{jang_2018}.}

Figure \ref{fig:trgb_comp} compares applications of eight different forms for TRGB detection from the literature similar to the comparison in NGC\,1365 made by \citet[][their figure 9]{jang_2018}.
The same photometry for NGC\,5011\,C,a dwarf galaxy in the Centaurus A group \citep{ngc5011C_data}, has been used in each panel. 
The histograms are slightly different owing to the prescriptions of the individual method; those that use direct binned LFs have $N$ labels and those that smooth and/or resample the LF have are labeled $\phi$.  
Comparison of the LF histograms, show that they look relatively similar irregardless of the smoothing, thus differences in the edge-response (red) can be largely attributed to the formulation of edge-detection algorithm.

The edge-detection response function is shown in red in each panel of Figure \ref{fig:trgb_comp}.
The form of the algorithm is given in Table \ref{tab:trgbdetect}. 
The \citet{lee93}, \citet{mf95}, \citet{jang_2017b}, and \citet{Madore_2009} kernels are similar and are applied to binned LFs, except that additional elements are progressively added into the kernel that act to ``smooth'' the response; this is seen by comparing Figures \ref{fig:trgb_comp}a, \ref{fig:trgb_comp}b,  \ref{fig:trgb_comp}g, and \ref{fig:trgb_comp}f that show a progressive smoothing of the edge-detection response.
The algorithms for \citet{sakai_1996}, \citet{mendez_2002}, and \citet{mag08} implement smoothing directly into the algorithm using various means of suppressing the noise; in particular, \citet{sakai_1996} uses adaptive binning and both \citet{mendez_2002} and \citet{mag08} use a poison-noise weight applied to a logarithmic form. 
The impact of these forms can be seen as Figure \ref{fig:trgb_comp}c largely shows the same features as the previous discrete forms, but slightly broader and smoother and both Figure \ref{fig:trgb_comp}d and \ref{fig:trgb_comp}e show a slight amplification of the smaller scale features due both their weighting and logarithmic forms.
Figure \ref{fig:trgb_comp}h that was proposed in \citet{hatt_2017} and used subsequently by \citet{jang_2018} and \citet{hatt_2018} uses a Guassian smoothing function on the LF but applies the simple Sobel kernel from \citet{lee93}, which results in an unambiguous single peaked edge-response.

The magnitude bin of maximum response for each panel are not significantly different from each other (especially considering the discrete forms use magnitude bins of 0.05~mag): $m_{F814W}^{TRGB}$ = 24.18, 24.03, 24.04, 24.02, 24.03, 24.03, 24.03, and 24.04 mag for panels (a) to (h), respectively.
The total range of th detection is 0.16~mag or 8\% in distance. 
The edge-response peak is asymmetric in many of the panels, but only in \citet{lee93} (Figure \ref{fig:trgb_comp}a) is the maximum response in the secondary peak; excluding this result, the range is 0.02~mag or 1\% variation in distance.
This is a more consistent result than for NGC\,1365, which was the example shown in \citet{jang_2018}. 
This LF showed a much larger degree of variation in the peak response for these same algorithms, albeit the authors ultimately concluded the results were not dissimilar within the statistical uncertainties. 
Thus, it is important to choose the algorithm carefully for a given application.

The uncertainties on the TRGB detection are particularly difficult to determine and there is a similar degree of variation in how this measurement is made as in the TRGB detection algorithm.
The strategies incorporate one or more of the following elements: boot-strap or jack-knife resampling of the LF, the bin-size, the mean photometric uncertainty at the tip, the width of the response function, the width of the kernel, and the signal-to-noise of the detection. 
Certainly each of these elements play a role in the underlying uncertainty, but each contributes differently in terms of random or systematic elements. 
One particular uncertainty that is often overlooked is the start point of the binning itself; even with bins of 0.05~mag, the TRGB result will change as the bins are shifted small amounts.

To counter this, \citet{hatt_2017} presents a detailed discussion of how the uncertainties, from the magnitudes and colors to the binning, are coupled in complex ways. 
As a result of this investigation, \citet{hatt_2017} developed end-to-end simulations that insert an artificial TRGB into their images, photometer it, and measure the TRGB in statistical fashion.
This technique combines many of the strategies listed, but adds an ability to measure both random and systematic uncertainty as the TRGB inserted into the images was known a priori. 
Using these simulations, they not only produce rigorous estimates of their uncertainties, but use the statistical distributions to select the best smoothing parameters for their LF. 
The \citeauthor{hatt_2017} methodology naturally incorporates the effects of incompleteness and crowding. 
This method has been applied thus far to five galaxies out to 20~Mpc distant with good results \citep{hatt_2017,jang_2018,hatt_2018}.

\begin{figure} 
\includegraphics[width=\textwidth]{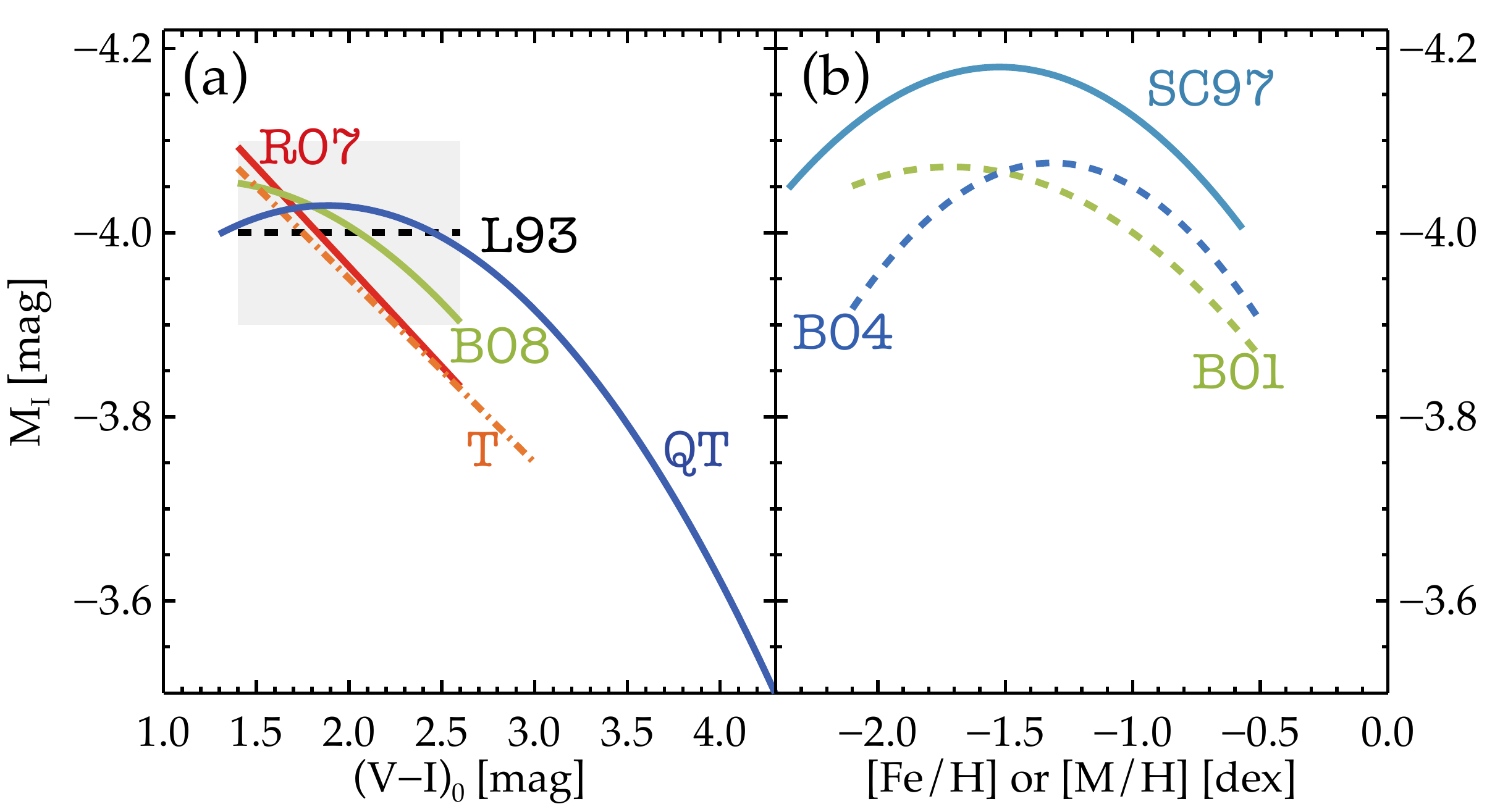} 
\caption{\label{fig:trgbcalib} Comparison of optical TRGB calibrations. 
(a) Calibrations of M$_{I}$ in terms of $(V-I)_{0}$. 
The uncertainties on the zero points are $\sim\pm$0.10~mag the span of which is shown by the gray shading, with the exception of With the exception of the QT calibration at $\pm$0.06 mag using multiple calibration paths \citep[for details see][]{jang_2017a}.
(b) Calibrations of M$_{I}$ in terms of [Fe/H] or [M/H], the uncertainty on the zeropoint of B04 and B01 is $\pm$0.12~mag and for SC97, which is from theoretical models, see Section \ref{sec:trgb_phys}.
The labels are as follows, in panel a: L93 -- \citet{lee93}; R07 -- \citet{rizzi07}, B08 -- \citet{Bellazzini_2008}, T: \citet{Madore_2009}, and QT: \citet{jang_2017a}; and in panel b: SC97 -- \citet{sc97}; B01 -- \citet{bfp}, and B04 \citet{bellazzini_2004}.}
\end{figure} 
\begin{table*} 
\centering
\caption{Calibrations of the Shape of the OPT-TRGB} \label{tab:trgbcalib}
\begin{tabular}{llc}
\hline \hline 
Reference & Calibration of $M_{I}$ & $\sigma_{ZP}$ \\
\hline
\multicolumn{3}{l}{Color Based Calibrations:} \\
\citet{lee93}   & -4.0 & $\pm$ 0.1 \\
\citet{rizzi07} & -4.05 + 0.217[($V-I$)$_{0}$ - 1.6] & none given \\
\citet{Bellazzini_2008} & -3.939 - 0.194($V-I)_{0}$ + 0.080($V-I$)$_{0}^2$ & none given\\
\citet{Madore_2009} & -4.05 + 0.2[($V-I$)$_{0}$ - 1.5] & none given \\
\citet{jang_2017b} & -4.015 - 0.007[($V-I$)$_{0}$ - 1.5] +0.091[($V-I$)$_{0}$ - 1.5]$^2$ & $\pm$ 0.058\\
\hline
\multicolumn{3}{l}{Metallicity Based Calibrations:} \\
\citet{sc97}  & -3.732 + 0.588[M/H] +0.193[M/H]$^2$ & Theoretical \\
\citet{bfp}  & -3.66 + 0.48[Fe/H] + 0.14[Fe/H]$^2$ & $\pm$0.12 \\
\citet{bellazzini_2004}  & -3.629 + 0.679[M/H] + 0.258[M/H]$^2$ & $\pm$0.12 \\
\hline \hline
\end{tabular}
\end{table*} 
%
\subsubsection{Techniques to ``Sharpen'' the Tip Edge} \label{trgb_sharp}

The absolute magnitude of the TRGB is often approximated to a single value: $M_{I}^{TRGB} \sim -$4.05 $\pm$ 0.10 mag \citep[see, e.g.,][and references therein]{lee93, sc97, tr, bfp}. 
However, as shown in Figures \ref{theory_1} and Figures \ref{theory_2}, outside of the old, metal-poor regime, $M_{I}^{TRGB}$ is not, in fact, constant. 
Fortunately, the age or metallicity effect is projected into the color of the star, e.g., more metal rich and/or younger stars are fainter and redder in $I$,$(V-I)$ CMDs than old their metal-poor counterparts.
To both boost the signal at the TRGB and to avoid spurious detections, many authors take this effect into consideration. 

The two primary ways to counteract the shape of the TRGB are as follows:  
\begin{enumerate}
\item \textit{Magnitude-Color Calibration:}~ Instead of determining a single-valued calibration, the calibration can be measured as a function of the mean $V-I$ color known as the $M_{I}^{TRGB}-(V-I)^{TRGB}$ relation. Here, the TRGB $(V-I)$ color basically accounts for the star formation history of the observed population. Typically, users of this technique will use the mean color of the RGB sequence to to estimate $M_{I}^{TRGB}$ for their system. An alternate formulation relates  $M_{I}^{TRGB}$ directly to the metallicity. Given that the metallicity has to be measured in a independently, preferably from spectroscopy, but often in reference to theoretical stellar tracks (isochrone fitting), this is not necessarily an advantage. 
\item \textit{T magnitude:} \citet{Madore_2009} suggested that, instead of fitting a more complex zero-point, to calibrate the slope out of the color-magnitude data. \citeauthor{Madore_2009} used theoretical tracks to produce an alternate magnitude system, known as the ``T'' for TRGB-magnitude system. This technique transforms the observed photometry into a system where the TRGB is flat -- effectively torquing the CMD \citep[a useful schematic can be found in figure 1 of][]{Madore_2009}. Then, a single calibration to the metal-poor portion of the TRGB can be applied to the data over the full range. This can be done independent of priors on metallicity. The \citet{Madore_2009} formulation is as follows:
\begin{equation}
T = I - \beta[(V - I)_{0} - \gamma], 
\end{equation}
\noindent where $\gamma$ is a fiducial color (preferably where the absolute zero point is determined) and $\beta$ was determined relative to theoretical models to be $\beta$ = 0.20 from \citet{bfp}, but values ranging from 0.15 \citep{mag08} to 0.22 \citep{rizzi07} have been used. 
\item \textit{QT magnitude:} More recently, \citet{jang_2017a} expanded upon this initial formulation proposing the $QT$ magnitude system, where the $Q$ stands for ``quadratic'' because the functional form is a quadratic in color. The QT magnitude takes on the following form:
\begin{equation}
QT = F814W - \alpha(color - \gamma)^2 - \beta(color - \gamma)
\end{equation}
\noindent where $\alpha$ = 0.159 $\pm$0.010, $\beta$= -0.047 $\pm$ 0.020, and $\gamma$ = 1.1 for the $F814W$, $F606W-F814W$ magnitude-color combination (similar to the ground based $I$, $V-I$). \citet{jang_2017a} provide calibrations in several common \emph{HST} filter combinations.
\end{enumerate}

Table \ref{tab:trgbcalib} summaries several of the color-magnitude and metallicity-magnitude calibrations of the optical TRGB and their uncertainties, if available. 
The calibrations are compared in Figure \ref{fig:trgbcalib}, where Figure \ref{fig:trgbcalib}a shows the $M_{I}$-$(V-I)$ calibrations and Figure \ref{fig:trgbcalib}b shows the $M_{I}$-$[Fe/H]$ calibrations.
The common single-value from \citet{lee93} and its approximate color-range is also shown for comparison and its uncertainty largely encompasses the $M_{I}$ span of the more complex forms for the blue-side of the RGB.
Moreover, \citet{hatt_2017} and \cite{jang_2018} both demonstrated that there was no formal different between the TRGB detections for in native, T or QT magnitudes for metal-poor systems.
As with the edge-detectors, the application of these different techniques depends largely on the situation; in particular, only when there are well measured colors, there are significant red populations, and the uncertainties due to the transform be estimated, is it advisable to adopt more complex forms of the absolute calibration.

\subsubsection{Contamination to the LF} \label{agb_effects}

Unfortunately, over its color-magnitude span, the RGB is not the only stellar sequence.
A particularly vexing containment are the asymptotic giant branch stars (AGB), that cover a magnitude range overlapping with the brightest portion of the RGB, and eventually, depending on the galaxy star formation history, can extend to much brighter magnitudes than the TRGB \citep[for more description of the AGB see][and references therein]{Habing_2003,rosenfield_2014,rosenfield_2016}. 
The AGB are particularly daunting because the RGB is their progenitor population and so, where there are strong RGB sequences, there is likely an AGB sequence, albeit at lower overall numbers. 

The start of the AGB sequence for stars more luminous than the RGB can be confused for the TRGB. 
This can be seen in Figure \ref{fig:trgb_comp}, especially for the \citet{mendez_2002} and \citet{mag08} algorithms, there are small increases in the LF and edge-response at $\sim$1~magnitude brighter than the TRGB. 
While in the NGC\,5011\,C case the effect is small, it is not so in galaxies with more recent star formation. 
An example being attempts to detect the TRGB in the Antennae galaxy (NGC\,4038/39) by \citet{saviane_2004,saviane_2008}, who found $\mu = 30.7\pm0.25$ (random) $\pm0.14$ (systematic)~mag. 
A later study, \citet{schweizer_2008}, took into account the AGB population and used fields outside of the main star forming disk to find $\mu = 31.51\pm0.12$ (random) $\pm 0.12$ (systematic)~mag, which has been independently confirmed by \citet{jang_2015}. 
The \citeauthor{saviane_2008} physical distance was 13.3 Mpc compared to \citeauthor{schweizer_2008} measuring 20 Mpc, which demonstrates the impact neglecting AGB stars can have.

Mitigation strategies for the AGB include: fitting the full form of the LF, including and AGB component, as in \citet[][among others]{mendez_2002} or working in regions of galaxies where the liklihood of significant contamination is small like in stellar halos \citep[see discussion in][]{beaton2016}. 
It is worth mentioning, that in regions that are dominated by younger stellar populations (like in disks), other populations like red super giants or highly extincted yellow super giants (Cepheids), can also cause difficulty measuring the TRGB confidently (see, e.g., in the LMC and SMC in Figure \ref{fig:cmd}). 
These latter populations show much more intrinsic variations as well as being embedded in dust (e.g. local changes in extinction) such that they are much more difficult to model. 
For the highest precision and accuracy, the TRGB is best applied to old population regimes.

\subsection{Case Study for the Optical TRGB}\label{ssec:trgb_practice}

\begin{figure} 
\centering
\includegraphics[width=\textwidth]{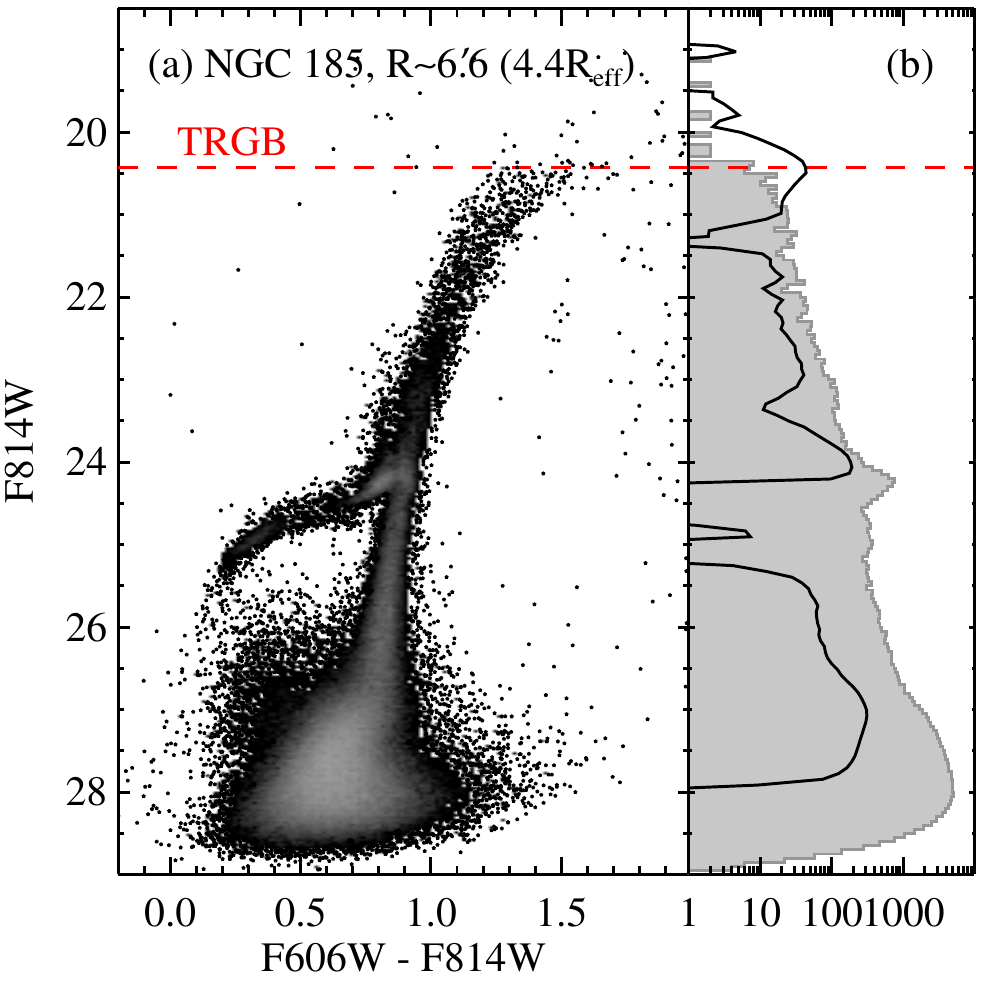} 
\caption{\label{fig_n185} Demonstration of the TRGB measurement process for the Local Group galaxy NGC\,185 using \emph{HST}+ACS data.
(a) F814W - (F606W-F814W) CMD of resolved stars in its outer region ($R$~\textgreater~4.4$R_{eff}$). The TRGB is marked by a dashed line. 
(b) F814W-band luminosity function of resolved stars (histogram) and corresponding edge-detection response (solid line). A strong edge-detection response is seen at the TRGB ($F814W_{TRGB}=20.43\pm0.03$ mag).}
\end{figure} 

Here we present an example of the TRGB distance measurement for the Local Group dwarf elliptical galaxy, NGC\,185 (E3 pec) using the dataset described in \citet[][Proposal ID:GO11724]{geh15}. 
Figure \ref{fig_n185}a shows the F814W$-$(F606W-F814W) CMD (similar to ground based $I$, $V-I$) for a halo field in NGC\,185 and it is evident that the bulk of the bright stars are Pop~II.
The CMD, itself, shows multiple stellar populations, including the main sequence, blue stragglers, HB, AGB, and RGB, but few unambiguously Pop~I sequences (e.g., compare to the LMC and SMC in Figures \ref{fig:cmd}a and \ref{fig:cmd}b). 
The TRGB is the brightest part of RGB sequence and is marked by a dashed line.
 
The apparent magnitude of the TRGB is determined from the I ($F814W$) luminosity function, which is given in Figure \ref{fig_n185}b. 
Because the TRGB is, effectively, the ending point of the RGB sequence, the LF should show a strong discontinuity at the TRGB. 
Here, we applied the edge-detection algorithm introduced in \citet{mag08} and plotted the response in Figure \ref{fig_n185}b as a solid line. 
The edge-detection response shows a strong peak at $F814W~=~20.43$~mag. 
Using bootstrap resampling of the LF and subsequent TRGB detection, the uncertainty on the TRGB measurement is $\pm~0.03$~mag or a 1.5\% statistical uncertainty for the TRGB measurement.

The distance modulus is then determined using a calibration of the TRGB absolute magnitude in the F814W filters and we adopt the \citet{rizzi07} calibration given in Table \ref{tab:trgbcalib}. 
In this formulation, the age-metallicity behavior of the TRGB is modeled using the color of the RGB star, effectively fitting a tilted line to the TRGB absolute magnitude (shown in Figure~\ref{theory_2}).
The median TRGB color in Figure \ref{fig_n185}a is $F606W$$-$$F814W$~=~1.43$\pm$0.02~mag.
Foreground extinction is estimated to be E(B-V)=0.162~mag, corresponding to $A_{F\rm814W}=0.281$ and $A_{F\rm606W}=0.455$~mag \citep{sch11}. 
Combining the absolute magnitude of the TRGB ($M_{F\rm814W}=-4.06\pm0.12$) with the apparent magnitude of the TRGB ($F814W=20.43\pm0.03$), we obtained a value for the distance modulus, $(m-M)_0=24.21\pm0.03 (\rm random)\pm0.12 (\rm systematic)$ mag. 

This value is in good agreement with those based on red clump stars, $(m-M)_0=24.51 \sim 24.64$ mag \citep{pie10}, and RRLs, $(m-M)_0=23.93 \sim 24.15$~mag \citep{sc97, sal98, tam08}.
The distance summary for NGC\,185 provided by \citet{degrijs_2014_m31} quotes the mean distance of 24.027$\pm$0.333~mag from 26 studies for the TRGB, 23.993$\pm$0.128~mag for the RR Lyrae from 8 studies, and 23.997$\pm$0.119~mag for the full sample of distances.

\begin{figure} 
\includegraphics[width=\columnwidth]{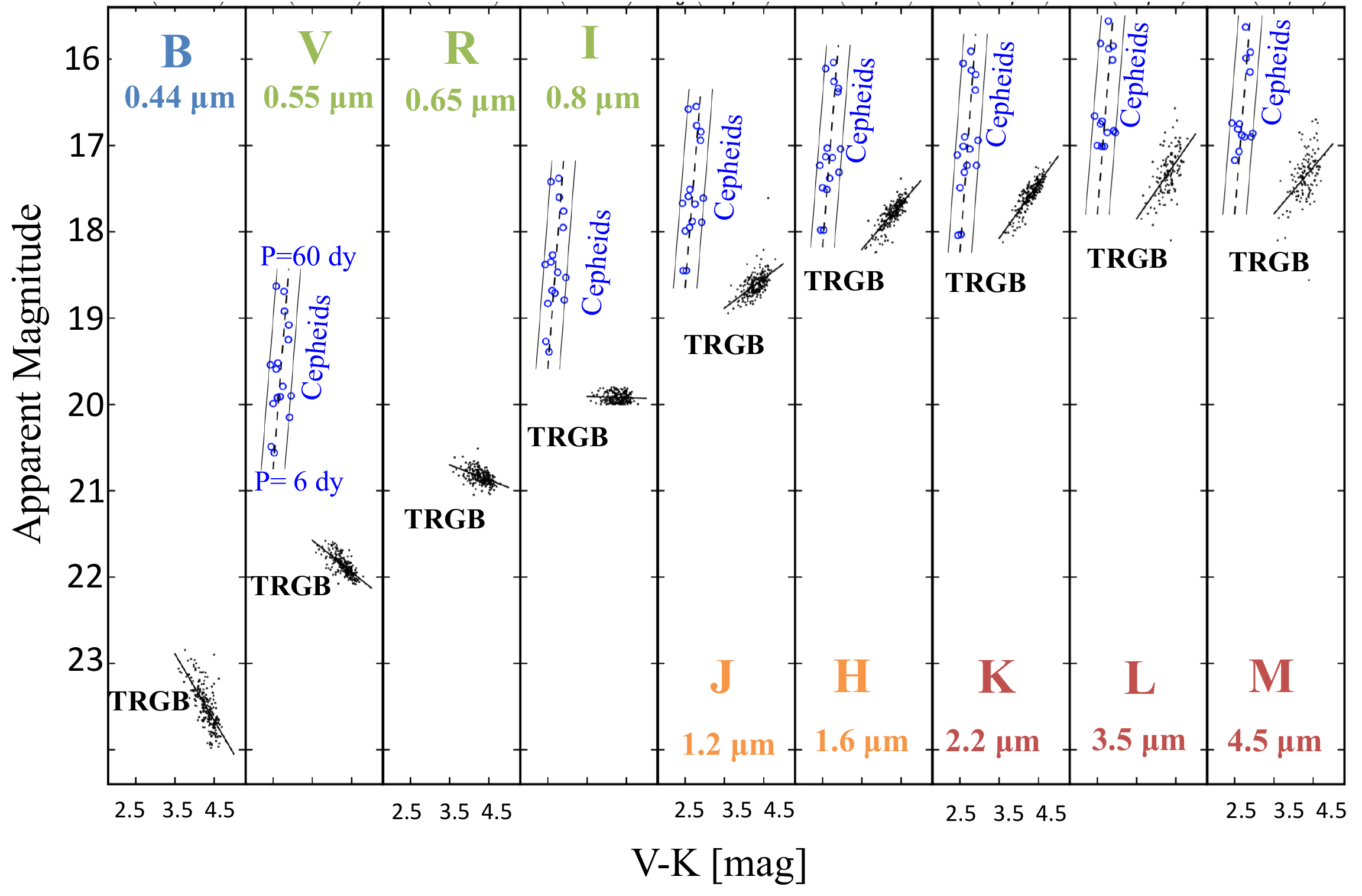}
\caption{\label{fig:carlsondiagram}
The TRGB morphology as a function of wavelength for the Local Group dwarf galaxy NGC\,6822. 
From left to right, the magnitudes in the $B$, $V$, $I$, $J$, $H$, $K$, $L$, and $M$ bands are plotted against the $V-K$ color.
TRGB stars are selected in the $I$, where the slope is flat, and the same stars are plotted in the other panels. 
For comparison, the Cepheid sequence in NGC\,6822 is plotted when the Cepheids could be identified in the data. Figure courtesy of Erika Carlson and using data from \citet{rich_2014}.}
\end{figure} 
%
\subsection{Developments for the IR-TRGB as a Distance Indicator}\label{ssec:trgb_practice_ir}

Figure \ref{fig:carlsondiagram} compare the TRGB (black points) to the CC sequence (blue circles) in the Local Group dwarf NGC\,6822.
Each panel uses a different photometric band plotted against the $V-K$ color, more specifically, (from left to right) $B$, $V$, $I$, $J$, $H$, $K$, $L$, and $M$. 
The TRGB stars are identified in the $I$ where the tip is relatively flat and the same stars are plotted in each of the panels.
The physical behavior of the TRGB described previously is seen empirically as the sloping tips in bands bluer and redder than $I$. 
While in the optical bands the CC population is significantly brighter than the TRGB, in the IR the TRGB is comparable to the shortest period stars (here $P$=10 days). 
Considering that TRGB is non-variable, there are significant gains in observational efficiency using the TRGB in the IR for distances. 
We will focus on the NIR, but note that discussions of MIR stellar populations can be found in \citet{Boyer_2015i,Boyer_2015ii}.

Of course, owing to the steep slope of the TRGB in the IR, the tools developed in the optical case study are not immediately applicable. 
\citet{dalcanton_2012}, \citet{wu14}, and \citet{gorski} address this issue by using TRGB detection at a fiducial mean color for the RGB sequence (this is not dissimilar from means of calibrating the redder optical  populations).
Using their large sample of nearby dwarf galaxies \citet{dalcanton_2012} note, however, that there is a range of mean color across their sample and they correlate this range with the mean metallicity of the galaxies (inferred from star formation histories of the same field).
Stated differently, it is not sufficient to calibrate the TRGB in the IR at one color, the slope of the TRGB needs to be determined. 

\begin{figure} 
\includegraphics[width=\textwidth]{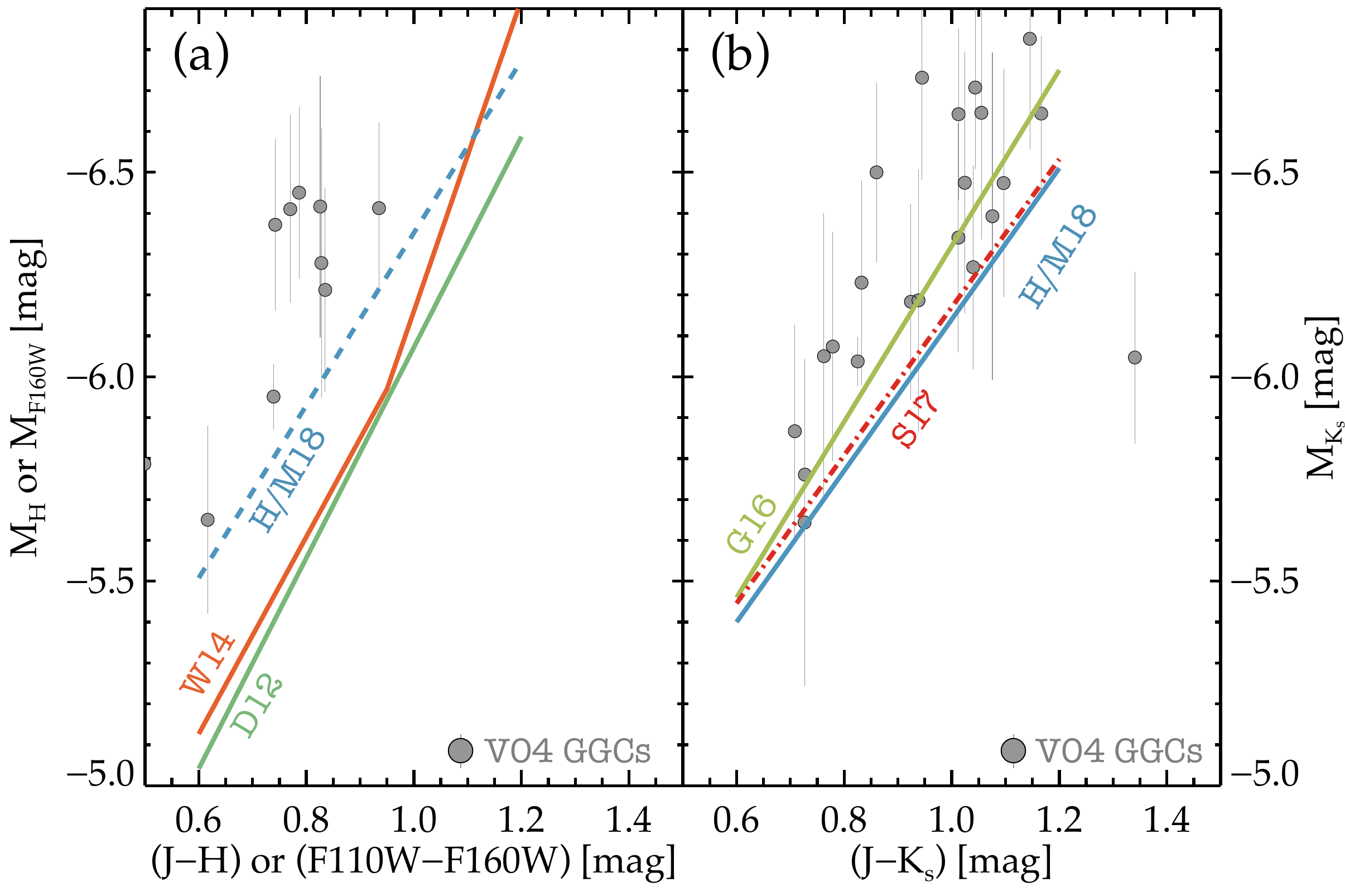} 
\caption{ \label{fig:irtrgbcalib} Comparison of IR-TRGB slope measurements.
(a) $M_{H}$ against $J-H$ for GGCs in \citet[][grey circles]{Valenti_2004} and the LMC zeropoint + IC\,1613 slopes from \citet{hoyt_2018} and \citet{madore_2018} (dashed blue line), respectively. 
$M_{F110W}$ is plotted against $F110W-F160W$ on the same panel (without transformation) and compares \citet[][orange]{wu14} and \citet[][green]{dalcanton_2012}. 
(b) $M_{K_{s}}$ against $J-K_{s}$ for GGCs in \citet[][grey circles]{Valenti_2004}, theoretical calibration from \citet[][dot-dash red]{serenelli_2017}, and LMC zeropoint + IC\,1613 slopes from \citet{hoyt_2018} and \citet{madore_2018} (solid blue line). 
There is a clear distinction between the ground 2MASS and space WFC3/IR filter sets, but generally calibrations agree reasonably well. 
Functional forms are given in Table \ref{tab:irtrgbcalib}.}
\end{figure} 

\begin{table*} 
\centering
\caption{Calibrations of the Shape of the IR-TRGB \label{tab:irtrgbcalib}} 
\begin{tabular}{lclc}
\hline \hline 
Reference & $\lambda$ & Calibration of $M_{\lambda}$ & Notes \\
\hline
\multicolumn{3}{l}{Color Based Calibrations, \emph{HST} Flight System:} \\
\citet{dalcanton_2012} & F160W & -3.496~-2.576~(F110W-F160W)       & rms=0.05 \\
\citet{wu14}           & F160W & -5.97~-2.41[(F110W-F160W) - 0.95] & blue \\
\multicolumn{2}{l}{ }          & -5.97~-3.81[(F110W-F160W) - 0.95] & red  \\
                       & F110W & -5.02~-1.41[(F110W-F160W) - 0.95] & blue \\
\multicolumn{2}{l}{ }          & -5.02~-2.81[(F110W-F160W) - 0.95] & red  \\
\citet{serenelli_2017} & F110W & -4.630 -9.525[(F110W-F160W) - 0.68] & blue, Theoretical \\
\multicolumn{2}{l}{ }          & -4.630 -1.511[(F110W-F160W) - 0.68] & red, Theoretical \\ 
                       & F160W & -5.310 -10.525[(F110W-F160W)- 0.68] & blue, Theoretical \\
\multicolumn{2}{l}{ }          & -5.310 - 2.511[(F110W-F160W) -0.68] & red, Theoretical \\
\hline 
\multicolumn{3}{l}{Color Based Calibrations, 2MASS System:} \\
\citet{gorski}          & K & -6.32 - 2.15 [(J-K$_s$) - 1.00] &  ZW84 scale \\
                        & J & -5.32 - 1.15 [(J-K$_s$) - 1.00] &  ZW84 scale \\
\citet{serenelli_2017}  & K & -6.17 - 1.81[(J-K$_s$) - 1.00] & theoretical \\
                        & J & -4.96 - 0.81[(J-K$_s$) - 0.76] & theoretical \\
\citet{madore_2018} and & J & $-$5.13($\pm0.01_{stat}\pm0.06_{sys}$) - 1.11($\pm$0.15) [(J-H)-0.80] & \\
\citet{hoyt_2018}       & H & $-$5.93($\pm0.01_{stat}\pm0.06_{sys}$) - 2.11($\pm$0.26) [(J-H)-0.80] & \\
                        & K & $-$6.13($\pm0.01_{stat}\pm0.06_{sys}$) - 2.41($\pm$0.36) [(J-H)-0.80] & \\
                        & J & $-$5.14($\pm0.01_{stat}\pm0.06_{sys}$) - 0.85($\pm0.12$) [(J-K)-1.00] & \\
                        & H & $-$5.94($\pm0.01_{stat}\pm0.06_{sys}$) - 1.62($\pm0.22$) [(J-K)-1.00] & \\
                        & K & $-$6.14($\pm0.01_{stat}\pm0.06_{sys}$) - 1.85($\pm0.27$) [(J-K)-1.00] & \\
\hline
\multicolumn{3}{l}{Metallicity Based Calibrations:} \\
\citet{Valenti_2004} & J & -5.67 - 0.31[Fe/H] & GGCs, CG97 scale \\
                     & H & -6.71 - 0.47[Fe/H] & GGCs, CG97 scale \\
                     & K & -6.98 - 0.58[Fe/H] & GGCs, CG97 scale \\
                     & M$_{bol}$ & -3.87 - 0.18[Fe/H] & GGCs, CG97 scale \\
                     & J & -5.64 - 0.32[M/H] & GGCs \\
                     & H & -6.66 - 0.49[M/H] & GGCs \\
                     & K & -6.92 - 0.62[M/H] & GGCs \\
                     & M$_{bol}$ & -3.85 - 0.19[Fe/H] & GGCs \\
\hline \hline
\end{tabular}
\end{table*} 

\begin{figure} 
\includegraphics[width=\columnwidth]{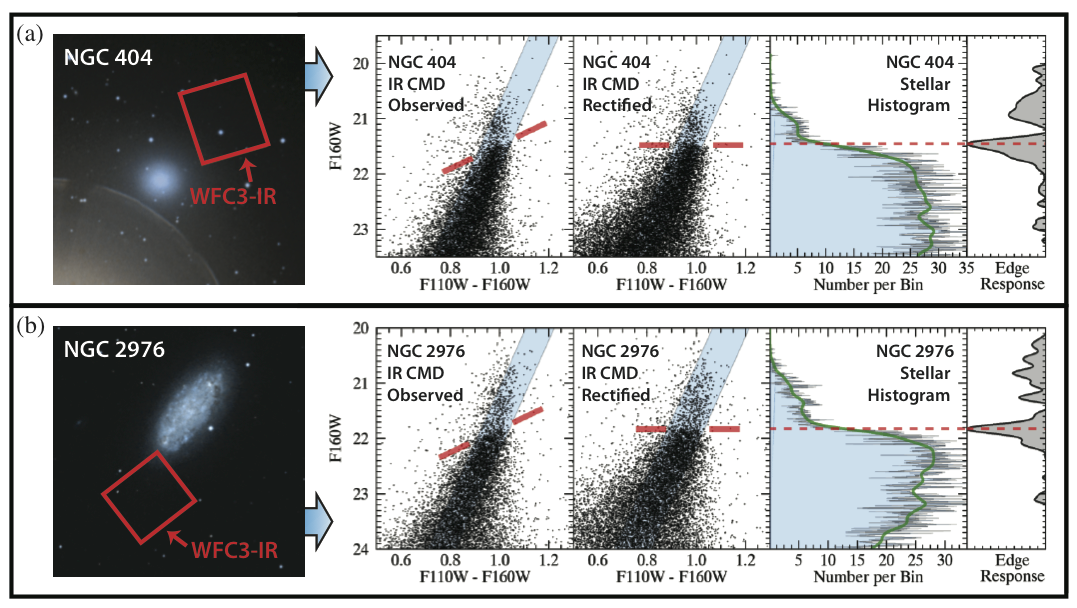} 
\caption{\label{ir_trgb} Use of the NIR-TRGB as a distance indicator applied to (a) NGC\,404 and (b) NGC\,2976.
The photometry is from \citet{dalcanton_2012}.
The WFC3-IR field used is shown in red for GALEX images of the galaxies in the left most panels.
The CMD is shown in the left central panels and a `rectified' CMD is shown in the next panel, from which a luminosity function is determined and an edge-detector employed. 
Figure courtesy of Mark Seibert and photometry courtesy of Julianne Dalcanton and Ben Williams.}
\end{figure} 

Several calibrations of the sloped IR-TRGB are given in Table \ref{tab:irtrgbcalib}.
The calibrations from both \citet{dalcanton_2012} and \citet{wu14} are in the \emph{HST} flight system and are tied to optical TRGB measurements. 
\citet{wu14} rely heavily on theoretical stellar tracks to guide their calibration.
\citet{gorski}, on the other hand, transform a calibration from GGCs by \citet{Valenti_2004} in 2MASS bandpasses to the the color-magnitude plane. Recently, \citet{madore_2018} and \citet{hatt_2018} combine a magnitude-color slope measurement from IC\,1613 with a zero-point determined in the LMC, both in 2MASS bandpasses.

These calibrations are compared in the panels of Figure \ref{fig:irtrgbcalib}, with Figure \ref{fig:irtrgbcalib}a showing $M_{H}$ ($M_{F160W}$) versus $J-H$ ($F110W-F160W$) and Figure \ref{fig:irtrgbcalib}b showing $M_{K_{s}}$ versus $J-K_{s}$. 
The calibrations previously described are all shown, with the individual GGC data from \citet{Valenti_2004} in both panels and a recent theoretical calibration from \citet{serenelli_2017} in Figure \ref{fig:irtrgbcalib}b.

First, there is a noticeable offset between the 2MASS and \emph{HST} filter systems shown in Figure \ref{fig:irtrgbcalib}a that is likely due to bandpass differences that are described in \citet{riess_2011_isr} and discussed by \citet{dalcanton_2012}. 
Both \citet{dalcanton_2012} and \citet{riess_2011_isr} use stellar models to transform between 2MASS and \emph{HST} flight systems, but inconsistencies still remain for these red stars whose models are more uncertain \citep[see][their figure 18 and associated discussion]{dalcanton_2012}.
Second, comparing similar filter systems, the results are not terribly dissimilar; \citet[W14,][]{wu14} and \cite[D12,][]{dalcanton_2012} largely agree, with the former have more objects at redder colors (they also employ very similar overall strategies). 
\citet{hoyt_2018} and \cite{madore_2018} (collectively called H/M18 in the figure) find results not dissimilar from the GGCS from \citet{Valenti_2004}, especially considering the TRGB in GGCs is not always well populated and AGB/RGB stars cannot be disentangled. 
\citet{gorski} agrees well with \citet{Valenti_2004}, as anticipated, as their relations are derived from those data.

Unfortunately, both \citet{dalcanton_2012} and \citet{wu14} predominantly use relatively distant objects for the calibration, for which precise distances are not necessarily known independent of TRGB analyses. 
\citet{gorski}, \citet{madore_2018}, \citet{hoyt_2018}, and \citet{Valenti_2004}, on the other hand, focus on the Local Group dwarfs for which independent high-precision distances are known and they find reasonable agreement between these works and their own distances. 
Of course the divide between filter systems still persists in these comparisons, as does the technique of application.

\citet{dalcanton_2012}, \citet{wu14}, and \citet{gorski} each measure the TRGB absolute magnitude at the mean-color of the RGB (similar to the procedure followed for NGC\,185 in Section \ref{ssec:trgb_practice}), which is slightly different from the measurements of \citet{hoyt_2018} and \citet{madore_2018} that measure the slope in the color-magnitude diagram. 
As in the optical, an alternate technique is to use a ``T'' magnitude that accounts for the color dependence. 
\citet{hoyt_2018} demonstrated this method to be extremely effective (precise and accurate), as they were able to see internal structure of the LMC by determining distances to individual zones across the LMC.
The alternate technique is demonstrated in Figures \ref{ir_trgb}a and \ref{ir_trgb}b for two galaxies in the \citet{dalcanton_2012} sample, NGC\,404 and NGC\,2976. 
Here, the sloped-TRGB is rectified into a ``T'' magnitude system, to which the same edge-detection tools from the optical are applied to determine the TRGB magnitude \citep[here following the methods of][]{hatt_2017}.

\subsection{Summary} \label{sec:trgb_future}

In this Section, we have described the TRGB as a standard candle from a theoretical and empirical perspective.
The TRGB is the terminus of the RGB sequence that occurs at the transition from H-shell burning to He-core burning. 
In the $I$-band, the TRGB is observed to have ea constant value of $M_{I}$=-4~mag with uncertainties at the 0.1~mag level for old, metal-poor stars. 
Many empirical and theoretical investigations have provided refinements to this value that are largely consistent, but only via secondary calibration methods; no direct calibration exists.

The TRGB method relies on detecting the star of the RGB sequence, which has been implement in a myriad of methods, some of which were compared directly in Figure \ref{fig:trgb_comp}. 
Generally, the choice of the most appropriate method depends on the scientific application and the data at hand, e.g., what filters are available, the number of sources populating the TRGB, and the contamination, among other concerns. 
Despite the diversity of the approaches summarized here, the advantage of using a relatively-luminous non-variable stellar population comes with dramatically higher observational efficiency than that of variable stars.

%
\section{Systematics and the Absolute Scale} \label{sec:sys}

The goal of this section is to compare the absolute scales amongst the old stellar population standard candles. 
An overall review of progress toward a self-consistent distance scale for all standard candles can be found in \citet[][]{degrijs_2017}, with significant works focusing on overall consistency for objects within $\sim$1~Mpc given in \citet{degrijs14,degrijs_2014_m31,degrijs15}.
A discussion of absolute scales, ultimately, rests onto two nuanced points: 
(i) full evaluation the systematics associated with the distance measurements, and 
(ii) a full understanding of the true independence of the scales utilized by each standard candles. 
To set the stage for this discussion, we open with a case study that compares distance measurements from CCs, RRL, and the TRGB in the Local Group galaxy IC\,1613 in Section \ref{sec:ic1613comp}. 
In Section \ref{sec:primary_absscale}, the ``primary'' methods to set the absolute scale are described with summaries of progress for each of the standard candles.  
Section \ref{sec:second_absscale} discusses the ``secondary'' methods to set the absolute scale, these being the adoption of a distance to a specific object and using the standard candle in that object.
With these efforts in mind, the systematic impacts of metallicity on the derived distances are discussed Section \ref{sec:metals}. 
Other systematic effects are discussed briefly in Section \ref{sec:othersys}.

\begin{figure}
\centering
\includegraphics[width=0.8\textwidth]{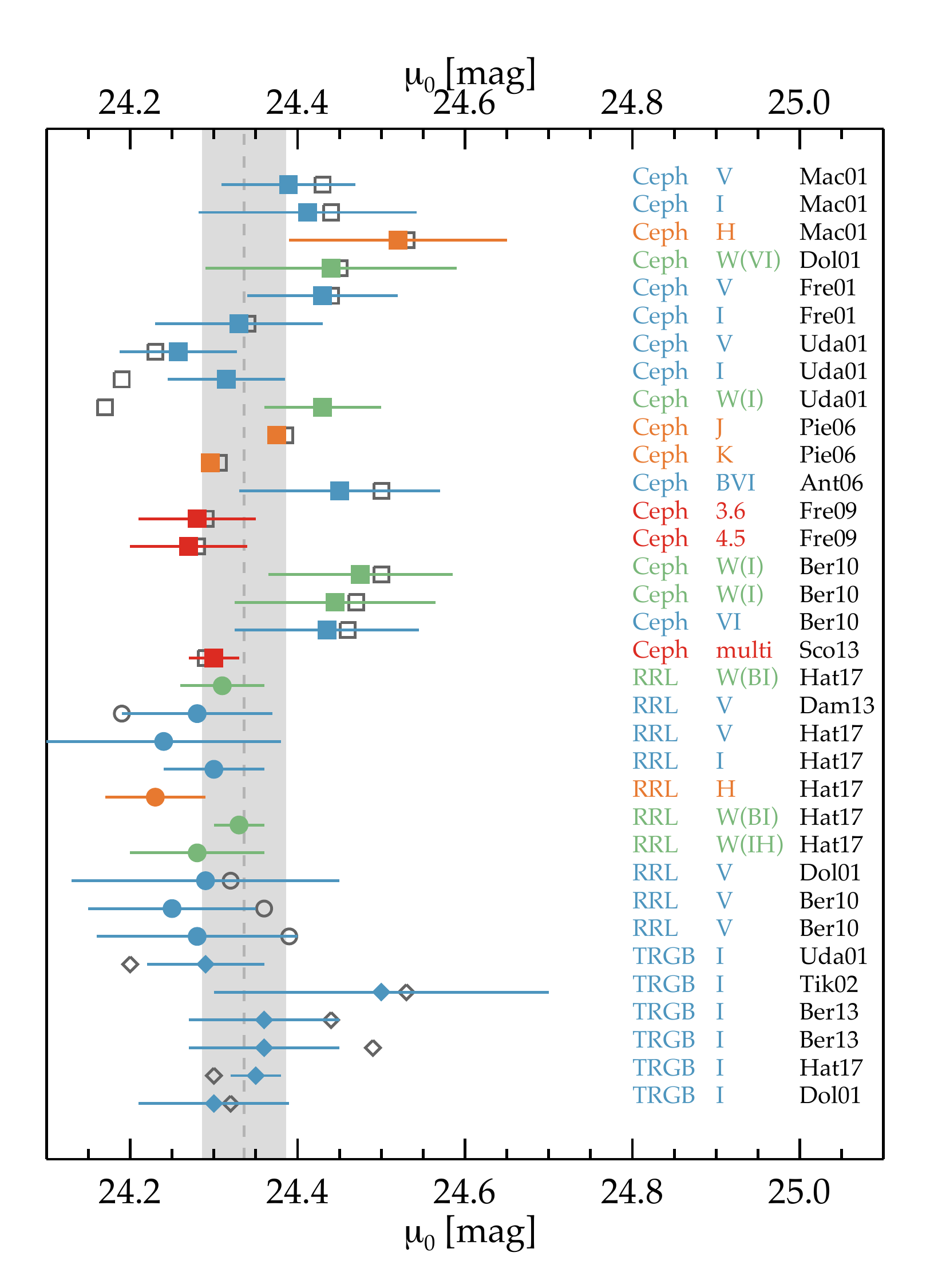}
\caption{\label{fig:ic1613} Comparison of distances to IC\,1613 since the \emph{HST} Key Project using the compendium of \citet[][their fig.~9, their table 4]{hatt_2017}. 
The original measurements are shown as open gray symbols, with Leavitt Law studies as squares, RRL as circles, and TRGB as diamonds. 
Each study has assumed slightly different zero points and foreground/internal extinctions.
The filled symbols display homogenized distance moduli for each study, with the symbols colored by the wavelength range; optical studies in blue, NIR studies in orange, MIR studies in red, and reddening-free Wesenheit studies in green. 
The consensus distance modulus from \citet{degrijs14} of 24.336$\pm$0.049$_{stat}$ mag is shown by the dashed line with a region of width $\pm$0.05 mag represents representative uncertainties for the absolute scales.
Overall this figure demonstrates agreement for the Pop~I and Pop~II scales and between RRL and TRGB.
The distance moduli are homogenized to an LMC distance modulus of 18.49 mag, a total E($B-V$)=0.08 mag for Cepheids) and E(B-V)=0.02 mag for RRL and TRGB, M$_{I}^{TRGB}$=4.00 mag, and M$_{V}^{RRL}$ = 0.63 mag; original uncertainties are plotted.}
\end{figure}
%

\subsection{Comparison Case Study in IC1613} \label{sec:ic1613comp}

We use the compendium of distances to IC\,1613 of \citet{hatt_2017} for CCs, RRLs, and optical TRGB standard candles; no new distances have been published since this paper \citep[while][uses IC\,1613 to calibrate the IR-TRGB, the distance is adopted from Hatt et al.]{madore_2018}. 
The references for these comparisons and the shorthand used in the plot are as follows: 
Mac01 -- \citet{macri01}, Dol01 -- \citet{dolphin01}, Fre01 -- \citet{freedman_2001}, Uda01 -- \citet{udalski01}, Tik02 -- \citet{tikhonov02}, Pie06 -- \citet{piet06}, Ant06 -- \citet{Antonello06}, Fre09 -- \citet{freedman09}, Ber10 --\citet{bernard10}, Sco13--\citet{scowcroft13}, Dam13 -- \citet{dambis13}, and Hat17 -- \citet{hatt_2017}. 
Only a single T2C candidate is known in IC\,1613 \citep{Majaess_2009} and, thus, we do not have a distance from T2Cs in this comparison case study; the study itself, however, still reveals the major points to be discussed in the later subsections. 

Figure \ref{fig:ic1613} shows independently determined distances to IC\,1613 since the \emph{HST} Key Project \citep{freedman_2001} ordered from top to bottom as CC studies, RRL studies, and TRGB studies.
The original distance moduli ($\mu_0$) are plotted with open gray symbols with the CC distances as boxes, RRL distances as circles and TRGB distances as diamonds. 
As is detailed explicitly by \citet[][their table 4]{hatt_2017}, the studies vary considerably in their absolute calibration: for the CCs the distance modulus to the LMC varies from 18.25 to 18.54~mag (10\% in distance), for RRLs the RRL zp varies from $M^{RRL}_{V}$=0.52 to 0.72 (10\% in distance), and the TRGB varies from $M^{TRGB}_{I}$=-3.91 to -4.13 mag (10\% in distance). 
Similar variations also occur in the Milky Way foreground extinction and the internal extinction for the CCs. 
Thus, in attempting to address if these Pop~II distance indicators set comparable distance scales, we must homogenize for these systematics to prevent the dispersion between studies dominating the comparison. 

Following \citet{hatt_2017}, we normalize to $\mu_{LMC}$=18.49 mag for the CCs, $M^{RRL}_V$=0.63 mag and no foreground extinction ($A_{V}\sim$0.0 mag); we deviate from \citeauthor{hatt_2017} by assuming an mean extinction internal to IC\,1613 of A$_{I}$=0.1 mag (for the CCs) and $M^{TRGB}_{I}$=-4.0 \citep[see discussion in][]{hoyt_2018,hatt_2018}.  
The homogenized distance moduli are shown by the filled symbols in Figure \ref{fig:ic1613} and these symbols are color coded by the wavelength regime of the study: blue for optical, orange for NIR, red for MIR, and green for reddening-free relations for the Cepheids and RRL. 
The overall dispersion within the standard candles and between the standard candles reduces.
The mean distance determined from \citet{degrijs_2014_m31} of 24.336$\pm$0.049$_{stat}$ is shown by the dashed gray line with the filled gray region showing $\pm$0.05~mag, which is representative the typical systematic uncertainty on the zero points. 
The measurements shown here are all within 2-$\sigma$ of this ``consensus distance,'' from which we conclude that we see ``overall'' consistency between CCs, RRLs, and the TRGB in IC\,1613. 
It is worth noting, however, that in terms of the systematic effects, IC\,1613 is a simple case: it a metal-poor dwarf galaxy with low internal extinction and low source crowding even in ground-based imaging. 

Before discussing these systematic effects in detail, it is worth noting that comparisons between RRL, CCs, T2Cs, and the TRGB are currently only possible for galaxies in the Local Group; only within the Local Group are populations of for all three distance indicators identified (e.g., only the TRGB has been used for distances outside of the Local Group).
These four distance indicators can only really be compared in earnest with systems of stellar masses on order of relatively massive dwarf galaxies; Figure \ref{fig:cmd} demonstrates the difference in the numbers of CCs, T2Cs, and RRLs between the LMC and SMC.  

Farther afield, \citet{lee_2012} provide comparisons between CCs and the TRGB in M\,101 (the closest host-galaxy for a modern SN~Ia).
\citeauthor{lee_2012} find both standard candles scatter over a large range of 0.7~mag (35\% in distance) and 0.4~mag (20\% in distance), respectively, despite each study quoting uncertainties at the $\sim$0.1~mag level (5\% in distance) --- \citeauthor{lee_2012} did not homogenize for zero points, but these will not fully account for the range. 
Recent high-precision TRGB studies such as \citet{jang_2017b}, \citet{jang_2018}, and \citet{hatt_2018} compare CCs and the TRGB for SN~Ia host galaxies, finding overall agreement, but the bulk of these galaxies have only a single measurement from a single team for each result.
Previous work during and after the \emph{HST} Key Project found similar agreement between scales, but the measurements at that time had larger uncertainties \citep[e.g.,][among others]{fer00,sak04,freedman_2010}.
With these examples in mind, we now proceed to discuss the systematics affecting the absolute scales and individual distance measurements.

\subsection{Absolute Scale through Primary Calibration Techniques} \label{sec:primary_absscale}

Primary calibrations are produced by using individual stars to which distances are determined using geometric methods. 
These come in three types: (i) trigonometric parallax, (ii) statistical parallax, and, for RRLs and T2Cs, (iii) pulsational parallax. 
Spectroscopic parallax is another means of determining the distance to an individual star, but it requires a primary calibration from another means and will not be discussed here \citep[e.g.,][]{adams_1916,sandage_2016}.
First, we describe these methods in Section \ref{sec:primarycalibmeth} and then discuss their application to the RRLs in Section \ref{sec:primarycalibrrl}, T2Cs in Section \ref{sec:primarycalibt2c}, and TRGB in Section \ref{sec:primarycalibtrgb}.

\subsubsection{The Methods} \label{sec:primarycalibmeth}

\subsubsection*{I. Trigonometric Parallax} 

First measured by \citet{bessel_1838}, trigonometric parallax is the apparent displacement of an object, relative to its background, due having viewed the source from two distinct positions. 
The size of the displacement is proportional to the separation between the two viewing positions and for astronomical applications, the typical baseline is the Earth's orbit, e.g., observations are taken $\sim$six months apart. 
From this setup, triangles can be constructed connecting the two viewpoint, the true location for object of interest, and its apparent displacement.
Thus, once positions are robustly determined relative to a background frame, the distance is computed using simple geometry. 
Ground based parallaxes are feasible only for the most nearby stars, and thus, the resolution and stability of space-based observatories are required. 
A history of the technique and an assessment of ground-based methods is given in \citet{Vasilevskis_1966,vanaltena_1983,upgren_1985}.

Major space-based parallax-focused missions are \emph{Hipparcos} \citep[][and references therein]{lindegren_1992,perryman_1997,hipparcos,newhipparcos} and, more recently, \emph{Gaia} \citep[][and references therein]{perryman_2001,prusti_2016}. 
The Fine Guidance Sensor (FGS) on \emph{HST} was designed for astrometric applications \citep[see e.g.,][and references therein]{Duncombe_1991} and been used for parallaxes \citep[for a comprehensive compilation and review of these measurements see][]{benedict_review} and, very recently using a primary \emph{HST} instrument, WFC3 \citep{brown_2018}. 

\subsubsection*{II. Statistical Parallax}
The method of statistical parallaxes is an old one, having been invented by Kapteyn and Kohlsch{\"u}tter in the early 20th century and first published in \citet{adams_2014}.\footnote{A detailed history of these methodological discoveries from the source material is given in \citet{sandage_2004} whereas a briefer summary is more accessible in \citet{sandage_2016}.} 
This technique determines a mean parallax to a group of stars whose proper motions, with the motion of the Sun removed, are analyzed.
The observed proper motion, in angular units, is inversely proportional to the distance to the object, so from the analysis of an ensemble of proper motions, the parallax can be determined for the class. 
A discussion of the mathematics of such a determination is given in \citet[][and references therein]{Popowski_1997}.
The method was employed often in the 20th century to obtain calibrations for the distance scale, but has fallen slightly out of use as more accurate and precise trigonometric parallaxes have become available.

\subsubsection*{III. Pulsation Parallaxes} 
The method was proposed by \citet{Wesselink_1946} based upon earlier suggestions by \citet{Baade_1926} and, as a result, has been coined the Baade-Wesselink method. 
The method combines radius measurements from radial velocity curves with temperature and measurements from light curves to infer the surface brightness, and thereby, intrinsic luminosity of a pulsational variable.
The method was founded upon the postulation that at points of equal color on the ascending and descending branches of the light curve, the temperature of the star should be the same, and in consequence, any difference in luminosity at these points is due to the fractional change in radius ($\Delta$R/R). 
The $B-V$ color--surface brightness relationship first provided in \citet{Wesselink_1969} completed the development phase.  
A limitation of the method is the need for a term $p$ coined the ``projection factor'' that is used to translate between the observed radial velocity and the true pulsational velocity --- the radial velocity measured is the motion of the atmospheric line projected along the line-of-sight and integrated across the stellar disk. 
Thus, $p$ is a multiplicative factor that can come from theoretical inferences or can be measured directly with interferometric methods \citep[see e.g., recent work for Cepheids see][among others]{Pilecki2018,Kervella2017,Breitfelder2016}.

\subsubsection{Primary Calibration of the RRLs} \label{sec:primarycalibrrl}

All three of the direct methods can be applied to RRLs and largely find agreement, albeit most methods produce absolute calibrations at the $\sim$0.1~mag level. 

\smallskip
\noindent \textit{Trigonometric Parallax:}~~
A selection of the major results for calibration of the RRLs via trigonometric parallax are summarized below. It is important to note that \emph{only} in the recent \emph{Gaia} data releases have both the period and metallicity impacts, in addition to the zeropoint, been measurable directly from parallaxes to individual stars. Thus, the future is bright for larger samples of parallaxes to produce a precise and accurate calibration of both the zero-point and the slope of the PLs. 

\begin{itemize}
\item \textbf{\emph{Hipparcos}:}~~Most recently, \citet{Feast_2008} used 142 RRL included in the \emph{Hipparcos} re-reduction \citep{newhipparcos} found $M_V^{RRL}$=0.59$\pm$0.10 mag (adjusted to [Fe/H]=-1.60 dex).
For the NIR, \citeauthor{Feast_2008} found $M_{K_{s}}^{RRL}$=-0.63$\pm$0.10 mag at a fiducial $\log(P)$=-0.252 and assuming PL$_{K_s}$ slope to be -2.41. 
Other studies using the original \emph{hipparcos} reduction include: \citet{fernley_1998}, \citet{Feast_1998}, \citet{Solano_1998}, \citet{Luri_1998}, \citet{Koen_1998},  \citet{Groenewegen1999}, and \citet{Reid_1999}, among others.

\item \textbf{\emph{HST}-FGS:}~~\citet{Benedict_2011} presented \emph{HST}-FGS parallaxes to 5 RRL (4 RRab and 1 RRc) to precisions of \textless~10\%. 
In \citet{Benedict_2011}, itself, the zero point is $M_{V}$=0.42 at [Fe/H]=-1.58, which is systematically lower than other techniques. 
Considering, however, the small sample size, the zero-point could be biased by evolutionary effects, e.g., even a single egregious star could bias the result.
While these parallaxes were the best for the RRL, most works employed a dual theoretical-empirical procedure to determine zeropoints, which are described in the case studies in Section \ref{sec:rrl_practice}, and typically yield results more consistent with other techniques.

\item \textbf{\emph{Gaia} Data Release 1 (TGAS):}~~The first \emph{Gaia} data release included parallaxes from the Tycho-Gaia Astrometric Solution (TGAS), which contained parallaxes for all 142 RRL in \emph{Hipparcos} with some additional stars in the \emph{Tycho} catalog. 
\citet{Neeley_2017} demonstrated use of 46 well characterized RRL in TGAS provided constraints on the zero point for the MIR PL at the same precision as the smaller, but more precise, sample from \emph{HST}-FGS.
Notably, \citet{Neeley_2017} argued that these data necessitated metallicity term to comply with theoretical predictions. 
\citet{Sesar_2017}, found similar results, but using larger samples and a Bayesian inference method for the $W1$ and $W2$ bands fitting for slope, metallicity, and zero-point; this paper found both consistent and inconsistent results as compared to other studies, but also issued caution on the reliability of the abundances.

\item \textbf{\emph{Gaia} Data Release 2 (DR2):}~~ The second \emph{Gaia} data release  included a \textless~10\% parallaxes for several thousands of RRL.
\citet{Muraveva_2018_pl} exploited 401 of these RRL that had sufficient characterization in the literature to derive full PLs in a Bayesian framework for the $V$, $G$, $K_s$ and $W1$ bands; here the limiting factor on the sample is the existence of non-\emph{Gaia} measurements including photometry, extinctions, abundances, and long-baseline periods.
\citet{Muraveva_2018_pl} find: $M_{V}$ = 0.66$\pm$0.06 mag, $M_{G}$ = 0.63$\pm$0.08 mag, $M_{K_{s}}$ = -0.37$\pm$0.11 mag, and $M_{W1}$ = -0.41$\pm$0.11 mag for a fidcucial $P$=0.5238 days and [Fe/H]=-1.5~dex.
Despite the larger sample and better parallaxes, the uncertainties on the zeropoints remain at the 0.1~mag level.
Additional work is likely forthcoming from other teams and employing other wavebands. 
\end{itemize}

\noindent \textit{Statistical Parallax:}~~Because RRLs have a small-to-negligible period dependence in the optical, the absolute magnitude can be approximated to a single value and statistical parallax can be applied. 
A summary of these efforts at that time is given in \citet{smith_1995} and since that time less effort has been employed for this measurement. 
\citet{Popowski_1998,Popowski_1998b,Popowski_1998c} employed a rigorous mathematical framework that incorporated observational biases directly to find
$M_V^{RRab}$ = 0.75$\pm$0.13 mag at [Fe/H] = -1.61 dex for the \citet{Layden_1996AJ} sample and $M_V^{RRab}$ = 0.77$\pm$0.13 at [Fe/H] = -1.60 dex for a sample of 147 kinematically-selected halo RRL. 
Using the larger ASAS all sky sample of variables, \citet{kollmeier_2013} produced the first calibration of the $M_V^{RRc}$ using statistical parallax, finding $M_V^{RRc}$=0.59$\pm$0.10 mag for a mean metallicity of [Fe/H] = –1.59 dex. 
Such methods do not perform as well where there is a period dependence, because the statistical samples must be restricted to period ranges, which reduces precision. 

\smallskip
\noindent \textit{Pulsation Parallax:}~~
As discussed in \citet{smith_1995}, the foundational assumption of the Baade-Wesselink method regarding equal colors implying equal temperatures fails for the RRL because, in part, of RRL shocks that perturb the color-curves in the optical.
\citet{longmore_1985} noted that $K$ observations provided better results for the RRLs and after that time the NIR results were used for such analyses.
Compilations of pulsation parallaxes can be found in \citet{liu90b}, \citet{cacciari_1992}, and \citet{skillen_1993}.
\citet{Muraveva_2015} used 23 pulsation parallaxes to derive the PL$_{K_s}$ relationship and found the zero-point to agree well with those determined from the LMC, but disagree with that determined from \citet{Benedict_2011}. 

\medskip
There is only one star common to each of these means of measuring the parallax: SU Dra.
The pulsation parallax of \citet{liu90b} finds a distance of 640 pc or $\pi$ = 1.56 mas with an uncertainty on the absolute magnitudes of 0.12~mag, that translates to a $\sim$12\% uncertainty on the parallax of $\sim$0.18~mas.
The \emph{Hipparcos} re-reduction found $\pi_{trig}$=0.20 $\pm$ 1.13 mas \citep{newhipparcos} compared to the initial result of $\pi_{trig}$=1.11 $\pm$ 1.15 mas \citet{perryman_1997}.
The \emph{Gaia} DR2 value is 1.4016 $\pm$ 0.0308 mas, whereas \citet{Benedict_2011} finds 1.42 $\pm$ 0.16 mas. 
From comparison of a single star, there is overall agreement within the uncertainties as was demonstrated for the \emph{body} of \emph{HST}-FGS parallaxes against \emph{Gaia} by \citet{benedict_review,benedict_2018}. 
Thus, interpreting differences between zero-points determined in different contexts may be better framed as comparisons between assumptions of the PL-form, sample size, and abundances, rather than the fidelity of the parallaxes themselves.

\subsubsection{Primary Calibration of the T2Cs}\label{sec:primarycalibt2c}

The body of work on the primary calibration of the T2Cs is much smaller than that of RRL and they are discussed with respect to two key stars: VY~Pyx and $\kappa$~Pav that have parallax measurements from multiple techniques. 

\citet{Feast_2008} used \textit{Hipparcos} trigonometric parallaxes for VY~Pyx (BL~Her type with $P=1.24$~days) and $\kappa$~Pav (W~Vir type with $P=9.09$~days), as well as pulsation-based parallaxes for a few T2Cs to see if their absolute magnitudes based on the parallaxes were consistent with the PL relations derived by \citet{Matsunaga_2006} for T2Cs in GGCs.
Their \textit{Hipparcos} parallaxes led to the absolute infrared magnitudes more-or-less predicted by \citet{Matsunaga_2006}, although the pulsation parallax of $\kappa$~Pav had a significant offset for which the authors gave cautions for both the pulsation and trigonometric parallax (which still had a large uncertainty).
\citet{Feast_2010} later suggested that $\kappa$~Pav may be of peculiar W~Vir type, which are, indeed, intrinsically brighter than expected by the PL relation for the W~Vir type. 

\citet{Benedict_2011} reported \emph{HST}-FGS parallaxes for the same two T2Cs, VY~Pyx and $\kappa$~Pav, with significantly smaller parallax uncertainties than those from \textit{Hipparcos}. 
Despite the lower uncertainties, these new measurements left the questions unresolved; using the \emph{HST}-FGS parallaxes, $\kappa$~Pav seems to follow the PL relation, but VY~Pyx, in turn, is fainter than anticipated.
\citet{Breitfelder_2015} found that the \emph{HST} parallax of $\kappa$~Pav leads to p-factor consistent with other Cepheids based on their interferometric and spectroscopic data.
\citet{Breitfelder_2015} also found no evidence of binarity for this object, which would have been a strong clue that the star was of a peculiar type. 
These new results suggest that $\kappa$~Pav is a normal W~Vir, not a perculiar W~Vir, and, thereby, should follow the PL relation of the W~Vir type. 
The nature of VY~Pyx, however, remained unclear.

\begin{figure} 
\centering
\includegraphics[width=0.8\textwidth]{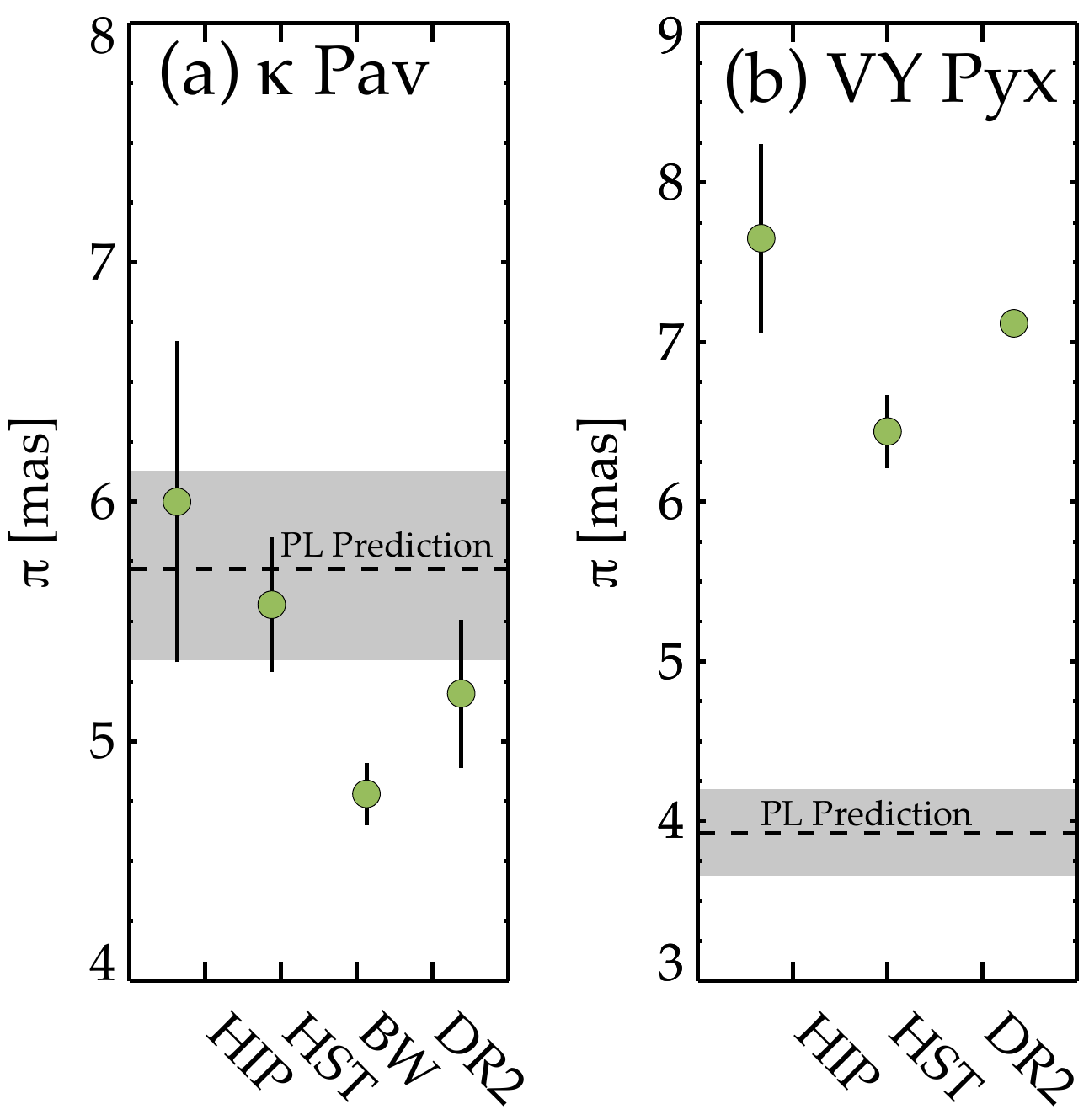}
\caption{\label{fig:pi_comp} Comparison of geometric distance measurements to the T2Cs
 (a) $\kappa$~Pav ($P$=9.09 dy) and (b) VY~Pyx ($P$=1.2 dy) to the predicted parallax from the \citet{Matsunaga_2006} K$_{s}$-PL relation. 
 The shaded region gives the scatter about the \citet{Matsunaga_2006} PL ($\pm$0.15~mag) from the GGCs. 
While $\kappa$~Pav is largely consistent with the PL, VY Pyx is quite deviant.}
\end{figure}

\emph{Gaia} DR2 offers more insight, finding for $\kappa$~Pav, $\pi$ = 5.1992$\pm$ 0.3094 mas and for VY~Pyx, $\pi$ = 7.1173$\pm$0.0265 mas. 
Figure \ref{fig:pi_comp} compares the parallax results from \emph{Hipparcos}, \emph{HST}-FGS, pulsation parallaxes, and \emph{Gaia}--DR2 for these two T2Cs.
The prediction from the PL of \citet{Matsunaga_2006} is plotted with the uncertainty of the zero-point in gray; the dashed line represents if the star is directly on the PL (e.g., no intrinsic dispersion). 
The trigonometric parallaxes for $\kappa$-Pav, Figure \ref{fig:pi_comp}a, are all consistent with \citet{Matsunaga_2006}, whereas the pulsation parallax is several sigma away. 
There is no pulsation parallax for VY~Pyx, but the trigonometric parallaxes in Figure \ref{fig:pi_comp}b are all inconsistent with the \citet{Matsunaga_2006}-PL. 
Thus, the nature and classification of VY~Pyx likely requires further study. 
As with the RRL, this comparison highlights the need for parallaxes to a {\it representative} sample of stars to infer PLs instead of relying on individual stars as a prototype for a class, e.g., the \emph{intrinsic dispersion within the class} may be large. 

\subsubsection{Primary Calibration of the TRGB} \label{sec:primarycalibtrgb}

At the time of writing, the authors are only aware of a single study to provide a primary calibration of the TRGB. 
\citet{tabur_2009} combined \emph{Hipparcos} and 2MASS data to provide a preliminary calibration of the TRGB absolute magnitude in $K_s$, finding $M_{K}$= -6.85$\pm$0.03~mag. 
This study, however, has a number of caveats: (i) for the bright stars populating the tip, most of the magnitudes were derived from DIRBE because such stars in 2MASS had large (0.2~mag) uncertainties due to saturation, 
(ii) the \emph{Hipparcos} parallaxes have extremely large errors (\textgreater\textgreater~20\%), and 
(iii) the authors only fit for a single color bin.
Moreover, the authors do not use a Bayesian approach to determining distances using parallaxes with large uncertainties as is recommended \citep[see e.g.,][]{cbj_2015}.
As mentioned in \citet{tabur_2009} with respect to $JHK$, the current limitation for direct measurement of the TRGB is the lack of homogeneous photometry for bright stars in the bands of interest for the TRGB -- primarily, $I$ and $JHK$.
Such work is challenging because these stars cannot be identified a priori as tip stars and the nearby stars with the most precise parallaxes are saturated in most modern surveys. 
Thus, even with the wealth of parallaxes from \emph{Gaia}-DR2, this measurement remains elusive.

\subsection{The Absolute Scale via Secondary Calibration} \label{sec:second_absscale}
Secondary calibration works in two steps: (i) establish the distance to an object, like a star cluster or nearby galaxy, using a geometric technique and (ii)  determine the absolute magnitude of the standard candle using samples from that object.
Thus, the absolute scale set has two components to the systematic uncertainty: (i) the total uncertainties from the geometric technique and (ii) the uncertainties from the standard candle. 
It is also possible to combine other standard candles to obtain a secondary calibration, which adds another layer of systematic uncertainty; for brevity, we will not include those results here though some discussion exist in the sections for each distance indicator.

In this volume, \citet{thevenin_2017} describe ``modern'' geometric techniques, which includes a number of different methods -- to which methods relying on gravitational waves can be added \citep[e.g,][]{gravitational_waves}, which are also described in this volume by \citet{czerny_2018}.
However, of those only two are applicable on the distance scales to directly calibrate the Pop~II distance indicators described here, (i) eclipsing binaries and (ii) mega-masers (only for the TRGB). 
As for the primary calibration, we give the general sense of the methods in Section \ref{sec:2ndscale_meth} before describe application to the RRLs in Section \ref{sec:2ndscale_rrl}, T2Cs in Section \ref{sec:2ndscale_t2c}, and TRGB in Section \ref{sec:2ndscale_trgb}. 

\subsubsection{Methods for Secondary Calibration} \label{sec:2ndscale_meth}

\subsubsection*{I. Eclipsing Binaries:} 

Eclipsing binaries (EBs) are a set of binary star systems where the binary orbit is sufficiently along the line-of-sight that the one star eclipses the other. 
Detached eclipsing binaries (DEBs) are a subset of EBs for which both stars are entirely within their respective Roche Lobes, meaning that the stars can be modeled as a simple two-body system. 
Using a combination of (i) light curve analyses for the inclination, relative radii, and relative luminosity and (ii) radial velocity curves, both masses and the separation of the binary can be determined in absolute units. 
The best DEBs for distances have two eclipses and one of the eclipses is ``complete'' or nearly so \citep[see discussion in][]{Kaluzny_2005}.
From the absolute separation, the radii of the two stars can be solved for, which, using a surface brightness--radius relation \citep[e.g.,][]{diBenedetto_1998,Graczyk_2017}, can be converted into a luminosity, and hence, a distance.
The precision of the surface brightness--radius relation is a function of the stellar type and luminosity class, with late-type binaries being those with more well-defined relations. 
The surface brightness--radius relation can only be calibrated for stars in the Solar neighborhood, which could impart age or metallicity biases, but so far the relationship shows only mild sensitivity to metallicity \citep{Houdashelt_2000AJ}. 

A summary of EB results in GGCs can be found in \citet{Kaluzny_2005}.
Since then EBs have been found in, monitored in, and distances derived to: 
NGC\,6362 \citep[14.74$\pm$0.04~mag][]{Kaluzny_2015}, 
M\,4 \citep[][see Section \ref{sec:rrl_practice}]{kaluzny_2013},
47\,Tuc \citep[13.35$\pm$0.08~mag][]{Thompson_2010}, 
NGC\,6397 \citep[11.65~mag][]{Rozyczka_2014}, 
$\omega$~Centauri \citep[14.05$\pm$0.11~mag][]{Thompson_2001},
and 
M\,55 \citep[13.94$\pm$0.05~mag][]{Kaluzny_2014} within the context of the Clusters AgeS Experiment (CASE), which is also monitoring a number of other nearby GGCs. 

An extragalactic census of EB results is given in \citet{bonanos_2013} and includes sources detected in NGC\,6822, IC\,1613, LMC, SMC, M\,31, and M\,33; distances have been determined from the EBs in all but NGC\,6822 and IC\,1613.
A 2\% distance to the LMC using eight DEBs was published in \citet{Pietrzynski_2013}, with modeling of an additional 12 systems recently published in \citet{Graczyk_2018} for a total of 20 DEBs in the LMC.
A distance to M\,33 with 1 DEB was given in \citet{Bonanos_2006}. A distance modulus of $\mu_{M33}$=24.92$\pm$0.12~mag (964$\pm$54 kpc) was found.
A distance to M\,31 using 2 DEBs was published in \citet{Vilardell_2010}.
Since then, several hundred EBs have been discovered in the M\,31 field \citep[e.g., 298 from][]{lee_2014}, of which only 11 are bright enough for radial-velocity follow-up (e.g., $V$\textless20.5~mag for 8-10m class telescopes) and consistent with being in M\,31. 
Eclipsing binaries provide an excellent means of calibrating zero points, but only in systems with well populated sequences and accompanying information on the metallicity effects.

\subsubsection*{II. Megamasers in Accretion Disks:} 
The Local Volume galaxy, NGC\,4258, hosts a megamaser in the accretion disk of its central black hole. 
The accretion disk is thought to be nearly edge-on and the emission from the megamaser traces the motion of the gas in the disk. 
Long-term monitoring of the megamaser emission produces a velocity field and with the assumption of Keplerian motion in the disk, a physical radius can be calculated and a geometric distance determined. 
Further methodological details can be found in this volume by \citet{thevenin_2017}.

Since its discovery, seven distances to NGC\,4258 have been published: 
28.66$\pm$0.47~mag \citep{Greenhill_1995},
29.03$\pm$0.29~mag \citep{miyoshi_1995},
29.29$\pm$0.08~mag \citep{Herrnstein_1999},
29.29$\pm$0.02~mag \citep{Humphreys_2008},
29.31~mag (no uncertainty was given) \citep{riess_2012},
29.40$\pm$0.05~mag \citep{Humphreys_2013},
29.39$\pm$0.06~mag \citep{riess_2016}.
While early results had large uncertainties as the methodology was being developed, the last four publications cite \textless 2.5\% uncertainties distance, but differ at the 5\% level. 
The physical distances span from 7.2~Mpc to 7.6~Mpc \citep{jang_2017b}. 

\subsubsection{Secondary Calibration of the RRLs} \label{sec:2ndscale_rrl}

RRL calibration in globular clusters has been performed in M\,4 in the NIR and MIR. 
Reasonable agreement between EBs and RRL distances from trigonometric parallaxes were found for M\,4. 
Many of the clusters with EB distances also have samples of RRL \citep{clement01}.\footnote{An impressive and detailed tabulation of all known variables in clusters, which includes RRL, T2Cs, and EBs among other types, has been updated by Christine Clement regularly for many years and is available at this URL: \url{http://www.astro.utoronto.ca/~cclement/cat/listngc.html}} 
A benefit of the GGCs as calibration objects is their small age and metallicity ranges, which allows comparisons to be made for these parameters. 
At the same time, however, the use of mono-abundance and mono-age relations is not applicable to mixed populations; albeit $\omega$ Centauri is one exception \citep[for detailed discussion see][]{braga_2018}.
Thus, so far use of geometric EB distances for direct calibration of RRL remain poorly utilized; whereas, non-geometric distances (isochrone fitting, mean HB magnitude, and others) are used more frequently. 

Performing direct calibration of the RRLs in the LMC and SMC is challenging due their line-of-sight depth and lack of clear information on how to address internal extinction; moreover, the RRL have $V\sim$\textgreater18~mag, which requires large telescopes for direct abundances. 
Nevertheless, \citet{Muraveva_2015} performed a calibration using 70 LMC RRL, finding it to agree well with pulsation parallaxes, but disagree with the \emph{HST}-FGS parallax sample. 
There are reasonably large samples of RRLs in IC\,1613, M\,33 and M\,31, typically detected via \emph{HST}, but these are largely still used to determine the distance to these objects rather than to calibrate the RRL themselves \citep[see e.g.,][among others]{Sarajedini_2006,Sarajedini_2009,Bernard_2010}.
Again, the RRL at the distance of M\,31 are at $V$\textgreater25~mag and are out of range for most spectroscopic facilities we well as being difficult to obtain sufficient sampling to apply photometric techniques for metallicity.

\subsubsection{Secondary Calibration of the T2C} \label{sec:2ndscale_t2c}

A tabulation of T2Cs in GGCs is given by \citet[][their table 2]{Matsunaga_2006} and it contains 46 individual T2Cs in 26 different clusters, of which only 12 clusters contain more than a single T2C. 
Unfortunately, the list of clusters \emph{with} T2Cs and \emph{with} geometric distances is disjoint; \citet{Matsunaga_2006} calibrated their relations to the magnitude of the horizontal branch. 
A large sample of T2Cs exist in the LMC and SMC and the geometric distances to these galaxies were the basis for relations derived in Section \ref{sec:t2cpl}.

\subsubsection{Secondary Calibration of the TRGB} \label{sec:2ndscale_trgb}

GGCs are not ideal objects to calibrate the TRGB because their stellar mass is sufficiently small that it is not guaranteed that there will be a true TRGB star. 
The EBs in extragalactic systems, however, make good sources of calibration; the LMC EB distance is used routinely, both \citet{gorski} and \citet{hoyt_2018} use the LMC to calibrate the IR-TRGB and \citet{jang_2017b} uses the LMC in the optical. 

The TRGB can be independently calibrated with NGC\,4258 using \emph{HST}.
The TRGB measurements are as follows:
$F814W^{TRGB}$ = 25.20$\pm$0.06~mag (WFPC2), $F814W^{TRGB}$ = 25.24$\pm$0.04~mag (ACS) \citep{mag08},
 25.25 (with 90\% confidence interval from 25.23 to 25.38) and  25.22 $\pm$0.09 mag \citep{Mouhcine_2005},
$F814W^{TRGB}$ = 25.39$\pm$0.11~mag \citep{Madore_2009}, and  
 F814W$_0^{TRGB}$ = 25.357$\pm$0.031~mag \citep{jang_2017a}. 
Thus, the individual TRGB measurements span from 25.20 to 25.357~mag, a span of 7\% in distance.
For this reason, \citet{jang_2017b} perform a through comparison using different photometry packages and different techniques for determining the point-spread-function finding differences between studies at the 0.05~mag level, which agrees with earlier comparisons in NGC\,4258 in \citet{mag08}. 
Taken in conjunction with the other measurement uncertainties discussed in Section \ref{ssec:trgb_detect}, the difference in the TRGB magnitudes can largely be explained.

Given the variation in TRGB detections and the variation in the distances themselves, whether or not a TRGB calibration in NGC\,4258 agrees with other techniques is largely dependent on which studies are compared. 
\citet{rizzi07} note that the \citet{Humphreys_2005} distance to NGC\,4258 calibrates the TRGB to be 2.3-$\sigma$ fainter than their determination from other sources. 
\citet{beaton2016} note the opposite effect via the distance from \citet{Humphreys_2013}, finding $M_{I}^{TRGB} = -4.16\pm0.06_{stat}\pm0.04_{sys}$~mag, which is systematically more luminous.
\citet{jang_2017b} ultimately find $M^{TRGB}_{F814W}$ = -4.03$\pm$0.07~mag using \citet{riess_2016}, which is consistent with other ground-based calibrations.

%
\subsection{Metallicity} \label{sec:metals}

Theoretical and empirical investigations suggest that metallicity impacts the absolute magnitude for each of our Pop~II standard candles. 
In many ways, age and metallicity are coupled. 
The impacts are two-fold: (i) metallicity impacts the physical structure of the star, which in turn, changes how it evolves and (ii) the metals in the atmosphere selectively absorb and thermally re-emit flux in specific bands. 
The impact of metallicity can be subdivided into two challenges: 
 (i) difficulties in obtaining direct and/or independent measurements, and 
 (ii) the use of distinct metallicity scales between calibration and application studies.
The former point is distance indicator specific and will be discussed in subsections, whereas the latter is more general and will be discussed here.

The two common metallicity scales for Galactic work are those of (i) Carretta-Gratton \citep{Carretta_1997} and (ii) Zinn-West \citep{zw84}. 
The \citeauthor{zw84} scale was established using a variety of ``integrated parameters'' in GGCs, including the integrated line index ($Q_{39}$) and the infrared Calcium Triplet (CaT), to set a scale for [Fe/H]; this choice was motivated by the desire to establish a homogeneous scale that could work for both nearby and distant objects, but comes at the cost of these integrated techniques not being always correlated directly with the quantity [Fe/H], itself \citep[e.g., see][for visualizations of how different elements do or do not correlate with Fe across the Galaxy]{Hayden_2015}.
\citeauthor{Carretta_1997} used a smaller sample of objects (24 GGCs), but used higher resolution spectroscopy to produce metallicity measurements from Fe lines as well as other elements that contribute to integrated techniques.

A comparison between \citeauthor{zw84} and \citeauthor{Carretta_1997} is given in \citet[][their figure 5]{Carretta_1997} that shows a non-linear correlation between the two studies in addition to a zero-point offset due to adjustments to the solar Fe abundance, finding:
\begin{equation}
[Fe/H]_{CG97} = -0.618 - 0.097[Fe/H]_{ZW} - 0.352[Fe/H]_{ZW}^2
\end{equation}
with a scatter of $\sigma$=0.08 dex and uncertainties on the coefficients of $\pm$0.083 for the zeropoint, $\pm$0.0189 for the linear term, and $\pm$0.067 for the quadratic term. 
As an example, a star with [Fe/H] = -1.8 dex on the \citeauthor{zw84} scale corresponds to [Fe/H] = -1.7 dex on the \citeauthor{Carretta_1997} scale, with the differences being larger on the metal-poor and metal-rich ends of the scale. 
Thus, the impact of using different metallicity scales may be complex. 

\subsubsection{Metallicity Impacts in RRL:} 

Solid theoretical \citep{Marconi15} and semi-empirical \citep{Neeley_2017} evidence show that the mean magnitudes of the RRLs in all the optical-NIR-MIR photometric bands are affected by their iron abundance---the more metal-poor the variable is, the brighter. 
The metallicity coefficient of the PL relations ranges from 0.14 to 0.19~mag~dex$^{-1}$ \citep{Marconi15,Neeley_2017} in the optical, NIR and  MIR, showing no clear trend with wavelength. 
It is, however, important to note that the intrinsic dispersions for the long-wavelength PLs is smaller than the theoretical separation between {$\sim$}0.5~dex populations, implying the effect should be more easily distinguished for the longer-wavelength data \citep[see discussions in][]{Neeley_2017}.
Thus, metallicity is fundamental for the distance determination using RRL.

Direct spectroscopic metallicity measurements in RRL are complicated due to the short timescales over which the atmospheres are changing.
Typical radial velocity amplitudes are $\sim$50 km/s \citep[see, e.g.,][and references therein]{sesar_2012} such that even short observations ($\sim$10 min) can be impacted by line blurring.
Thus, spectroscopic observations are either conducted at specific phases or use techniques employed at specific phases to infer the metallicity.
Some common alternate techniques include:
\begin{itemize} 
\item \textbf{$\Delta$S method:} ~~ \citet{preston1959} developed a technique that infers metallicity using a combination hydrogen lines to infer the spectral type (e.g., temperature) and Ca II H and K lines to infer the metallicity known as the $\Delta$S method. The method is typically only applied at minimum light. Narrow-band filters can be used for the specific lines used in this technique \citep[see][]{rey00}. The recalibration by \citet{Clementini_1995} using high resolution spectra, which was revisited by \citet{Carretta_1997} to find (on their scale):
\begin{equation}
[Fe/H] = -0.187(\pm0.011)~\Delta S~-~0.088(\pm0.04) ~~ (\sigma=0.269)
\end{equation} 
\item \textbf{Fourier coefficients:} \citet[and references therein]{kovacs_1995,jurcsik_1996,smolec_2005} demonstrate a correlation between Fourier coefficients, in particular, $\phi_{31}$, and metallicity. The \citet{smolec_2005} calibration is tied to the \citet{zw84} metallicity scale and is based on 28 stars:
\begin{equation}
[Fe/H]_{phot} =  -3.142 (\pm0.646)  -4.902(\pm0.375) + 0.824(\pm0.104)\phi_{31}
\end{equation}
\item \citet{Wallerstein_2002} demonstrated a technique based on the Calcium triplet (CaT), which is commonly used in moderate resolution spectroscopy for red giants.
\end{itemize}

Direct measurements, however, can be performed if care is taken in the data acquisition and subsequent analysis. 
Of particular note is a series of papers that use high cadence spectroscopic datasets obtained for the full cycle of several RRL. 
\citet{for_2011a,for_2011b} use an iterative procedure to estimate the temperature for each spectrum and the perform direct spectral analyses for numerous chemical species. 
Both \citet{govea_2014} and \citet{Sneden_2018} continue this work for RRc variables with continued success. 
These direct measurements show relative consistency for stellar parameters and abundances derived self-consistently from the spectrum at all phases. 
Comparison of these parameters to other, more common methods have not yet been produced, but are important.  

\subsubsection{Metallicity Impacts in T2Cs:}

The metallicity effects are less well constrained for T2Cs. 
A major concern, however, is the proposed shifting of $P$-boundaries between the sub-types apparent in the LMC and SMC. 
The shift in sub-population classification would be a major systematic given the differences in the PLs derived with different sub-samples. 

The first chemical abundance analysis for BL~Her and W~Vir type T2Cs was presented in \citet{maas07} for 19 field stars, finding anomalous chemical compositions similar to previous studies of RV~Tau stars by \citet{Giridhar_2005}. 
Subsequent analyses have focused on using spectral features and chemical abundances as means to cleanly separate the sub-populations; examples include, 
(i) \citet{Kovtyukh_2018a} see differences in the metal-poor populations for short period T2Cs ($P$ \textless 3 day), 
(ii) \citet{Kovtyukh_2018b} that compares W~Vir abundances to those of the disk, suggesting the W~Vir stars to be He-enhanced, and, 
(iii) \citet{Lemasle_2015} that compare W~Vir stars to CCs. 
On the whole, the spectroscopic studies are more focused on the revealing the evolutionary origin of these stars rather than their use as distance indicators. 
Along this line-of-thinking, \citet{Welch_2012} note that additional T2C spectroscopic studies, both for radial velocity monitoring and for chemical abundance measurements, could yield significant insight into binary evolution formation channels. 

\subsubsection{Metallicity Impacts in TRGB:} 

The impact of metallicity on the TRGB was discussed in Section \ref{sec:trgb}. 
While theoretical studies see the effects of metallicity cleanly for bolometric quantities, there are uncertainties when translating these effects into observable quantities.
Empirical studies, on the other hand, can measure the color-magnitude behavior, but largely infer the metallicity dependence from the mean metallicity of the system determined from other means \citep[e.g., as in][]{Valenti_2004,dalcanton_2012}.
However, the age and metallicity effects are difficult to disentangle.
A benefit, however, is that the color-magnitude relationship for the TRGB can be constrained empirically from high signal-to-noise observations irregardless of the physical origin \citep[e.g., as in][]{jang_2017a,madore_2018}.
If the color-magnitude relationship holds in systems comprised of different stellar populations and independently determined distances such that systematics can be estimated \citep[e.g., as in][]{jang_2017a}, then impacts can be managed in application to distances.

\subsection{Other Systematics} \label{sec:othersys}

We list briefly other systematic terms and strategies used to compensate for them.
\begin{enumerate}

\item \textit{Milky Way Extinction:}~~The foreground extinction needs to be determined. For extra-galactic objects, the Milky Way foreground is well-enough constrained by \citet{schlegel_1998} maps \citep[noting the recalibration by][]{sch11}. Internal to the Milky Way, the line-of-sight extinction is more difficult to constrain. 
Means of addressing this issue include: (i) RJCE method \citep{majewski_2011}, (ii) models \citep[e.g.,][]{Drimmel_2003}, (iii) use of Wesenheit formulations \citep[e.g.,][]{madore_1982}, and (iv) using longer wavelengths where extinction has a lower total impact. 
\item \textit{Internal Extinction:}~~Tracers that are located inside external galaxies have an additional component from the extinction due to the external galaxy. This is difficult to constrain independently. Means of addressing this issue include: (i) use of Wesenheit magnitudes \citep{madore_1982}, (ii) using longer wavelengths where extinction has a lower impact, (iii) using multi-wavelength observations to fit for total extinction directly, (iv) avoiding galactic structures where extinction is likely to be a significant effect (e.g., stellar halos). These techniques has been employed extensively for CCs, with the volume of literature growing for the RRLs.
\item \textit{Non Universality of the Extinction Law:}~~ There is some evidence that ``Cardelli's Law'' is not universal, in particular in regions of high gas density (see discussions regarding M\,4 in Sections \ref{sec:rrlcase_nir} and \ref{sec:rrlcase_mir}). Wesenheit magnitudes are not useful in this situation because they have to assume a form of the extinction law. Thus, only observations at longer wavelengths or observations where dust is not anticipated (e.g., stellar halos) are robust to this systematic.
\item \textit{Crowding or Blending:} Measurements undertaken in crowded fields have to contend with flux-contamination from neighboring sources. There are algorithms that can attempt to account for this effect. Photometry packages have been designed with measurement techniques, the following packages are notable: 
(i) the {\sc DAOPHOT} family of routines does a particularly good job in crowded regions \citep{Stetson1987,Stetson1994}, (ii) {\sc DoPhot} \citep{Schechter_1993}, and (iii) {\sc Dolphot} \citep{dolphin_2000}. 
\citet[][and references therein]{riess_2016} use a modeling approach that uses higher resolution optical images to guide source isolation at lower resolutions (typically, longer wavelengths), but this assumes that the same sources are detected in both bands and, may, neglect the fact that different stellar populations dominate the light in optical and IR bandpasses. Template fitting may help to identify variable stars that have abnormal shapes due to contamination.  
\textbf{A recent instance of crowding strongly biasing a distance measurement is discussed by \citet{vanDokkum_2018}, who demonstrates how blended stars may have led \citet{Trujillo_2018} to infer the wrong distance to the galaxy NGC1052–DF2, a galaxy suspected of having a low dark matter mass \citep{vanDokkum_nature}.} Thus, the best means of avoiding crowding/blending is to avoid high-density regions of galaxies to which Pop~II distance indicators are particularly well suited.
\item \textit{Evolutionary Effects:} These effects largely include mixing stars of the same type that are in different evolutionary states due to evolution within the class (e.g., first and second pass HB stars), evolution over time (changes to He abundance), and the possibility that some pulsators come from binary evolution formation paths. These types of effects can either add scatter to a population when their numbers are small or strongly bias a distance when their numbers are comparable to non-evolved sub-types. Sculptor (Section \ref{sec:rrlcase_opt}) and $\omega$-Centauri \citep{braga_2018} with their wide metallicity and age spreads are good examples of where evolutionary effects may come into play to skew the means; in populations, however, these outliers can often be identified. Strong biases are more likely when working with individual field stars, where inferring evolutionary state is more complex and, indeed, this is cited for reasons why individual parallaxes might disagree with other forms of calibration (Section \ref{sec:primarycalibmeth}). Addressing systematics of this nature can only be accomplished by calibrating the standard candles with large statistical samples and in a variety of environments where various impacts are more or less likely to be a factor (e.g., comparison of calibrations determined using field stars, GGCs, and dwarf galaxies).
\item \textit{Contamination:} Samples can be contaminated by non-standard candles. For the TRGB, contamination can come from AGB stars or from younger red, luminous populations. These effects are mitigated by applying the technique in Pop~II dominated regions or by increasing uncertainties. Both RRL and T2C populations can be contaminated by variables with similar periods (indeed the T2Cs and RRL overlap with other variable types) or by unidentified atypical pulsators (e.g., pW~VIr, multi-periodic RRLs, Blashko stars). Moreover, both period determination and variable classification are difficult with sparsely sampled data. It is not uncommon for variables to be reclassified with additional data \citep[e.g., see notes in][]{Clement_2001}.
\end{enumerate}
%

\section{Summary and Future Prospects} \label{sec:future}

We have developed complementary physical and empirical portraits of three Pop~II standard candles: the RRLs, the T2Cs, and the TRGB. 
In particular, this chapter has put emphasis the history and evolution of these stars as distance indicators, with a particular focus on multi-wavelength use of these tracers.
In the course of preparing this chapter, parallels between these Pop~II distance indicators have become apparent.
There are several ``themes'' that can be used to discuss our current understanding of these standard candles and these ``themes'', in turn, relate directly into the future potential of these objects in the coming years. 
Thus, we will summarize the chapter and comment on the future prospects in tandem.

First, each of these distance indicators has a deep history dating back to the early days of stellar population research; indeed, often to before the term ``stellar populations'' had been coined \citep[e.g.,][]{baade_1944,baade_review}. 
As a consequence, empirical evidence and theoretical understanding have leap-frogged with each other over time.
These Pop~II standard candles have each played a significant role in our understanding of the structure of our Galaxy and of the Universe and have had their role in refining the distance scale. 
Moving forward, it is vital to not forget the formation of the historical nomenclature and the drivers of the historical thinking that led exploration (or a lack of exploration) for these distance indicators. 
Stated differently, it is important to look back to understand when and why assumptions were made or not made as we work with ever better datasets in which previously ``negligible'' effects may be transformed into major hurdles. In our historical discussions, we hope to have illuminated some of these previously hidden effects.

Second, none of these Pop~II standard candles have a robust geometric foundation.
While the RRL and T2C have some parallaxes from targeted studies, these are few in number and cannot fully encompass the intrinsic star-to-star variance nor account for systematic differences (e.g., those due to metallicity or age). 
Moreover, without a large geometric parallax sample, it is impossible to measure all of the desired quantities for even the simplest calibration (e.g., slopes and zero points).
Instead, it is typical to combine theoretical, empirical, and semi-empirical techniques to determine distances to take into account systematics; more specifically, making combinations of slopes and zero-points either measured in different objects or that are predicted from theoretical modeling. 
While these are incredibly logical to the experienced practitioners, it has the impact of being confusing and inelegant to the non-expert, while also potentially imparting systematic effects between studies. 

With the pending data releases from \emph{Gaia}, however, this theme has a bright future. 
The end-of-mission predictions for \emph{Gaia} are 10\% parallax uncertainties at a distance of 10~kpc \citep{debruijne_2014}, which will provide an unprecedented sample of stars from which to establish a geometric foundation for each of our Pop~II indicators. 
In addition, \emph{Gaia} will provide precise spectro-photometry for all sources, variability information, and, for bright ($G$~\textless~15) sources, spectroscopic parameters (including, chemical abundances and temperatures). 
The enormity of the \emph{Gaia} dataset is simply mind-boggling and has the potential to reveal not just the practical application of these tools, but also elucidate aspects of stellar theory. 
During the preparation of this manuscript, \emph{Gaia} DR2 occurred and some of the early results have been included herein, which is undoubtedly only a small fraction of those to occur.

Third, the most pressing questions in both theory and in practice are related to the need to fundamentally understand the impact of metallicity, age, and evolutionary effects for using  these stars as standard candles. 
Addressing these issues will come on two fronts: (i) observations of individual field stars and (ii) observations of populations in star clusters or nearby dwarf galaxies. 
Thus, addressing this issue will likely require more than just \emph{Gaia} generated datasets.  

For the field stars, diving into stellar parameters (temperature, $\log$(g), and chemical abundances) will likely involve the full pantheon of large scale spectroscopic surveys, including but not limited to APOGEE \citep{majewski_2017}, RAVE \citep{Steinmetz_2006}, LAMOST \citep{Zhao2012}, GALAH \citep{martell_2017}, and their numerous successors. 
These surveys, however, have to properly target and process data for stars whose fundamental parameters or observed radial velocity are changing on timescales as short as 10 minutes (for the RR Lyrae). 
Thus, standard processing may not be sufficient to provide high fidelity measurements.
These types of data will fundamentally improve models, while also helping to push the precision of these standard candles to new ends. 
Moreover, fully complementary insight can be gleaned from precision high cadence photometry, in particular asteroseismic analyses enabled from \emph{Kepler} \citep{kepler}, \emph{K2} \citep{k2}, and \emph{TESS} \citep{tess}. 
These photometric datasets provide an alternate means of determining some stellar parameters via asteroseismology, while also providing short and long timescale variability information to better identify contaminating populations. 

For clusters and satellites, many of these large scale programs do not always have sufficient sensitivity and resolution. 
Here the greatest insight is likely to be gained by levering insights from the field star analyses to larger aperture facilities, from \emph{JWST} in space, to efficient multi-plexing on 5- to 10-meter class and high spatial and spectral resolution studies enabled on 30-meter class ground-based facilities. 
Moreover, continued long-term monitoring of variables, as with OGLE \citep[e.g.,][]{Soszynski_2008}, Cluster AgeS Experiment \citep[e.g.,][]{Kaluzny_2005}, and the homogeneous photometry series \citep[e.g.,][]{Stetson2014}, provide additional characterization and classification diagnostics for these variables. 
With the ability to determine ages and star formation histories, clusters and nearby galaxies are ideal to study population-effects, while also being able to cleanly resolve individual stars.

Fourth, these standard candles are reaching to unprecedented precisions and accuracies that enable powerful scientific explorations, from the structure of our Galaxy, our galactic neighborhood, and ultimately to the measurement of fundamental cosmological parameters.
By utilizing IR observations, the impact of extinction has been minimized, and, for all of the distance indicators described here, provide smaller intrinsic variations, which in turn make these distance indicators more powerful. 
While few percent distances are incredible (especially to individual RRL or T2C), it is also a realm where every \textit{source of uncertainty matters}, from differences in the detailed definition of filter sets, to the unintentional inclusion of peculiar variables, to the potential for multiple formation channels, to the deviations from a standard extinction law. 
Stated differently, \textit{every detail matters}.
Moreover, an important consideration is the measure of the intrinsic dispersion in these relationships or for a given class of standard candle as a fundamental limit to their precision. 
For instance, {\it how stable is the TRGB magnitude?}, {\it how flat is the horizontal branch?}, \textbf{ {\it how standard are our standard candles?}, and {\it how homogeneous are our samples?} }
These are fundamental aspects to the measurement of distances, and quantities that can be estimated, but whose measure is nuanced and complex. 

The stage for better exploiting the Pop~II standard candles is set, in particular with the strong inter-relationship between the theory and observation. 
With the dominance of IR-focused observational facilities in the future, there is great potential for the RRL, T2Cs, and TRGB, all of which become more precise and more powerful in the IR regime. 
Coupled with the potential to probe nearly all galactic structures and galaxies of all Hubble types, and nearly all total luminosities, the Pop~II standard candles stand to sharpen our view of the Universe. 

\begin{acknowledgement}
The authors warmly acknowledge the hospitality of ISSI-BJ for an engaging conference and the conference organizer Richard De Grjis for his leadership in completing these chapters.
We further thank the anonymous referees for useful comments and perspectives that have improved and broadened this review.
RLB acknowledges many insightful discussions with Barry Madore, Wendy Freedman, and the Carnegie-Chicago program team, as well as helpful conversations and data for figures from Mark Seibert, Erika Carlson, Andrew Monson, Victoria Scowcroft, Julianne Dalcanton, and Ben Williams. 
Support for this work was provided by NASA through Hubble Fellowship grant \#51386.01 awarded to RLB by the Space Telescope Science Institute, which is operated by the Association of  Universities for Research in Astronomy, Inc., for NASA, under contract NAS 5-26555. 
GB thanks A Severo Ochoa research grant at the Instituto de Astrofisica de Canarias, where part of this manuscript was written.
VFB warmly thank Prof.~R.P.~Kudritzski for many useful discussions concerning stellar atmospheres and non-LTE effects in giant stars and Dr.~P.B~ Stetson for his superb-quality photometric reduction work, which has been a key component of the work presented here. 
VFB acknowledges PRIN-INAF 2011 "Tracing the formation and evolution of the
Galactic halo with VST" (P.I.: M. Marconi), PRIN-MIUR (2010LY5N2T) "Chemical and dynamical evolution of the Milky Way and Local Group galaxies" (P.I.: F. Matteucci).
VFB thank the Japan Society for the Promotion of Science for a research grant (L15518) and the support from FIRB 2013 (grant: RBFR13J716). We also thank the Education and Science Ministry of Spain 
(grants AYA201016717). VFB finally acknowledge the financial support from the PO FSE Abruzzo 2007-2013 through the grant ``Spectrophotometric characterization of stellar populations in Local Group dwarf galaxies", prot.89/2014/OACTe/D (PI: S. Cassisi).
GF has been supported by the Futuro in Ricerca 2013 (grant RBFR13J716).
CEMV and MM acknowledges support from the Spanish Ministry of Economy and Competitiveness (MINECO) under the grant (project reference AYA2014-56795-P)
NM is grateful to Grant-in-Aid (KAKENHI No.~26287028) from the Japan Society for the Promotion of Science.
\end{acknowledgement}

\clearpage

%
\end{document}